\documentclass[12pt]{article}

\usepackage{array} 
\usepackage{epsfig}
\usepackage{amssymb}
\usepackage{amsmath}                          
\usepackage{graphics,graphpap,color}

\renewcommand{\thefootnote}{\fnsymbol{footnote}}

\newcommand{\db}{\bar{d}}
\newcommand{\ub}{\bar{u}}

\newcommand{\bra}{\langle}
\newcommand{\ket}{\rangle}
\newcommand{\ds}{\displaystyle}
\newcommand{\fplpi}{f_+^{B\to\pi}}

\begin{document}

\begin{titlepage}
\begin{flushright}
\begin{tabular}{l}
IPPP/04/23 \\
DCPT/04/46 \\
TPI-MINN-04/25 \\
\end{tabular}
\end{flushright}
\vskip1.5cm
\begin{center}
   {\Large \bf \boldmath 
    New Results on $B\to\pi,K,\eta$ Decay Formfactors\\[5pt] from
    \text{Li}ght-Cone Sum Rules}
    \vskip1.3cm {\sc
Patricia Ball\footnote{Patricia.Ball@durham.ac.uk}$^{,1}$ and 
    Roman Zwicky\footnote{zwicky@physics.umn.edu}}$^{,2}$
  \vskip0.2cm
        $^1$ IPPP, Department of Physics, 
University of Durham, Durham DH1 3LE, UK \\
\vskip0.1cm 
$^2$ William I. Fine Theoretical Physics Institute, \\ University of
Minnesota, Minneapolis, MN 55455, USA  
\vskip2cm

\vskip4.5cm

{\large\bf Abstract:\\[10pt]} \parbox[t]{\textwidth}{ 
We present an improved calculation of $B\to$ light pseudoscalar 
formfactors from light-cone sum rules, including one-loop radiative
corrections to twist-2 and twist-3 contributions, and leading order
twist-4 corrections. The total theoretical uncertainty of our
results at zero momentum transfer is 10 to 13\% and can be improved,
at least in part, by
reducing the uncertainty of hadronic input parameters, in
particular those describing the twist-2 distribution amplitudes of
the $\pi$, $K$ and $\eta$. We present our results in a way which details the
dependence of the formfactors on these parameters and facilitates the
incorporation of future updates of their values from
e.g.\ lattice calculations.
}

\vfill

{\em Submitted to Physical Review D}
\end{center}
\end{titlepage}

\setcounter{footnote}{0}
\renewcommand{\thefootnote}{\arabic{footnote}}

\newpage

\section{Introduction}\label{sec:intro}
This paper aims to give 
a new and more precise determination of the
decay formfactors of $B$ mesons into light pseudoscalar mesons, i.e.\
$\pi$, $K$ and $\eta$. We do not include the $\eta'$ which is too
heavy to be treated in this framework. 
The calculation uses the method of QCD sum
rules on the light-cone, which in the past has been rather
successfully applied to various problems in heavy-meson physics,
cf.~Refs.~\cite{protz,maxcit,PB98,ballroman,gBBpi}\footnote{See also
  Ref.~\cite{LCSRs:reviews} for reviews.}; 
an outline of the method will be given
below. Our calculation improves on our previous papers
\cite{PB98,ballroman} by 
\begin{itemize}
\item including radiative corrections to twist-3 contributions to
  one-loop accuracy, for all formfactors;
\item a precisely defined method for fixing the sum rule specific
  parameters;
\item using updated values for input parameters;
\item a careful analysis of the uncertainties of the formfactors at
  zero momentum transfer;
\item a new parametrization of the dependence of the formfactors on
  momentum transfer, which is consistent with the constraints from
analyticity and  heavy-quark expansion;
\item detailing the dependence of formfactors on nonperturbative
  hadronic parameters describing the $\pi$, $K$, $\eta$ mesons, the
  so-called Gegenbauer moments, which facilitates the incorporation of future
  updates of their numerical values and also allows a consistent treatment
  of their effect on nonleptonic decays treated in QCD factorisation.
\end{itemize}
The motivation for this study is twofold and related to the overall
aim of $B$ physics to provide precision determinations of quark flavor
mixing parameters in the Standard Model. Quark flavor mixing is
governed by the unitary CKM matrix which depends on four parameters: three
angles and one phase. The constraints from unitarity can be visualized
by the so-called unitarity triangles (UT); the one that is relevant
for $B$ physics is under intense experimental study. 
The over-determination of the sides
and angles of this triangle from a multitude of processes will answer
the question whether there is new physics in flavor-changing
processes and where it manifests itself. One of the sides of the UT is
given by the ratio of CKM matrix elements $|V_{ub}/V_{cb}|$.
$|V_{cb}|$ is known to about 2\% accuracy from both inclusive and
exclusive $b\to c\ell \nu$ transitions \cite{CKMWS}, whereas the
present error
on $|V_{ub}|$ is much larger and around $15\%$. Its reduction requires
an improvement of experimental statistics, which is under way at the $B$
factories BaBar and Belle, but also and in particular an improvement
of the theoretical prediction for associated semileptonic spectra and
decay rates. This is the first motivation for our study of the
$B\to\pi$ decay formfactor $\fplpi$, which, in conjunction with
alternative calculations, in particular from lattice \cite{lattice},
will help to reduce the uncertainty from exclusive semileptonic 
determinations of $|V_{ub}|$. Secondly, formfactors of general
$B\to\,$light meson transitions are also needed as ingredients in the
analysis of nonleptonic two-body $B$ decays, e.g.\ $B\to K\pi$, 
in the framework of QCD factorization
\cite{QCDfac}, again with the objective to extract CKM parameters. 
 One issue calling for particular attention in this context
is the effect of SU(3) breaking, which enters
both the formfactors and the $K$ and $\eta$ meson distribution
amplitudes figuring in the factorization analysis. We would like to
stress here that the implementation of SU(3) breaking in the
light-cone sum rules approach to formfactors is {\em precisely} the
same as in QCD factorization and is encoded in the difference between
$\pi$, $K$ and $\eta$ distribution amplitudes, so that the use of
formfactors calculated from light-cone sum rules together with the
corresponding meson distribution amplitudes in factorization
formulas allows a unified
and controlled approach to the assessment of SU(3) breaking effects in
nonleptonic $B$ decays.

As we shall detail below, 
QCD sum rules on the light-cone allow the calculation of formfactors
in a kinematic regime where the final state meson has large energy in
the rest-system of the decaying $B$,
$E\gg \Lambda_{\rm QCD}$. This is in contrast to lattice calculations
which presently are available only for $B\to\pi$ and 
$q^2>15\,$GeV$^2$, due to the restriction to $\pi$ energies smaller
than the inverse lattice spacing.\footnote{This situation may change
  in the future
  with the successful implementation of ``moving NRQCD''
  \cite{davies}, where the $B$ decays while moving ``backwards'',
  which gives access to smaller values of $q^2$ without increasing
 the  discretisation error.} First unquenched results are underway 
\cite{davies,shigemitsu}, which, once published,
 will allow one to exploit the
complementarity of lattice simulations and light-cone sum rules in more detail.

The physics underlying $B$ decays into light mesons at large momentum
transfer can be understood qualitatively in the framework of hard
exclusive QCD processes, pioneered by Brodsky and Lepage et al.\ 
\cite{pQCD}. The hard scale in $B$ decays is $m_b$ and one can show
that to leading order in $1/m_b$ the decay is described by two
different parton configurations: one 
where all quarks have large momenta and the momentum transfer
happens via the exchange of a hard gluon, the so-called hard-gluon
exchange, and a second one where one quark is soft and does
interact with the other partons only via soft-gluon exchange, the
so-called soft or Feynman-mechanism. The consistent treatment of both
effects in a framework based on factorization, i.e.\ the clean separation of
perturbatively calculable hard contributions from nonperturbative
``wave functions'', is highly nontrivial and has spurred the
development of SCET, an effective field theory which aims to separate
the two relevant large mass scales $m_b$ and $\sqrt{m_b\Lambda_{\rm
    QCD}}$ in a systematic way \cite{SCET}. 
In this approach formfactors can indeed be split into
a calculable factorizable part which roughly corresponds to the hard-gluon
exchange contributions, and a nonfactorizable one, which includes the soft
contributions and cannot be calculated within the SCET
framework \cite{fuck}. 
Predictions obtained in this approach then typically aim to eliminate the soft
part and take the form of relations
between two or more formfactors whose difference is expressed in terms
of factorizable contributions. 

The above discussion highlights the need for a calculational
method that allows numerical predictions while treating both 
hard and soft contributions on the same footing. It is precisely QCD
sum rules on the light-cone (LCSRs) that accomplish this task.
LCSRs can be viewed as an extension of the original 
method of QCD sum rules devised by
Shifman, Vainshtein and Zakharov (SVZ) \cite{SVZ}, which was
designed to determine properties of
ground-state hadrons at
zero or low momentum transfer, to the regime of large
momentum transfer. QCD sum rules combine the
concepts of operator product expansion, dispersive representations of
correlation functions and quark-hadron duality in an ingenuous way
that allows the calculation of the properties of non-excited 
hadron-states with a very reasonable theoretical uncertainty. 
In the context of weak-decay formfactors, the basic quantity is the 
correlation function of the weak current and a 
current with
the quantum numbers of the $B$ meson, evaluated between the vacuum and
a light meson.
For large (negative) virtualities of these currents, the
correlation function is, in coordinate-space, dominated by distances
close to the light-cone and can be discussed in the framework of
light-cone expansion. In contrast to the short-distance expansion
employed by conventional QCD sum rules \`a la SVZ where
nonperturbative effects are encoded in vacuum expectation values
of local operators with
vacuum quantum numbers, the condensates, LCSRs
rely on the factorization of the underlying correlation function into
genuinely nonperturbative and universal hadron distribution amplitudes (DAs)
$\phi$ which are convoluted with process-dependent amplitudes $T$.
The latter are the analogues of the Wilson-coefficients in the
short-distance expansion and can be
calculated in perturbation theory. The light-cone expansion then reads, 
schematically:
\begin{equation}\label{eq:schemat}
\mbox{correlation function~}\sim \sum_n T^{(n)}\otimes \phi^{(n)}.
\end{equation}
The sum runs over contributions with increasing twist, labelled by
$n$, which are suppressed by
increasing powers of, roughly speaking, the virtualities of the
involved currents.
The same correlation function can, on the other hand, be written as a
dispersion-relation, in the virtuality of the current coupling to the
$B$ meson. Equating dispersion-representation and the
light-cone expansion, and separating the $B$ meson contribution from
that of higher one- and multi-particle states using quark-hadron
duality, one obtains a relation
for the formfactor describing the decay $B\to\,$light meson.

Our paper is organized as follows: in Sec.~2 we define all relevant
quantities, in particular correlation functions and meson distribution
amplitudes. In Sec.~3 we outline our calculations and derive
the light-cone sum rules. In Sec.~4 we present our numerical
results and Sec.~5 contains a summary and conclusions. Detailed
expressions for distribution amplitudes and explicit formulas
for the light-cone sum rules are given in the appendices.

\section{Definitions}

The formfactors $f_+^P$, $f_0^P$ and $f_T^P$ which are relevant for
the $B\to P$ transition, where $P$ stands for $\pi$, $K$ or $\eta$,
are defined as 
follows:\footnote{The following  notations are frequently used in the 
literature: $f_+=F_1$ and $f_0=F_0$.}
\begin{alignat}{2}
\label{eq:fplus}
& \bra P(p)|V^P_{\mu}|B(p_B) \ket &\,=\,& \big\{(p+p_B)_{\mu}-
\frac{m_B^2-m_P^2}{q^2}\,q_{\mu}\big\}
f^P_+(q^2)+\big\{\frac{m_B^2-m_P^2}{q^2}\,q_{\mu}\big\}  
\,f^P_0(q^2), \\\label{eq:fT}
& \bra P(p)|J^{P,\sigma}_{\mu}|B(p_B)\ket &\,=\,& 
\frac{i}{m_B+m_P}\big\{q^2(p+p_B)_{\mu}-
(m_B^2-m_P^2)q_{\mu}\big\} f_T^P(q^2,\mu),
\end{alignat}
where \mbox{$V^{\pi,\eta}_{\mu} = \ub \gamma_{\mu} b$} is the
standard weak current, $V^K_\mu$ is given by 
\mbox{$V^{K}_{\mu} = \bar s \gamma_{\mu} b$} and 
$J_{\mu}^{\pi(\eta),\sigma} = \db \sigma_{\mu\nu}q^{\nu} b$,
$J_{\mu}^{K,\sigma} = \bar s \sigma_{\mu\nu}q^{\nu} b$ are
penguin currents. The momentum transfer is given by $q=p_B-p$ and
the physical range in $q^2$ is \mbox{$0 
\leq q^2 \leq (m_B-m_{P})^2$.} The formfactors $f^P_+$ and $f^P_0$
are independent 
of the renormalization scale $\mu$ since $V_{\mu}$ is
a physical current, in contrast to the penguin current $J_{\mu}^{\sigma}$ . 
Note that $f^P_+(0)=f^P_0(0)$ which is a
consequence of  the parametrization chosen in  
Eq.~\eqref{eq:fplus}. We assume SU(2) isospin symmetry
throughout this work, i.e.\ we do not distinguish
$\bar B^0\to\pi^+$ and $B^-\to\pi^0$ formfactors etc.

In the semileptonic decay $B \to \pi l\nu_l$ the formfactor $f^\pi_0$
 enters proportional to the
lepton mass $m_l^2$ and hence is irrelevant for light leptons 
($l=e,\mu$), where only $f^\pi_+$ matters.
The semileptonic decay can be used to determine the size of 
the CKM matrix element $|V_{ub}|$ from the spectrum
\begin{equation}\label{eq:spectrum}
\frac{d\Gamma}{dq^2}(B \to \pi l \nu_l) = \frac{G_F^2 |V_{ub}|^2}{
192\pi^3m_B^3}\lambda(q^2)^{3/2} |f^\pi_+(q^2)|^2 \quad,
\end{equation}
where $\lambda(x) = (x+m_B^2-m_{\pi}^2)^2-4m_B^2m_{\pi}^2$. The
formfactor $f^\pi_0$ will be relevant in and
can be measured from the decay $B \to \pi \tau
\nu_{\tau}$. $f^\pi_T$ is relevant
for the rare decay $B \to \pi l^+l^-$, where the penguin current features 
in the effective Hamiltonian of the process.   

Our starting point for calculating the formfactors $f_{+,0}^\pi$ 
is the correlation function
\begin{eqnarray}
\label{eq:corr}
\Pi_{\mu}(q,p_B) &=& i\int d^4x e^{i q\cdot y} \bra \pi(p)| 
T V_{\mu}(x) j_B^{\dagger}(0) |0 \ket \\
&=& \Pi_+(q^2,p_B^2)(p+p_B)_{\mu} + \Pi_-(q^2,p_B^2) q_{\mu} \quad ,\nonumber
\end{eqnarray}
where $j_B= m_b \db i\gamma_5 b$ is the interpolating field for the
$B$ meson. For the calculation of  $f^\pi_T$, $V_{\mu}$ has to be replaced by 
$J_{\mu}^{\sigma}$. For virtualities
\begin{equation}\label{eq:virt}
m_b^2-p_B^2 \geq O(\Lambda_{\rm QCD}m_b), \qquad
m_b^2-q^2 \geq O(\Lambda_{\rm QCD}m_b),
\end{equation}
the correlation function \eqref{eq:corr}
is dominated by light-like distances
and therefore accessible to an expansion around the light-cone. 
The above conditions can be understood by demanding that the exponential 
factor in \eqref{eq:corr} vary only slowly. The light-cone expansion
is performed by integrating out the transverse and ``minus'' degrees
of freedom and leaving only the longitudinal momenta of the partons as
relevant degrees of freedom. The integration over transverse momenta
is done up to a cutoff, $\mu_{\rm IR}$, all momenta below which are
included in a so-called hadron distribution amplitude $\phi$, whereas
larger transverse momenta are calculated in perturbation theory. The
correlation function is hence decomposed, or factorized, in
perturbative contributions $T$ and nonperturbative contributions
$\phi$, which both depend on the longitudinal parton momenta and the
factorization scale $\mu_{\rm IR}$. If the $\pi$ is an effective
quark-antiquark bound state, as is the case to leading order in the
light-cone expansion, we can write the corresponding longitudinal
momenta as $up$ and $(1-u)p$, $p$ being the momentum of the $\pi$. 
The schematic relation
\eqref{eq:schemat} can then be written 
in more explicit form as
\begin{equation}
\label{eq:lcexp}
\Pi_+(q^2,p_B^2) = \sum_n \int_0^1 du \,
T^{(n)}(u,q^2,p_B^2,\mu_{\text{IR}}) \phi^{(n)}(u,\mu_{\text{IR}}).
\end{equation}
As $\Pi_+$ itself is independent of the arbitrary scale
$\mu_{\text{IR}}$, the scale-dependence of $T^{(n)}$ and $\phi^{(n)}$ 
must cancel
each other.\footnote{If there are more than one contributions of a
  given twist, they will mix under a change of the factorization scale
$\mu_{\rm IR}$ and it is only in the sum of all such contributions
  that the residual $\mu_{\rm IR}$ dependence cancels.} 
If $\phi^{(n)}$ describes the meson in a two-parton state, it is
called a two-particle distribution amplitude (DA), if it describes 
a three-parton, 
i.e.\ quark-antiquark-gluon state, it is called three-particle DA. In
the latter case the integration over $u$ gets replaced by an
integration over two independent momentum fractions, say $\alpha_1$
and $\alpha_2$.
Eq.~(\ref{eq:lcexp}) is called a ``collinear'' factorization formula,
as the momenta of the partons in the $\pi$ are collinear with the
$\pi$'s momentum,
and its validity actually has to be verified. We will come back to
that issue in the next section.

Let us now define the distribution amplitudes to be used in this
paper. Again we
only quote formulas for the $\pi$ meson, those for the $K$ and 
$\eta$ are analogous. All definitions and  formulas are well-known
 and can be found in Ref.~\cite{wavefunctions}. 
In general, the distribution amplitudes we are
interested in are related
to nonlocal matrix elements of type 
$$\bra 0 | \bar u(x) \Gamma [x,-x] d(-x)|\pi(p)\rangle \quad \mbox{or}\quad
\bra 0 | \bar u(x) [x,vx]\Gamma G^a_{\mu\nu}(vx)\lambda^a/2 [vx,-x] 
d(-x)|\pi(p)\rangle.
$$
$x$ is light-like or close to light-like and the light-cone expansion
is an expansion in $x^2$; $v$ is a number between 0 and 1 and $\Gamma$
a combination of Dirac matrices. The expressions $[x,-x]$ etc.\ denote 
Wilson lines that are needed to render the matrix elements, and hence
the DAs, gauge-invariant. One usually works in the convenient
Fock-Schwinger gauge $x^\mu A^a_\mu(x)\lambda^a/2 = 0$, where all Wilson
lines are just $\bf 1 $; 
we will suppress them from now on. The DAs are
ordered by twist, i.e.\ the difference between spin and dimension of
the corresponding operators. We will include DAs of twist-2 (the
leading twist), 3 and 4. 
The leading-twist DA $\phi_\pi$ is defined as
\begin{eqnarray}
\langle 0 | \bar u(x)\gamma_\mu\gamma_5 d(-x)|\pi^-(p)\rangle\
  &=&  
  if_\pi p_\mu \int_0^1 du \, e^{i\zeta p\cdot x} \left[ \phi_\pi(u) +
  \frac{1}{4}\, m_\pi^2 x^2 {\mathbb A}(u)\right] \nonumber \\ &+& i f_\pi
  \frac{m_\pi^2}{px}\, x_\mu \int_0^1 du \, e^{i\zeta p\cdot x}\,
g_\pi(u) + O(x_\mu x^2)\label{eq:22}
\end{eqnarray}
with $\zeta \equiv 2u-1$ and $p^2=0$. The above matrix element also
contains two twist-4 DAs, $g_\pi$ and $\mathbb A$. The variable $u$
can be interpreted as the momentum fraction carried by the quark (as
opposed to the antiquark) in the meson.

There are two two-particle twist-3 DAs, $\phi_p$ and $\phi_\sigma$,
which are defined as
\begin{eqnarray}
\label{eq:32a}
\bra 0| \ub(x)i\gamma_5 d(-x) |\pi(p) \ket &=& \mu_{\pi}^2 
\int_0^1 du\, e^{i\zeta p\cdot x} \phi_{p}(u)\,,\\
\label{eq:32b}
\bra 0| \ub(x)i\sigma_{\mu\nu}\gamma_5 d(-x) |\pi(p) \ket & = & 
-\frac{i}{3}\,\mu_{\pi}^2
(1- \rho_{\pi}^2) (p_{\mu}x_{\nu}-x_{\mu}p_{\nu})
\int_0^1 du\, e^{i\zeta p\cdot x}
\phi_{\sigma}(u)\,,
\end{eqnarray}
where $\mu_{\pi}^2 \equiv f_{\pi}m_{\pi}^2/(m_u+m_d)$ and
$\rho_{\pi}^2 \equiv (m_u+m_d)^2/m_{\pi}^2$. 

The precise definitions of three-particle DAs are a bit cumbersome and
given in App.~\ref{app:A}. The salient
feature is that there is one three-particle DA of twist-3 and four of
twist-4.

Although we have introduced not less than 10 different DAs, which are
all nonperturbative quantities, it may seem, at first glance, that
light-cone sum rules do not retain much predictive power.
Fortunately, however, it turns out 
that the DAs are highly constrained functions
which can be analysed in the framework of conformal expansion, a
topic being discussed in App.~\ref{app:A}. The main result is that,
to next-to-leading order in conformal expansion, which is sufficient
for the accuracy we are aiming at, all 10 DAs can be expressed in
terms of 7 independent hadronic parameters.

This completes the definitions necessary for the calculation of
formfactors.

\begin{figure}
$$\epsffile{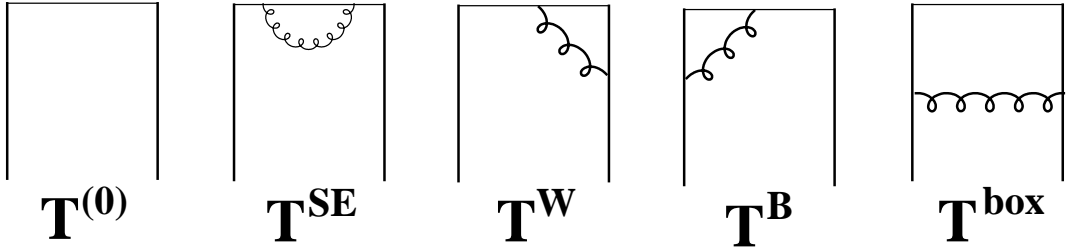}$$
\caption[]{Perturbative contributions to the correlation function
  $\Pi$. The external quarks are on-shell with momenta $up$ and
  $(1-u)p$, respectively.}\label{fig:1}
\end{figure}

\section{The Sum Rules}

The diagrams to be calculated to $O(\alpha_s)$ for two-particle DAs 
are shown in Fig.~\ref{fig:1}. The quark
(antiquark) is collinear with the light meson and carries momentum
$up$ ($(1-u)p$). 
Quarks are projected onto the corresponding distribution amplitudes
using the completeness relation
\begin{eqnarray*}
\bar u_a d_b  &=&  \frac{1}{4}\, ({\bf 1})_{ba} (\bar u d) -
\frac{1}{4} \, (i\gamma_5)_{ba} (\bar u i\gamma_5 d) + \frac{1}{4}\,
(\gamma_\mu)_{ba} (\bar u \gamma^\mu d) - \frac{1}{4}\,
(\gamma_\mu\gamma_5)_{ba} (\bar u \gamma^\mu \gamma_5 d)\\
&&{} +
\underbrace{\frac{1}{8}\,(\sigma_{\mu\nu})_{ba} (\bar u
  \sigma^{\mu\nu} d)}_{\equiv
  -\frac{1}{8}\,(\sigma_{\mu\nu}i\gamma_5)_{ba}
(\bar u \sigma^{\mu\nu}i\gamma_5 d)}.
\end{eqnarray*}
The diagrams are calculated in momentum space. The terms in $x_\mu$ in
the contribution of $\phi_\sigma$, Eq.~(\ref{eq:32b}), are rewritten
in terms of derivatives
$$
x_\mu \to -i\,\frac{\partial}{\partial(up)_\mu}.
$$
In the previous section we mentioned that the fact that $\Pi$ can be
written in factorized form can not be taken for granted, but requires
proof. We do not attempt to give a proof to all orders in
$\alpha_s$, although that should be possible using the techniques of
SCET, but restrict ourselves to $O(\alpha_s)$ in twist-2, to all
orders in the conformal expansion, and to 
$O(\alpha_s)$ and leading order in the
conformal expansion for twist-3. The proof 
essentially relies on the cancellation of
singularities, of which there are several possible types: infrared and
ultraviolet singularities arising from loop calculations and so-called
soft singularities which occur when the integral over $u$ in
Eq.~(\ref{eq:lcexp}) does diverge at the endpoints. The latter
divergences have actually posed a severe problem in early attempts to treat
$f_+^\pi$ in QCD factorization: in Ref.~\cite{Brodsky} only the hard
gluon exchange was included, which yields a logarithmic divergence for
the parton configuration where the $u$ quark emerging from the weak
decay carries essentially all pion momentum. As we understand now,
this divergence disappears when contributions from the
Feynman-mechanism are added. In our case, it turns out that all
  $T$ are regular at the endpoints $u=0,1$, so there are {\em no soft
divergences, independent of the end-point behavior of the distribution
amplitudes.} As for infrared and ultraviolet singularities,
they can be treated  in
dimensional regularisation. Using the lowest-order expression of the
Brodsky-Lepage evolution kernel for $\phi_\pi$ derived in \cite{pQCD}, 
we have followed the strategy outlined in \cite{Braaten} to check
that the infrared
divergences precisely cancel those contained in the bare DA
$\phi_\pi^{\rm\scriptstyle bare}$. As for twist-3, the evolution
kernel is not known, so we have only checked the cancellation of
infrared divergences of the lowest order term in the conformal
expansion, whose divergent behavior is well known -- in fact, only the
one-loop renormalisation of the quark condensate is needed. The
ultraviolet divergences cancel for $f_+$ and $f_0$, which are
physical formfactors and hence do not depend on the ultraviolet
renormalisation scale; for $f_T$, we reproduce the well-known one-loop
anomalous dimension.

We then have
used the explicit expressions for the twist-2 and 3 two-particle
DAs given in App.~\ref{app:A} to perform the integration over $u$
analytically. Actually it is not the correlation function $\Pi$ itself that
is needed, but its imaginary part, see below. $\Pi$ has a cut in
$p_B^2$ starting at $m_b^2$ and taking the imaginary part after
integration over $u$ is straightforward. The strategy outlined here
is different from the procedure we followed in our previous
papers \cite{PB98,ballroman}, where we took the imaginary part before 
integrating
over $u$. This latter procedure resulted in expressions with a very
complicated analytical structure which made it impossible to give
explicit formulas for the imaginary parts. With our new
procedure we obtain lenghty, but not very complicated expressions; the
complete set of spectral densities $\rho=({\rm Im}\,\Pi)/\pi$ 
for the sum rule for the
formfactor $f_+$ is given in App.~\ref{app:B}.

Armed with the spectral densities, we can derive the LCSR for
e.g.\ the formfactor $f_+$. The basic quantity is $\Pi_+$, which
is calculated in two ways. In light-cone expansion, it can be written
in dispersive representation  as
\begin{equation}\label{xy}
\Pi^{\rm LC}_+(p_B^2,q^2) = \int_{m_b^2}^\infty ds\,\frac{\rho^{\rm
    LC}_+(s,q^2)}{s-p_B^2}
\end{equation}
with the explicit expression for the spectral density $\rho_+^{\rm
  LC}(s)$ given in App.~\ref{app:B}.
This expression has to be compared to the physical correlation function,
which also features a cut in $p_B^2$, starting at $m_B^2$:
\begin{equation}
\Pi^{\rm phys}_+(p_B^2,q^2) = \int_{m_B^2}^\infty ds\,\frac{\rho^{\rm
    phys}_+(s,q^2)}{s-p_B^2};
\end{equation}
the spectral density is given by hadronic contributions and reads
\begin{equation}\label{xz}
\rho^{\rm phys}_+(s,q^2) = f_B m_B^2 f_+(q^2)\delta(s-m_B^2) +
\rho^{\rm\scriptstyle higher-mass~states}_+(s,q^2).
\end{equation}
Here $f_B$ is the $B$ meson decay constant defined as
\begin{equation}\label{fb}
\langle 0 |\bar q \gamma_\mu\gamma_5 b | B\rangle = if_B p_\mu
\quad{\rm or} \quad (m_b+m_q) \langle 0 |\bar q i\gamma_5 b | B\rangle
= m_B^2 f_B.
\end{equation}
To obtain a light-cone sum rule for $f_+$, one equates the two
expressions for $\Pi_+$ and uses quark-hadron duality to approximate
\begin{equation}\label{dunno}
\rho^{\rm\scriptstyle higher-mass~states}_+(s,q^2) \approx \rho_+^{\rm
  LC}(s,q^2)\Theta(s-s_0),
\end{equation}
where $s_0$, the so-called continuum threshold is a parameter to be
  determined within the sum rule approach itself. In principle one
  could now write a sum rule
$$\Pi_+^{\rm\scriptstyle phys}(p_B^2,q^2) = \Pi_+^{\rm LC}(p_B^2,q^2)$$
and determine $f_+$ from it. However, in order to suppress the impact
of the approximation (\ref{dunno}), one subjects
 both sides of the equation to a Borel
  transformation
$$\frac{1}{s-p_B^2}\to 
\hat{B}\,\frac{1}{s-p_B^2} = \frac{1}{M^2}\, \exp(-s/M^2)$$
which ensures that contributions from higher-mass states be sufficiently
suppressed and improves the convergence of the OPE. 
We then obtain 
\begin{equation}\label{srx}
e^{-m_B^2/M^2} m_B^2f_B\;f_+(q^2) = \int_{m_b^2}^{s_0} ds\, e^{-s/M^2} 
\rho_+^{\rm LC}(s,q^2). \quad
\end{equation}
This is the final sum rule for $f_+$; expressions for the other
formfactors are obtained analogously. The task now is to find sets of
parameters $M^2$ (the Borel parameter) and $s_0$ (the continuum
threshold) such that the resulting formfactor does not
depend too much on the precise values of these parameters; in
addition the continuum contribution, that is the part of the
dispersive integral from $s_0$ to $\infty$ that has been subtracted
from both sides of (\ref{srx}), should not be too large, say less than
30\% of the total dispersive integral.

\section{Numerics}\label{sec:4}

In this section we obtain numerical results from the sum rules
(\ref{srx}). The section is organised as follows: in
Sec.~\ref{sec:borelfix} we explain how we determine the sum rule
specific parameters, i.e.\ the Borel parameter $M^2$ and the continuum
threshold $s_0$. We also determine $f_B$, which is a necessary
ingredient in (\ref{srx}). In Sec.~\ref{sec:hadronic} we explain in
more detail how we fix the hadronic input parameters, in particular
the Gegenbauer moments $a_{1,2,4}$ that describe the final state
mesons. In Sec.~\ref{sec:4.3} we calculate the formfactors at $q^2=0$
and discuss their uncertainties. 
In Sec.~\ref{sec:fitting} we present the formfactors for
central input-values of the parameters and provide a simple
parametrization valid in the full kinematical regime of $q^2$.
The results for $q^2=0$ are collected in Tab.~\ref{tab:final} and
Eq.~(\ref{eq:resq2at0}), central results for arbitrary $q^2$ in
Tab.~\ref{tab:fitpars}. More detailed results that allow one
to determine the formfactors for arbitrary values of $m_b$ and the
Gegenbauer moments $a_{1,2,4}$ are
collected in App.~\ref{app:C}.

\subsection{Fixing the Borel Parameter and the Continuum Threshold}
\label{sec:borelfix}
We illustrate our procedure to determine $M^2$ and $s_0$ with the
comparatively simple example of $f_B$, the $B$ decay constant defined
in (\ref{fb}). This example is actually of immediate practical use, as
$f_B$ enters our determination of the formfactors from
Eq.$\,$(\ref{srx}). Since
it is not known from experiment, its value has to be taken from
theoretical calculations -- which basically means either lattice
determinations \cite{fBlatt} or (local) QCD sum rules
\cite{SRfB,fBSR}. To ensure consistency of our calculations, we use
the values of $f_B$ as determined from QCD sum rules to $O(\alpha_s)$
accuracy \cite{SRfB}. The reason for this choice is twofold: firstly,
it is well-known that the use of $f_B$ from sum rules reduces the
dependence of the formfactors on input-parameters, in particular $m_b$
\cite{protz}; secondly, $O(\alpha_s^2)$ corrections to $f_B$ turn out to
be rather large \cite{fBSR}, which was anticipated in the
second reference in \cite{SRfB}, where it was argued that these
corrections are dominated by Coulombic
corrections. Precisely the same corrections also enter the light-cone
expansion of the correlation function $\Pi$, but will largely cancel
in the ratio $f_+\sim\Pi/f_B$. In conclusion, we expect a cancellation
of both large radiative corrections and parameter dependence in the
formfactors when $f_B$ is replaced by its sum rule to $O(\alpha_s)$
accuracy; we do not expect the resulting numerical values of $f_B$ to be
``good'' predictions for that quantity. 

The sum rule for $f_B$ reads \cite{SRfB}\footnote{The contribution of
  the gluon condensate
is not sizable and we therefore neglect it.}
\begin{eqnarray}\label{eq:aa}
f_B^2 m_B^2 e^{-\frac{m_B^2}{M^2}} = \int_{m_b^2}^{s_0} ds\,\rho^{\rm pert}(s)
e^{-\frac{s}{M^2}} + C_{\bar qq} \bra\bar q q \ket + C_{\bar q Gq} 
\bra\bar q \sigma gG q \ket
\equiv \int_{m_b^2}^{s_0} ds\,\rho^{\rm tot}(s)
e^{-\frac{s}{M^2}}.
\end{eqnarray}
The $C$ are the Wilson coefficients multiplying the condensates,
for which we use the following numerical values at $\mu=1\,{\rm GeV}$: 
\begin{equation}\label{eq:conds}\bra\bar q
q\ket = -(0.24\pm0.01)^3\,\text{GeV}^3 \quad\mbox{and} \quad 
{\bra \bar q \sigma gG q \ket}= 0.8\,\text{GeV}^2\bra
\bar q q \ket.
\end{equation}
The condensates (and $\alpha_s$) are actually evaluated at the scale
$M^2$. The criteria for determining $M^2$ and $s_0$ are often not
stated very precisely. In the present context, with many different
formfactors to calculate, which entails the need for a well-defined
procedure to determine the input-parameters for each of them, 
we decide to opt for a precisely defined method to fix the pair
$(M^2,s_0)$ and impose
the following criteria on the sum rule for $f_B$ (and, later on, the
formfactors):
\begin{itemize}
\item the derivative of the logarithm of Eq.$\,$\eqref{eq:aa} with
  respect to $1/M^2$ gives a sum rule for the $B$ meson mass $m_B$:
 $$m_B^2 = \int_{m_b^2}^{s_0} ds\,s\,
\rho^{\rm tot}(s)/\int_{m_b^2}^{s_0} ds\,\rho^{\rm tot}(s).$$
We require this sum rule to be fullfilled to high accuracy $\sim 0.1\%$.
\item the sum rule for $f_B$ is required to exhibit an extremum for a given
  pair $(M^2,s_0)$.
\end{itemize}
These criteria define a set of parameters for each value of $m_b$,
which are collected in Tab.~\ref{tab:fB}, together with the resulting
$f_B$. For all these parameter sets the continuum
contribution (i.e.\ the integral $\int_{s_0}^\infty$) is between
25\% and 30\% of the $B$ contribution and hence well under control.

For the formfactors $f^\pi_+$, $f^\pi_0$ and $f^\pi_T$ we follow the
same procedure 
which results in different values of $M^2$ and $s_0$ for formfactors and $f_B$.
For $K$ and $\eta$ we use 
the same values for the Borel parameter and the continuum threshold.
{}From the explicit formulas of the tree-level sum rules for the
formfactors quoted in e.g.\ the 3rd reference in
\cite{protz}, one finds that the effective Borel parameter is
$uM_{\rm LC}^2$ rather than $M_{\rm LC}^2$.\footnote{We denote the
  Borel parameter of the LCSR
  \eqref{srx} by $M_{\rm LC}^2$ and the Borel parameter of the SR
  \eqref{eq:aa} by $M^2$.} In order to keep this product constant, we rescale
the Borel parameter by $\bra u \ket^{-1}$ by
\begin{equation*}
\bra u \ket(q^2) \equiv \int_{u_0}^{\infty} du \;u \; \phi_{\pi}(u) 
e^{-\frac{m_b^2-(1-u)q^2}{u M^2}} /
\int_{u_0}^{\infty} du\;\phi_{\pi}(u) e^{-\frac{m_b^2-(1-u)q^2}{u
    M^2}}\,, \; u_0=\frac{m_b^2-q^2}{s_0-q^2},
\end{equation*}
resulting in the approximate values
$\bra u \ket (0\,\text{GeV}^2)=0.86$ and 
$\bra u \ket (14\,\text{GeV}^2)=0.77$.
Pa\-ra\-me\-tri\-sing the relation
between the Borel parameters by
\begin{equation}\label{eq:borels}
M_{\rm LC}^2 \equiv c_c M^2/\bra u \ket,
\end{equation}
we obtain the values and
continuum thresholds given in Tab.~\ref{tab:fB}. 

\begin{table}
\begin{center}
\begin{tabular}{|l|cccc||cccc|}
\hline
 & $m_b$ & $s_0$ & $M^2$ & $f_B$ &
 $s_0^+\approx s_0^0$ & $c_c^+$ & $s_0^T$ & $c_c^T$  \\\hline
set 1 & 4.85 & 33.8 & 3.8 & 0.150 & 33.3 & 2.00 & 33.6 & 2.4\\
set 2 & 4.80 & 34.2 & 4.1 & 0.162 & 33.9 & 2.25 & 34.3 & 2.5\\
set 3 & 4.75 & 34.6 & 4.4 & 0.174 & 34.5 & 2.50 & 35.1 & 2.6\\
set 4 & 4.60 & 35.7 & 5.1 & 0.210 & 36.8 & 3.00 & 37.8 & 2.9\\\hline
\end{tabular}
\end{center}
\caption[]{Parameter sets for $f_B$ and $f(0)$; we use the same values
  of $c_c$ and $s_0$ for $\pi$, $K$ and $\eta$. $m_b$ and $f_B$
  are given in GeV, $s_0$ and $M^2$ in GeV$^2$.}\label{tab:fB}
\end{table} 


\subsection{Hadronic Input Parameters}
\label{sec:hadronic}

The hadronic parameters needed are, for each meson, 
7 parameters characterising the twist-2, 3 and 4 distribution
amplitudes to NLO in the conformal expansion, cf.\ App.~\ref{app:A},
the decay constants of the $\pi$, $K$ and $\eta$ and $B$, 
the factorization scale $\mu_{\rm IR}$, the $b$ quark mass $m_b$
and the strong coupling $\alpha_s$. As for the latter, we fix
$\alpha_s(m_Z)=0.118$ and use NLO evolution down to the required
scale. The quark mass parameter entering our formulas is the
one-loop pole mass $m_b$ for which we use  $m_b = (4.80\pm
0.05)\, \text{GeV}$ (cf.\ Table 6 in the recent review
\cite{LCSRs:reviews} 
and references therein).
We also include results for $m_b = 4.6\,$GeV. The infrared
factorization scale separating contributions to be included in DAs and
perturbatively calculable terms is chosen to be $\mu_{\rm IR}^2 =
m_B^2 - m_b^2$, which also sets the scale of $\alpha_s$; we will
discuss the residual scale-dependence of our results below. The decay
constants for the $\pi$ and $K$ are very well known experimentally;
for the $\eta$ the situation is complicated due to $\eta$--$\eta'$
mixing. We use the following values:
\begin{equation}
f_\pi = 131\,{\rm MeV},\quad f_K = 160\,{\rm MeV},\quad f_\eta =
130\,{\rm MeV}.
\end{equation}
$f_B$ has been discussed in the previous subsection.

As for the meson DAs, we quote the preferred values for the twist-3
and 4 parameters in Tab.~\ref{tab:t3}; the form factors are not too
sensitive to their precise values. The situation is different,
however, for the Gegenbauer moments $a_{1,2,4}(\mu)$ parametrizing the
twist-2 DAs $\phi_{\pi,K,\eta}$, and so we shall
discuss in a bit more detail what is presently known about these
parameters. 

Both theoretical calculations and experimental determinations focus
mainly on the $\pi$ DA (for which all odd Gegenbauer moments vanish 
due to G-parity; in particular $a_1^\pi=0$). 
The probably earliest calculation of the
lowest Gegenbauer moment $a_2$
 was done by Chernyak and Zhitnitsky (CZ), yielding
\cite{CZreport}
$$a_2^{\rm CZ}(0.5\,{\rm GeV}) = 2/3.$$
This result was obtained from local QCD sum rules, 
where $a_n$ is extracted from the correlation function of the 
(local) interpolating field $\bar u \gamma_\nu\gamma_5
(\stackrel{\leftrightarrow}{D}\cdot x)^n d$, where $x$ defines the
light-cone, $x^2=0$, and the usual interpolating current for the
$\pi$, $\bar u \gamma_\mu\gamma_5 d$. The price to pay
for the expansion of an intrinsically nonlocal quantity like
$\phi_\pi$ in contributions of 
local operators is an increasing sensitivity to
nonperturbative effects, i.e.\ the precise values of the
condensates. As the coefficients of the condensates in the sum rule
for $a_n$ increase with powers of $n$ and, for sufficiently large $n$,
dominate over the
perturbative contributions, it is clear that this method is
inappropriate for calculating high moments, but one might expect it to
be reliable at least for the lowest moment with 
$n=2$. 

The DA obtained by CZ has the remarkable feature that
$\phi_\pi(1/2,0.5\,{\rm GeV}) = 0$, which is of course an artifact
of neglecting all contributions from $a_{n\geq 4}$.
It was subsequently shown by Braun and Filyanov (BF) \cite{BF}
that both the pion-nucleon-nucleon coupling $g_{\pi N N}$ and its
 mesonic equivalent $g_{\rho\omega\pi}$, when calculated from LCSRs,
 require a value of $\phi_\pi(1/2)$ significantly different from 0 (albeit
 at a slightly different scale):
\begin{equation}\label{phionehalf}
\phi_\pi(1/2,1\,{\rm GeV}) = 1.2\pm 0.3 = \frac{3}{2} - \frac{9}{4}\,
a_2(1\,{\rm GeV}) + \frac{45}{16}\,a_4(1\,{\rm GeV}) + \dots,
\end{equation}
where the dots denote neglected terms in $a_{n\geq 6}$. 
The large error is due to a large sensitivity of this result to
twist-4 corrections to the sum rules. BF also redetermined $a_2$,
using the same procedure as CZ, and combining their result, which is
consistent with $a_2^{\rm CZ}$, with the above
constraint from $\phi_\pi(1/2)$, they obtained
$$a^{\rm BF}_2(1\,{\rm GeV}) = 0.44,\qquad a^{\rm BF}_4(1\,{\rm GeV}) =
0.25.$$

An alternative calculation aims to cure the problem of increasing
condensate contributions by resumming them into nonlocal condensates
\cite{MR}. The Gegenbauer moments in this approach are mostly
sensitive to the ratio 
$$\lambda_q^2 = \langle \bar q \sigma g G q\rangle/(2\langle \bar q
q\rangle) = (0.4\pm0.1)\,{\rm GeV}^2\quad (\mu = 1\,{\rm GeV})
$$
and have moderate to small values. The most recent paper on that
topic, Ref.~\cite{BMS2}, quotes
\begin{equation}\label{stefanis}
a_2(1.16\,{\rm GeV}) = 0.19,\quad a_4(1.16\,{\rm GeV}) = -0.13,\quad
a_{6,8,10}\sim 10^{-3}.
\end{equation}

There are not too many lattice calculations of moments of the $\pi$
DA. The fairly old values quoted in \cite{lattDA} 
for the 2nd moment suffer from large uncertainties. 
This quantity has been investigated again recently \cite{sachrajda},
but the results, obtained in quenched approximation, are still
preliminary.

Alternative determinations of Gegenbauer moments rely on the analysis
of experimental data, in particular the pion-photon transition
formfactor $\gamma+\gamma^*\to\pi$, measured at CLEO and Cello, and the 
electromagnetic formfactor of the pion. The results of these analyses
are typically either determinations of $a_2$ (setting $a_{n\geq 4}$ to
0) or constraints on a linear combination of $a_2$ and $a_4$ (setting
$a_{n\geq 6}$ to 0).\footnote{In principle it is possible to
  determine $a_2$, $a_4$ and even higher moments separately from the
  $Q^2$ dependence of their respective contributions. However, such
  an analysis requires accurate measurements of the formfactors
  over a large enough range of $Q^2$, which are presently not available. See
  also Ref.~\cite{bijnens}, in particular Fig.~4.} 
These determinations are limited by mainly two
problems: large experimental errors and the contamination by poorly
known twist-4 and higher effects, which are usually estimated from QCD
sum rules. As
for the pion-photon transition formfactor, which has been measured by
CLEO and Cello, the technique used to extract $a_2$ and
$a_4$ has been pioneered by Khodjamirian \cite{akho}, refined by 
Schmedding and
Yakovlev \cite{SY}, with subsequent further refinements by Bakulev,
Mikhailov and Stefanis \cite{BMS1}. The upshot is that for not too
small $Q^2$ the pion-photon
transition is mostly sensitive to a like-sign combination of
$a_2$ and $a_4$. Summarizing the analyses of this process, we conclude from
Tab.~I in \cite{BMS2} that 
\begin{equation}\label{eq:constraint2}
a_2(1\,{\rm GeV}) + a_4(1\,{\rm GeV}) = 0.1\pm 0.1
\end{equation}
is a fair reflection of the current state of knowledge of $a_{2,4}$
from that process.

As for the pion electromagnetic formfactor, the authors of Ref.~\cite{BKM}
unfortunately only obtain a value for $a_2$ and set $a_4$ to
zero. A very recent analysis of that formfactor, Ref.~\cite{BMS2},
concludes that calculations using the nonlocal-condensate model are in
good agreement with data.

\begin{figure}
$$\epsfxsize=0.5\textwidth\epsffile{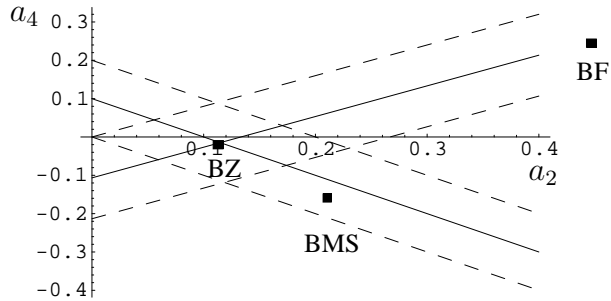}$$
\caption[]{$a_2(1\,{\rm GeV})$ and $a_4(1\,{\rm GeV})$ as determined
  from the constraints (\ref{phionehalf}) and (\ref{eq:constraint2}).
Solid line: central values, dashed lines: uncertainties. The black
  square labeled BZ denotes the central values used in this paper,
  Eq.~(\ref{eq:center}), BMS the prediction of the nonlocal condensate model,
  Eq.~(\ref{stefanis}), rescaled to $\mu=1\,$GeV, and BF is the central
  value obtained in Ref.~\cite{wavefunctions}.}\label{fig:a2a4}
\end{figure}

So what then do we actually know about $a_2$ and $a_4$? 
It seems to us
that, taking everything together, and with due consideration of the
respective strengths and weaknesses of different approaches, the most
reliable constraints for these quantities are (\ref{phionehalf}) and
(\ref{eq:constraint2}). These two constraints contain 
opposite-sign combinations of $a_2$ and $a_4$ and hence are about
equally sensitive to both parameters. 
The resulting
allowed area for $a_2$ and $a_4$ 
is shown in Fig.~\ref{fig:a2a4}; its center is at
\begin{equation}\label{eq:center}
\begin{array}{r@{\ =\ }l@{\quad}r@{\ =\ }l}
a_2(1\,{\rm GeV}) & 0.115, & a_4(1\,{\rm GeV}) & -0.015,\\[5pt]
a_2(2.2\,{\rm GeV}) & 0.080, & a_4(2.2\,{\rm GeV}) & -0.0089.  
\end{array}
\end{equation}
These are the central values we will use in our calculation of
formfactors. The figure also shows that the remaining
uncertainties are still considerable. Anticipating a future better
determination of these parameters, from lattice or else, we will 
present our final results in such a way as to facilitate the inclusion of any
shift in these values. Since much less is known about the Gegenbauer 
moments of the other pseudoscalar mesons $K$ and $\eta$, 
we resort to SU(3) symmetry and use {\em the same}
Gegenbauer moments.

Eq.~(\ref{eq:center}) and Fig.~\ref{fig:a2a4} confirm 
the findings of previous analyses that 
the CZ DA is strongly disfavored; 
the same applies to the values obtained by BF and 
to the local QCD sum
rule for $a_2$, which favors a large positive $a_2\sim 0.4$. 
One explanation for the failure of the corresponding QCD sum rule
could be that already the case $n=2$ may be too ``nonlocal'' for sum
rules to work. Another one could be that the treatment of $a_1$ and
other resonances contributing to that sum rule may be insufficient. We
leave a further discussion of that question to future work. The
result from sum rules with nonlocal condensates \cite{MR,BMS2,BMS1}, 
shown as black square in Fig.~\ref{fig:a2a4},
is also outside the favored area in Fig.~\ref{fig:a2a4}, which is
mainly due to the large value of $|a_4|$. 
It would definitely be very interesting to see
all these results and constraints on $a_{2,4}$ be supplemented by 
 lattice determinations. 

The only parameter left to discuss is $a_1$ for the $K$ meson (by which we
understand an $s\bar q$ bound state), which is a
G-parity breaking parameter. Here the
situation is even worse than for $a_{2,4}$, 
as neither size nor even sign of that
quantity are reliably known. The facts at hand are the following: the
intuitive expectation is that $a_1$ (i.e.\ the moment with a
weight-function proportional to $2u-1$) should be positive, as the DA is
expected to be slightly tilted towards larger values of $u$ which is the
momentum fraction carried by the (heavier) $s$ quark in the meson --
the heavier the quark, the more the DA is expected to peak at large
$u$, the extreme case being a $b\bar q$ bound state whose DA should be
close to $\delta(1-u)$.  The
(tree-level) QCD sum rule calculation in \cite{CZreport} seemed to
confirm intuition, but was challenged, when Ref.~\cite{SU(3)breaking}
found a sign-mistake in that calculation and, including
two-loop radiative corrections, obtained a {\em negative} sign for
$a_1^K$.  For this paper, we first
decided to stick to
that result and use the central value 
$a_1^K(1\,{\rm GeV}) = -0.18$. It turned out, however, that this value
tends to produce formfactors with an unfavorable
$q^2$-dependence.\footnote{That is: formfactors not very compatible
with the parametrization discussed in Sec.~\ref{sec:fitting}, which is
based on generic analytic properties of the formfactors.} We
therefore decided to revert to the original result by CZ
\cite{CZreport} and use
\begin{equation}\label{a1def}
a_1^K(1\,{\rm GeV}) = 0.17\: \leftrightarrow\:
a_1^K(2.2\,{\rm GeV}) = 0.135.
\end{equation}
The conclusion from that inconclusive situation can only be
that a second opinion has to be sought, and we urge our colleagues
from the lattice community to take up the challenge and provide the
first-ever lattice determination of $a_1^K$.
For the time being, we will present our results in a way
that makes it possible to obtain the formfactors also for different
values of $a_1^K$.

\subsection{\boldmath Results for $q^2=0$}
\label{sec:4.3}

\begin{table}[tb]
\addtolength{\arraycolsep}{3pt}
\renewcommand{\arraystretch}{1.3}
$$
\begin{array}{|l|lllllllc|}
\hline
 & \text{set 1} & \text{set 2} & \text{set 3} & \text{set 4} &
 \Delta_{as} & \Delta _{a_2,a_4} & \Delta & \Delta_{a_1} \\
\hline
f^{\pi}_+(0) & 0.250 & 0.258 & 0.263 & 0.274 & 0.023 & 0.019 & 0.030 & - \\
f^{\pi}_T(0) & 0.244 & 0.253 & 0.260 & 0.273 & 0.013 & 0.022 & 0.026 & - \\
\hline
f^{K}_+(0) &  0.324 & 0.331 & 0.335 & 0.339 & 0.033 & 0.023 & 0.040 & 0.25\delta_{a_1} \\
f^{K}_T(0) &  0.347 & 0.358 & 0.367 & 0.381 & 0.022 & 0.027 & 0.035 & 0.31\delta_{a_1} \\
\hline
f^{\eta}_+(0) & 0.269 & 0.275 & 0.278 & 0.286 & 0.029 & 0.019 & 0.035 & - \\
f^{\eta}_T(0) & 0.277 & 0.285 & 0.292 & 0.305 & 0.018 & 0.022 & 0.028 & - \\
\hline
\end{array}
\addtolength{\arraycolsep}{-3pt}
\renewcommand{\arraystretch}{1.}
$$
\caption[]{Final central values of the formfactors at $q^2=0$ for the 
parameter sets of Tab.~\ref{tab:fB}. $f_0(0)\equiv f_+(0)$. The errors
$\Delta_{as}$, $\Delta_{a_2,a_4}$ and $\Delta_{a_1}$
are described in the text. $\Delta$ is defined as 
$\Delta=(\Delta_{as}^2+\Delta_{a_2,a_4}^2)^{1/2}$ and $\delta_{a_1}$ as
$\delta_{a_1}=a_1(\text{1GeV})-0.17$. Note that $\delta_{a_1}$ 
carries information on the sign of $a_1$ and can become
negative.}\label{tab:final}
\end{table}

\begin{figure}[p]
$$
\begin{array}{@{}l@{}}
\epsfxsize=0.95\textwidth\epsffile{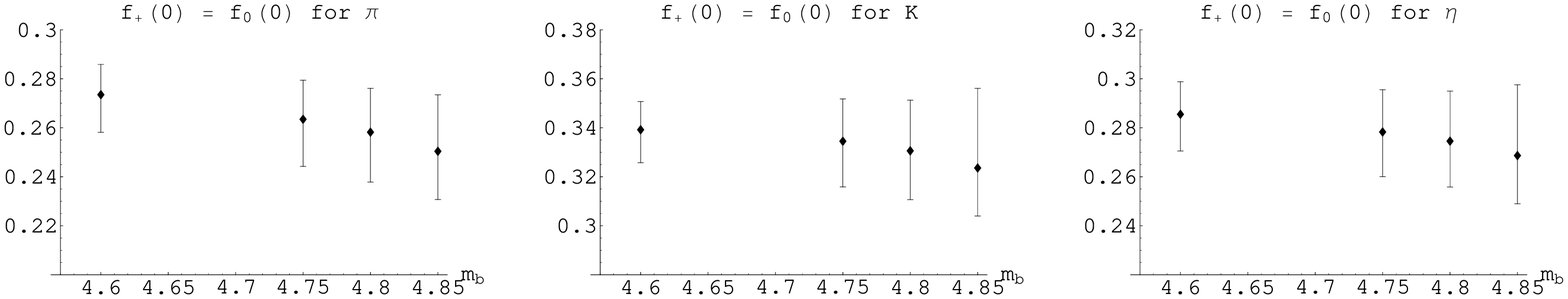}\\
\epsfxsize=0.95\textwidth\epsffile{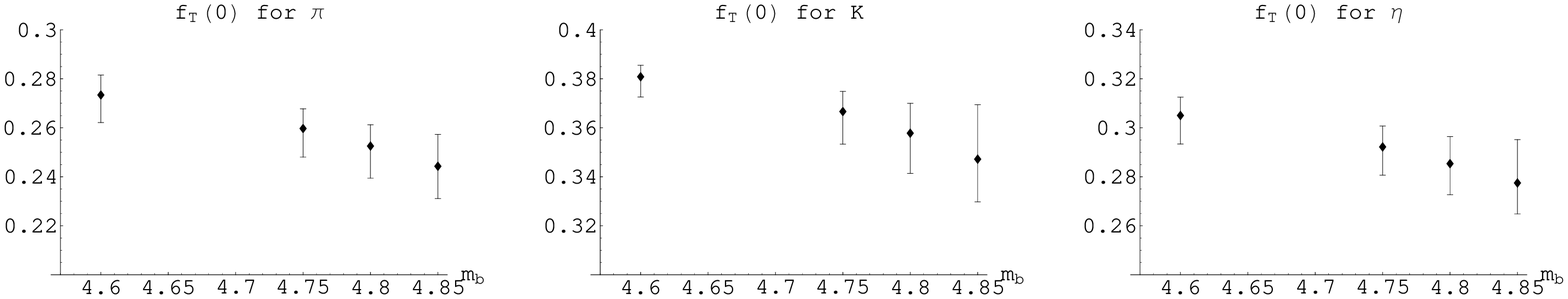}
\end{array}
$$
\vskip-0.4cm
\caption[]{Central values of the formfactors $f(0)$ and uncertainties 
$\Delta$. Numbers from Tab.~\ref{tab:final}.}\label{fig:error}
$$
\epsfxsize=0.45\textwidth\epsffile{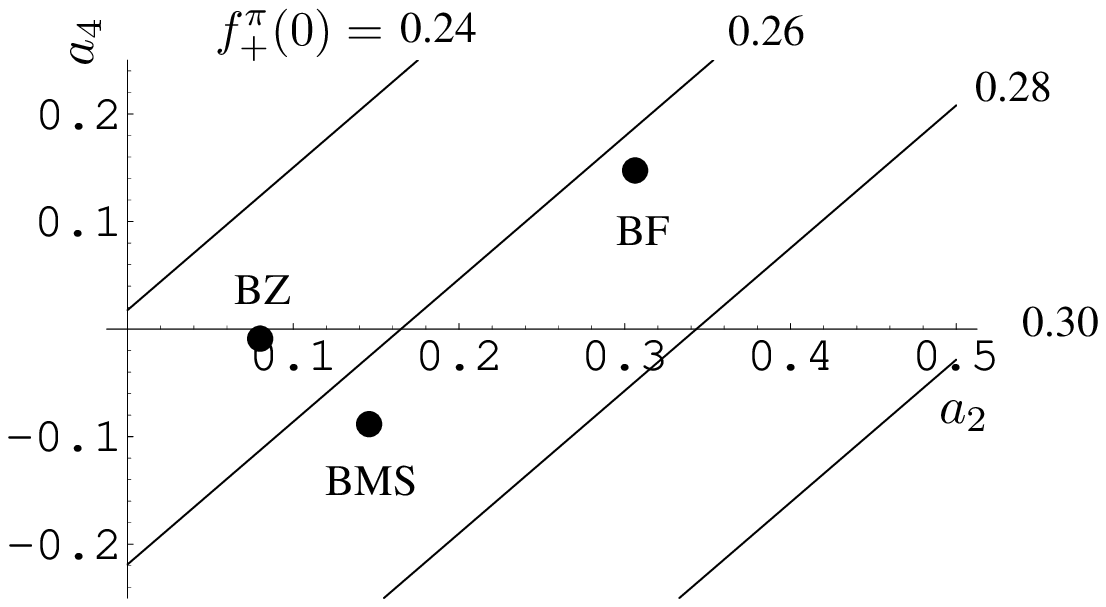}
$$
\vskip-0.4cm
\caption[]{Dependence of $f_+^\pi(0)$ on $a_2(\mu_{\rm IR})$ and
  $a_4(\mu_{\rm IR})$, for parameter set 2. The lines are lines of
  constant $f_+^\pi(0)$. The dot labeled BZ denotes our preferred
  values 
of $a_{2,4}$, BMS the values from the nonlocal condensate model and BF
  from the sum rule calculations of Ref.~\cite{wavefunctions}.}\label{fig:x}
$$\epsfxsize=0.45\textwidth\epsffile{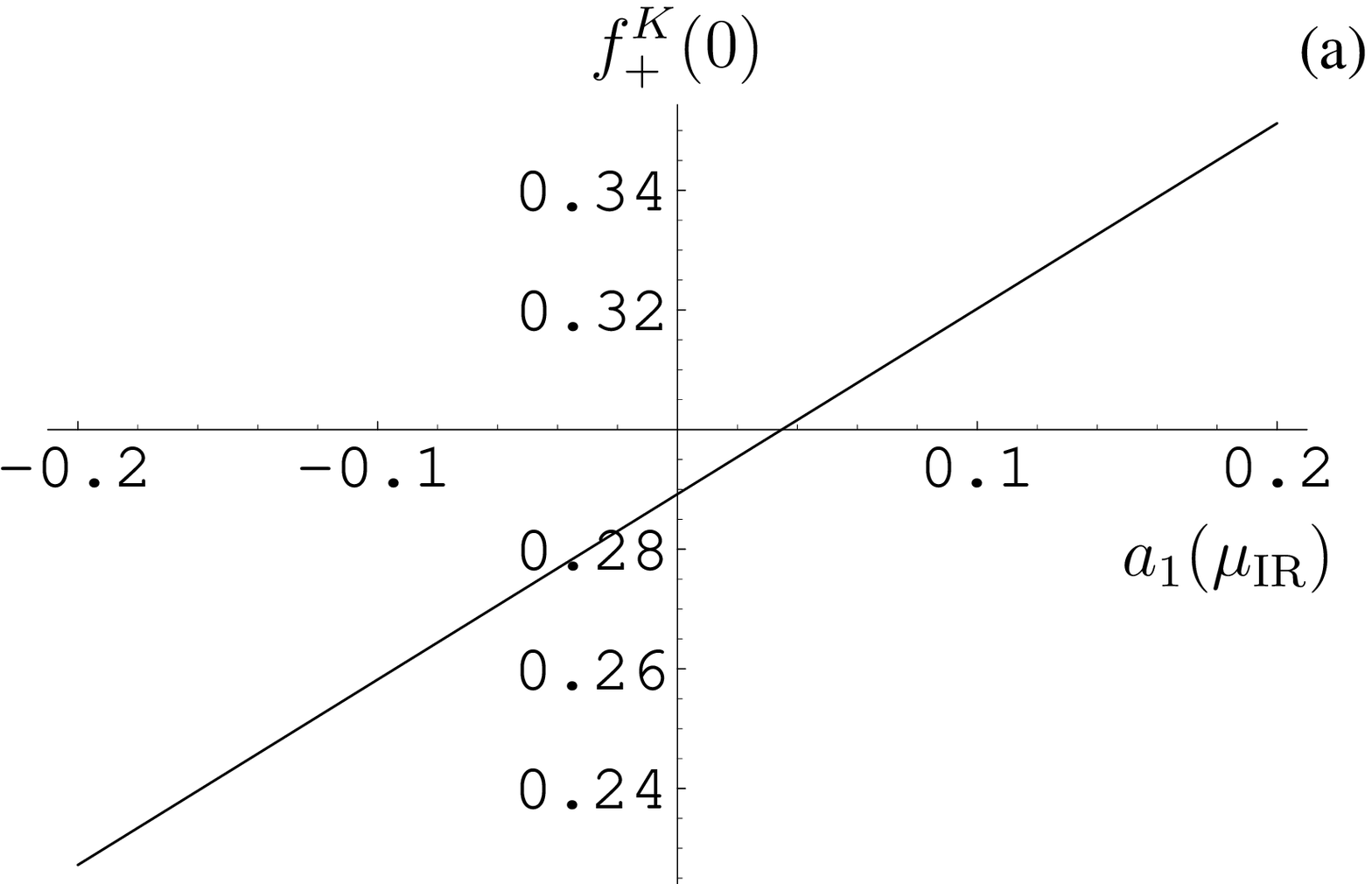}\qquad 
\epsfxsize=0.45\textwidth\epsffile{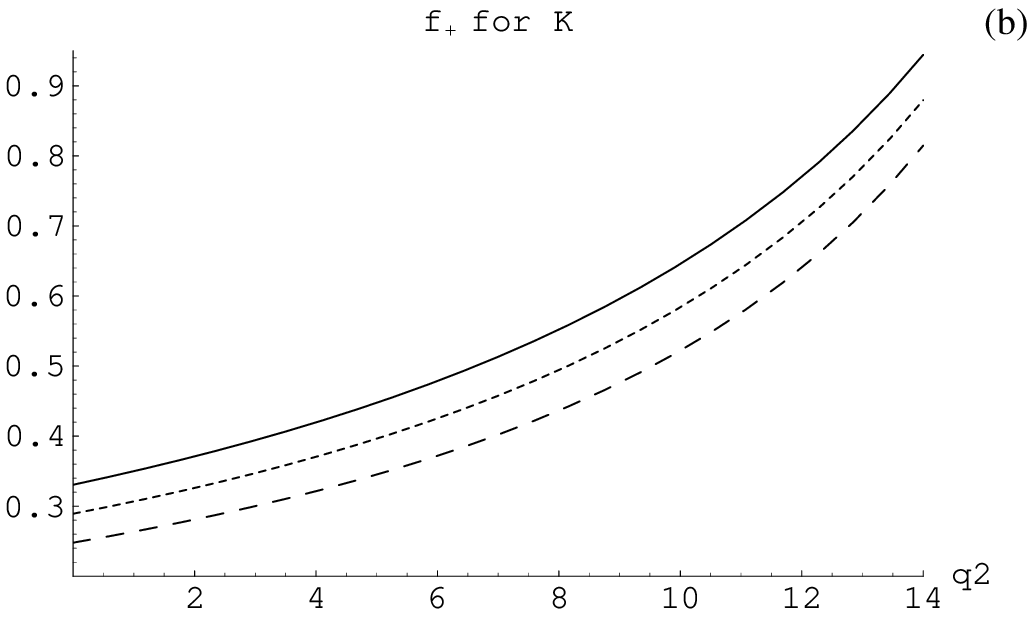}$$
\vskip-0.4cm
\caption[]{(a) Dependence of $f_+^K(0)$ on the Gegenbauer moment
$a_1(\mu_{\rm IR})$. (b) $f_+^K(q^2)$ as function of $q^2$ for
  different values of $a_1$: solid line: $a_1^K(1\,{\rm GeV}) = 0.17$,
  short dashes: $a_1^K(1\,{\rm GeV}) = 0$,
  long dashes: $a_1^K(1\,{\rm GeV}) = -0.18$. Input parameters: 
set 2.}\label{fig:a1}
\end{figure}

\begin{figure}[tb]
$$\epsfxsize=0.45\textwidth\epsffile{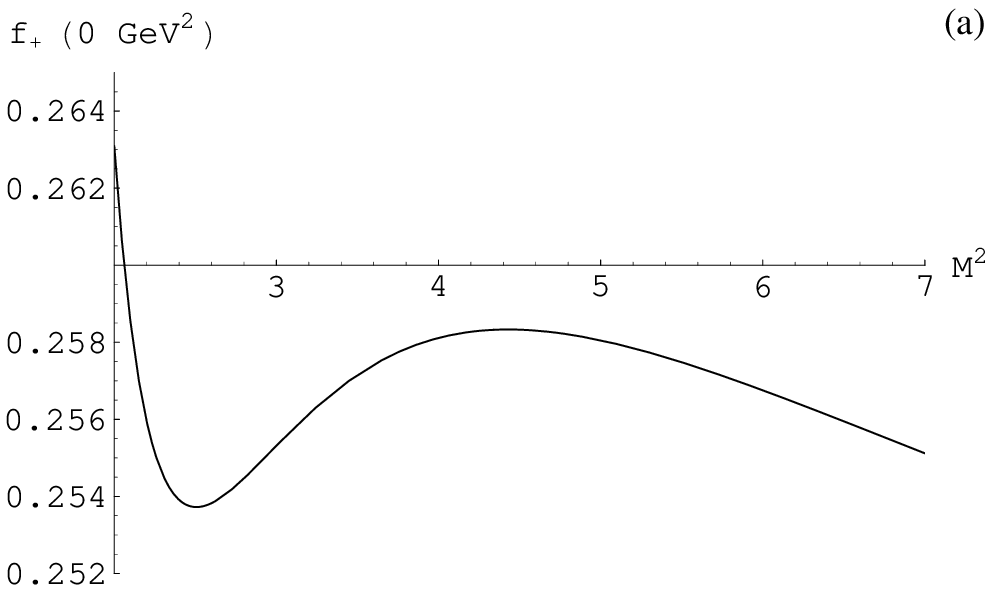} \qquad
\epsfxsize=0.45\textwidth\epsffile{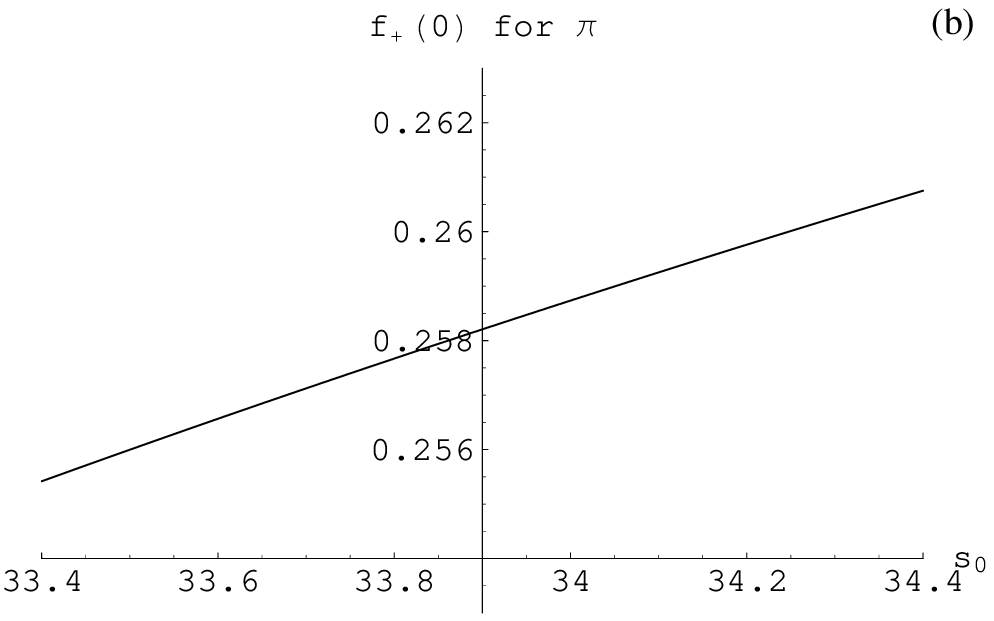}$$
\caption[]{Dependence of $f_+^\pi(0)$ on (a) the Borel parameter
$M^2$ and (b) the continuum threshold $s_0$. 
Input parameters: set 2 in Tab.~\ref{tab:fB}.}\label{fig:SRdep}
$$\epsfxsize=0.45\textwidth\epsffile{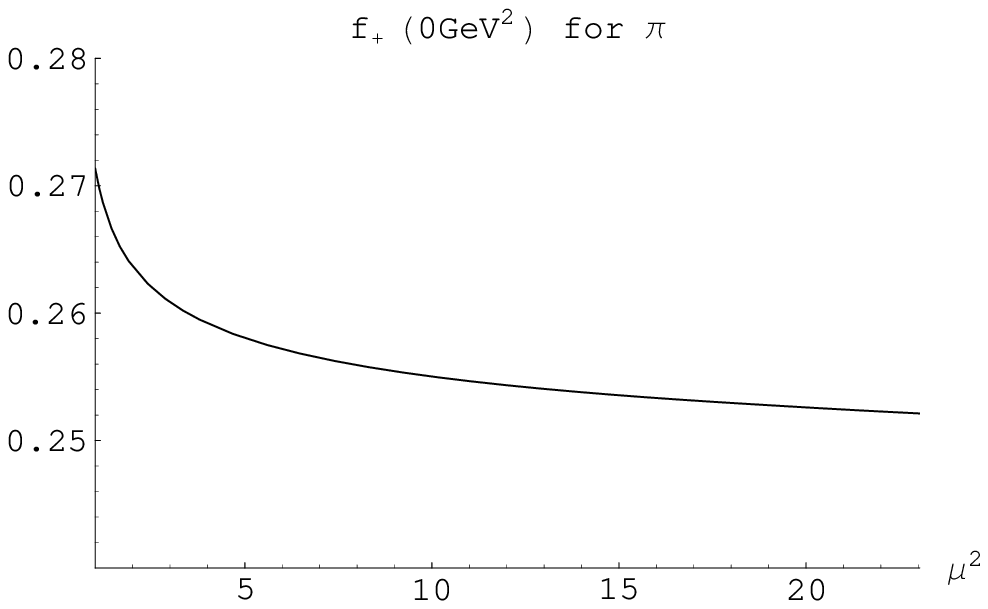}$$
\caption[]{Dependence of $f_+^{\pi}(0)$ on 
the factorization scale $\mu_{\rm IR}$. Same input
parameters as in Fig.~\ref{fig:SRdep}.}\label{fig:mu2} 
\end{figure}

Let us first discuss the sum rule results for $q^2= 0$. They are
collected in Tab.~\ref{tab:final}, 
for all 4 parameter sets from Tab.~\ref{tab:fB}.\footnote{
$f_0(0)$ is not included as $f_0(0)\equiv f_+(0)$.}  Including the
uncertainty of $m_b$, $m_b = (4.80\pm 0.05)\,$GeV, the final central
values and uncertainties of the formfactors are given in
Eq.~(\ref{eq:resq2at0}).

The formfactors are calculated from Eq.~(\ref{srx})
using the parameter sets given in Tab.~\ref{tab:fB}
and the hadronic input parameters given in Eqs.~(\ref{eq:center})
and (\ref{a1def}) and
Tab.~\ref{tab:t3}. 
The dependence of the formfactors on $m_b$, i.e.\ the set, is shown in
Fig.~\ref{fig:error}. It is evident that the residual dependence of
$f(0)$ on $m_b$ is much smaller than the one of $f_B$ in  
Tab.~\ref{tab:final}, which confirms our expectation that the
calculation of
$f_B$ from a sum rule reduces the parameter dependence of the
formfactors. $f_+^\pi(0)$ depends sensitively on $a_2$ and $a_4$ as
illustrated in Fig.~\ref{fig:x}. The formfactors show moderate SU(3) breaking
between $\pi$ and $\eta$, which is due to terms in the LCSRs
proportional to the meson mass. For $K$, the situation is 
different, and we observe a strong enhancement of the formfactor due
to the combination of two effects: the fact that $f_K$ is larger than
$f_\pi$ and the positive contribution of the Gegenbauer moment $a_1$
to the formfactor. As
discussed in the previous subsection, the numerical value of $a_1$,
and even its sign, is not precisely known. Fig.~\ref{fig:a1}(a)
illustrates the dependence of $f_+^K(0)$ on $a_1$, which is quite
strong. Fig.~\ref{fig:a1}(b) shows the dependence of $f_+^K(q^2)$ on
$q^2$ for different values of $a_1$. It is evident that 
$a_1$ mainly determines the normalisation of the formfactor, but has
only minor impact on its shape. The
uncertainty of $f_+^K(0)$ induced by $a_1$ will be discussed below.
The dependence of $f_+^\pi(0)$ on the sum rule parameters $M^2$ and
$s_0$ is illustrated in Fig.~\ref{fig:SRdep} and is very mild, thanks
to the optimized criteria for choosing $M^2$ and $s_0$ outlined in
Sec.~\ref{sec:borelfix}. The behavior of the other formfactors is
very similar. In Fig.~\ref{fig:mu2} we show the variation of 
$f_+^\pi(0)$ with a change of the factorization scale 
$\mu_{\rm IR}$ in the large range
$1\,{\rm GeV}\leq \mu_{\rm IR}\leq m_b$. The curve is remarkably flat
which can be understood from the fact that radiative corrections
cancel to a certain extent between $\Pi_+$ and $f_B$ and that large
logarithms of type $\ln m_b/\mu_{\rm IR}$ occur only at subleading order in the
conformal expansion of the DAs, which is numerically suppressed with
respect to the leading ($\mu_{\rm IR}$-independent) term, and at subleading
twist, which is also suppressed.

Let us now turn to the uncertainties of the formfactors induced by a
variation of the input parameters.
It is convenient to split the formfactors into contributions from
different Gegenbauer moments, 
\begin{equation}
\label{eq:form}
f(q^2) = f^{as}(q^2)+a_1 f^{a_1}(q^2)+\left\{
a_2 f^{a_2}(q^2) +a_4 f^{a_4}(q^2)\right\},
\end{equation}
where $f^{as}$ contains the contributions to the formfactors from the
asymptotic DA and also all higher-twist effects from 
three-particle quark-quark-gluon matrix elements. Explicit expressions for
the functions $f^{as,a_1,a_2,a_4}$ can be obtained from
Tab.~\ref{tab:afit} in App.~\ref{app:C}; in particular $f^{a_i}(0)$ is
just given by the parameters $a$ in that table. We calculate
separately the
uncertainties $\Delta_{as,a_1}$ of the first and second term and the
combined uncertainty $\Delta_{a_2,a_4}$ of the term in curly brackets.
We start with $\Delta_{as}$. To estimate its value
 we vary the following quantities:
\begin{itemize}
\item the threshold $s_0$ by $\pm 0.5\,\text{GeV}^2$;
\item the Borel parameter
$M^2$ in Eq.~\eqref{eq:borels} by $\pm 1.2\,\text{GeV}^2$;
\item the infrared factorization scale $\mu_{\rm IR}^2=m_B^2-m_b^2$ 
by $\pm 2\,\text{GeV}^2$;
\item the quark condensate and the mixed condensate as indicated in 
Eq.~\eqref{eq:conds};
\item the twist-3 matrix-element $\eta_3$ by $\pm 50\%$.
\end{itemize}
$m_b$ is kept fixed and we calculate the uncertainty separately for each
parameter set; for a given formfactor, 
$\Delta_{as}$ is then the largest uncertainty of the 4 sets. 
The errors are correlated and we therefore scan the five-parameter
space for the largest deviations from the central values. 
The resulting $\Delta_{as}$ are given in Tab.~\ref{tab:final}.

The uncertainty of $f^K(0)$ induced by $a_1$ is dominated by $a_1$
itself, so we do not attempt to determine the uncertainty of $f^{a_1}$
arising from varying $M^2$, $s_0$ etc., but just take the maximum value of
$f^{a_1}(0)\equiv a$ from
Tab.~\ref{tab:afit} in App.~\ref{app:C} and multiply it by
$\delta_1 = 
a_1(1\,\rm{GeV}) - 0.17$ and the leading-order scaling factor 
from ${\rm 1 GeV}$ to $\mu_{\rm IR}$, 
which gives the entry labeled $\Delta_{a_1}$
in Tab.~\ref{tab:final}.

As the allowed input values of $a_2$ and $a_4$ are correlated and
given by the rhomboid shown in 
Fig.~\ref{fig:a2a4}, we only determine the combined uncertainty 
$\Delta_{a_2,a_4}$ arising from the corresponding 
variation of the Gegenbauer moments, separately for each parameter
set. The resulting uncertainties
depend strongly on the precise values of $M^2$ and $s_0$, so for a
conservative estimate of the uncertainty we scan the full 7-parameter
space in $a_2$, $a_4$, $M^2$ etc.\ 
and quote the largest deviation from the central value as
uncertainty, which yields the
$\Delta_{a_2,a_4}$ quoted in Tab.~\ref{tab:final}.
Taking everything together, and including the variation of $m_b=(4.80\pm
0.05)\,$GeV in the error estimate, adding errors in quadrature, we
find ($\delta_{a_1}$ is defined in Tab.~\ref{tab:final}):
\begin{equation}
\boxed{ \label{eq:resq2at0}
\renewcommand{\arraystretch}{1.2}\addtolength{\arraycolsep}{2pt}
\begin{array}[b]{lcl@{\qquad}lcl}
f_+^\pi(0) &=& 0.258\pm 0.031, & f_T^\pi(0)& =& 0.253\pm0.028,\\
f_+^K(0)& = &0.331\pm 0.041 + 0.25\delta_{a_1}, & f_T^K(0)& =&
0.358\pm0.037 + 0.31\delta_{a_1},\\
f_+^\eta(0)& =& 0.275\pm 0.036, & f_T^\eta(0)& =& 0.285\pm0.029.
\end{array}
\renewcommand{\arraystretch}{1}\addtolength{\arraycolsep}{-2pt}}
\end{equation}
These are our final results for the formfactors at $q^2=0$. For
$f^{\pi,\eta}$ the total theoretical uncertainty is 10\% to 13\%, for
$f^K$ it is 12\%, plus the uncertainty in $a_1$, which hopefully
will be clarified through an independent calculation in the not too
far future. These uncertainties
include a variation of both the external input parameters and the sum
rule specific parameters, but they do not include an additional ``systematic''
uncertainty of the sum rule method itself. To a certain extent, this
intrinsic sum rule uncertainty is included by the variation of the sum
rule specific parameters $M^2$ and $s_0$, which sets the minimum
uncertainty of the result: all external hadronic parameters fixed,
this variation induces a $\sim7\%$ uncertainty of $f_+^\pi(0)$ quoted 
in Eq.~\eqref{eq:resq2at0}. Realistically, one may hope to reduce the
$\sim12\%$ uncertainty quoted to $\sim 10\%$ by reducing the errors on the
Gegenbauer moments $a_{2,4}$ by a factor of 2. Further improvement
will then have to come from a better control over higher-twist matrix
elements, dominated by the quark condensate and the quark-quark-gluon
matrix element $\eta_3$ discussed in App.~\ref{app:A}.

\subsection{\boldmath Results for $q^2\neq 0$, Fits and Extrapolations}
\label{sec:fitting}

In this subsection we calculate the $q^2$-dependence of the 
formfactors for central values of the input parameters and cast them
into a three-parameter parametrization that is valid for all $q^2$. The
results are given in Tab.~\ref{tab:fitpars} which is to be used
together with Eq.~(\ref{eq:para}). The fit parameters
for other sets of input parameters are given in App.~\ref{app:C}. We
refrain from a complete analysis of the uncertainty of the
$q^2$-dependence of the formfactors, but just mention that it is
likely to be smaller than that at $q^2=0$, which is indicated by a
decrease of the spread between the formfactors calculated from
the different parameter sets in Tab.~\ref{tab:fB}, cf.\ Fig.~\ref{fig:piplot}.

\begin{figure}[tb]
$$
\begin{array}{@{}c@{\quad}c}
\epsfxsize=0.45\textwidth\epsffile{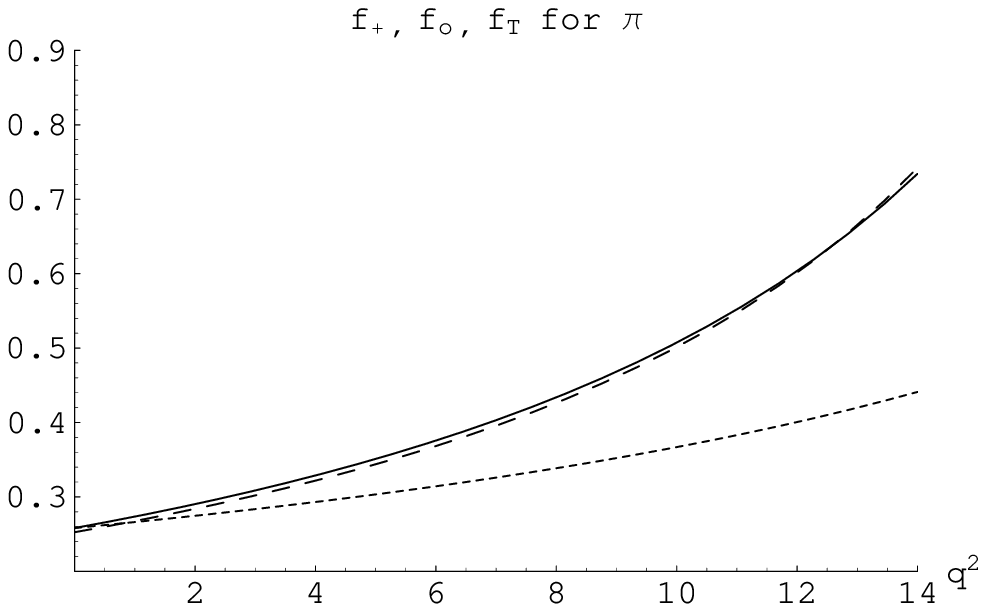} & 
\epsfxsize=0.45\textwidth\epsffile{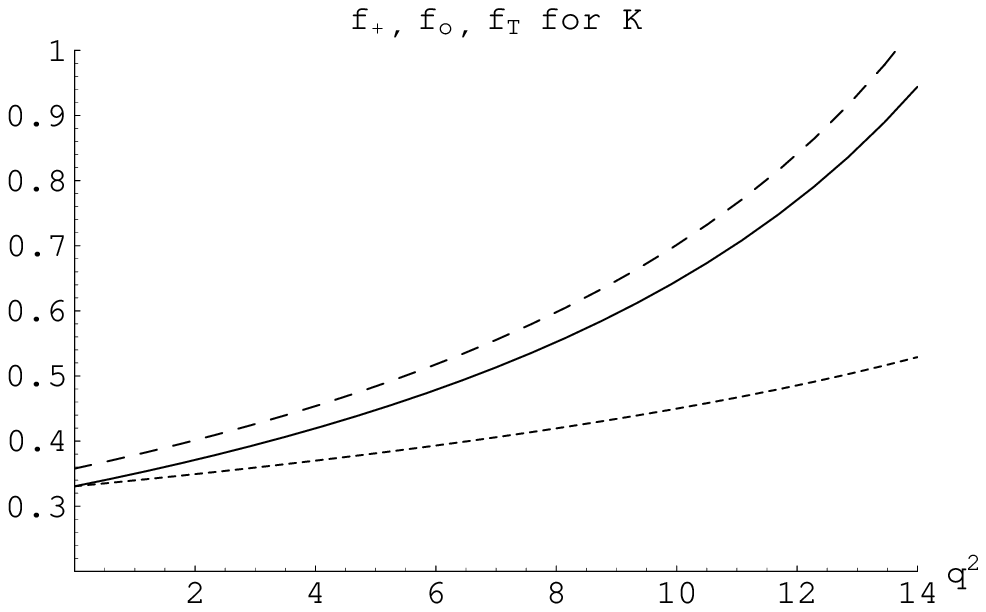}\\[5pt]
\epsfxsize=0.45\textwidth\epsffile{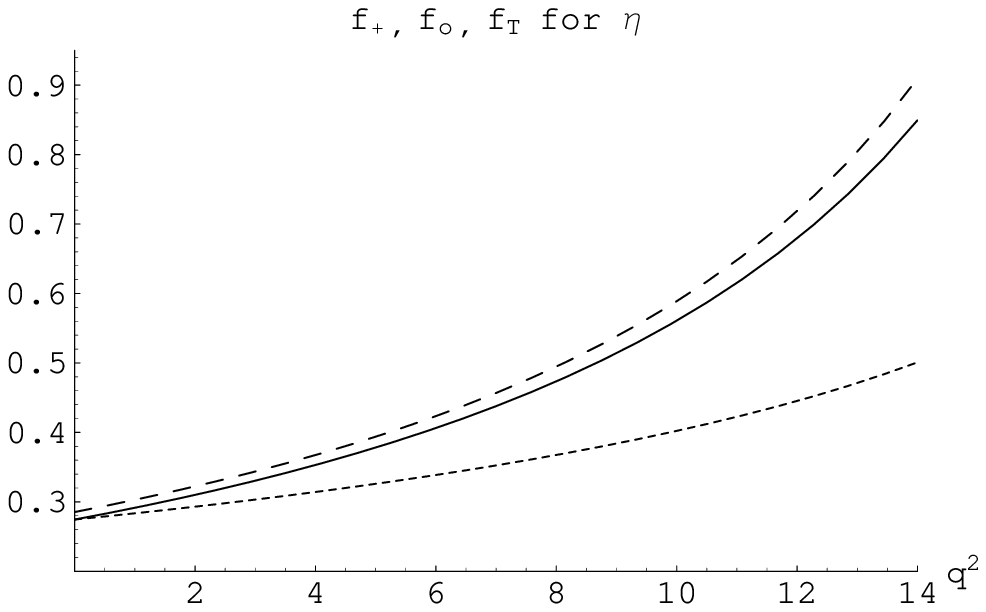}
\end{array}
$$
\caption[]{$f_+$ (solid lines), $f_0$ (short dashes) and $f_T$ (long
  dashes) as functions of $q^2$ for $\pi$, $K$
and $\eta$. The renormalisation scale of $f_T$ is chosen to be
$m_b$. Input parameters: set 2 in
  Tab.~\protect{\ref{tab:fB}}.}\label{fig:2}
\end{figure}

\begin{figure}[t]
$$
\epsfxsize=0.45\textwidth\epsffile{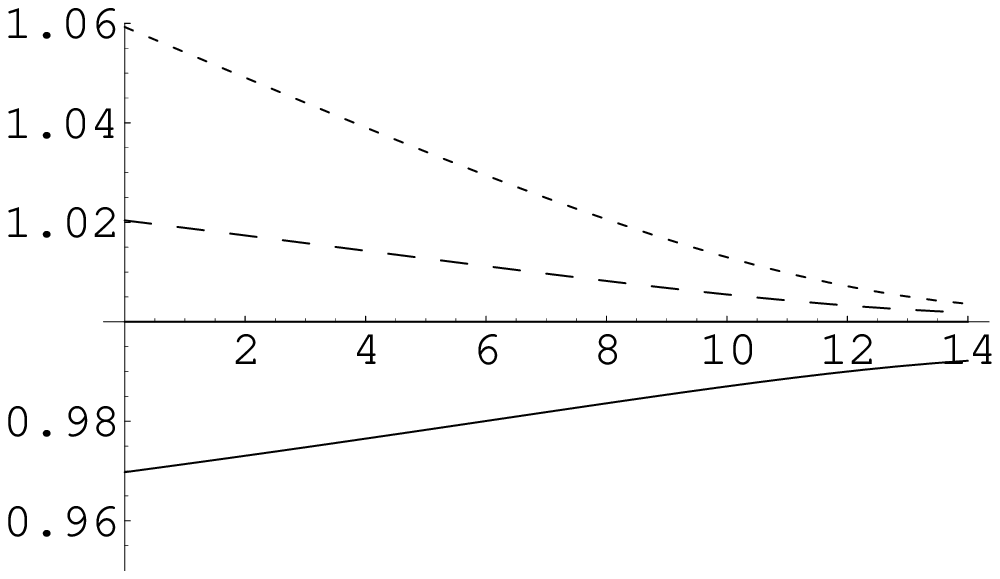}
$$
\caption[]{Ratio of $f^{\pi{\rm (set {\it i})}}_+(q^2)/
f^{\pi{\rm (set 2)}}_+(q^2)$ as function of $q^2$. Solid line: set 1;
long dashes: set 3; short dashes: set 4.}\label{fig:piplot} 
\end{figure}

The validity of the LCSR approach is restricted to the kinematical
regime of large meson energies, $E_P \gg \Lambda_{\rm QCD}$, which via
the relation 
$$q^2 = m_B^2 - 2 m_B E_P$$
implies a restriction to small and moderate $q^2$; specifically, we
evaluate the sum rules only for $0\leq q^2\leq 14\,{\rm GeV}^2$.
The resulting formfactors
are plotted in Fig.~\ref{fig:2}, using the parameter set 2 in Tab.~\ref{tab:fB}
and the hadronic input parameters given in Eqs.~(\ref{eq:center}) and
 (\ref{a1def}) and Tab.~\ref{tab:t3}. As expected
from LEET \cite{LEET}, $f_+$ and $f_T$ nearly coincide. Although this
agreement is expected to be best for small $q^2$, i.e.\ large energies
of the light meson, it is seen to hold for all $q^2$. From the LCSR
point of view, this agreement is due to the fact that the leading
twist-2 contributions to the corresponding correlation functions
coincide at tree-level. The figure also shows that the
$q^2$-dependence of $f_0$ is weaker than that of the other
formfactors. This is can be understood from the fact that, 
if $f_+$ is represented as a
dispersion relation over hadronic states, these states have quantum
numbers $J^P=1^-$ and hence zero orbital angular momentum, whereas for
$f_0$ the quantum number is $J^P=0^+$ 
and thus the coupling of these states or, in the language of potential-models, 
their wave-function at the origin, is suppressed as
it corresponds to a state with orbital angular momentum $L=1$. 
Fig.~\ref{fig:2} also shows sizable SU(3) breaking for
the $K$, but a moderate one for $\eta$, which is due to the same effects 
discussed for the formfactors at $q^2=0$. In Fig.~\ref{fig:piplot} we
show $f_+^\pi(q^2)$ as function of $q^2$, calculated for sets 1, 3 and
4 and normalized to set 2. It is evident that the uncertainties
induced by $m_b$, which amount to 6\% at $q^2=0$, become less
important for larger $q^2$, so that for instance the branching ratio
of the semileptonic decay $B\to\pi e \nu$ will be less dependent on
the precise value of $m_b$ than $f_+^\pi(0)$.


One of the main goals of this paper is to give simple expressions for 
the formfactors in the full physical
regime $0\leq q^2\leq (m_B-m_P)^2\approx 23\,{\rm GeV}^2$. We thus have to
find a parametrization that 
\begin{itemize}
\item reproduces the data below $14\,{\rm GeV}^2$ with good accuracy; 
\item provides an extrapolation to $q^2>14\,{\rm GeV}^2$ that is 
consistent with the expected analytical properties of the formfactors
and reproduces the lowest-lying resonance
(pole) with $J^P=1^-$ for  $f_+$ and $f_T$.\footnote{For $f_0$, the
lowest pole with quantum numbers $0^+$ lies above the two-particle threshold
starting at $(m_B+m_P)^2$ and hence is not expected to feature prominently.} 
\end{itemize}
It is actually not very difficult to find good fits:
the parametrization 
\begin{equation}\label{eq:paraold}
f(q^2) = \frac{f(0)}{1-a_F q^2/m_B^2 - b_F (q^2/m_B^2)^2}
\end{equation}
advocated in previous works, e.g.\ \cite{ballroman}, is one example
for an excellent fit to the results of the
sum rules for $q^2<14\,{\rm GeV}^2$. In the present context, however,
it turns out to be unsuitable as it produces, for $f_+^\pi$, 
a pole at $q^2\approx 23\,{\rm GeV}^2$, which is below the physical pole
at $q^2 = m_{B^*}^2 = (5.32\,{\rm GeV})^2$. In our previous
paper \cite{ballroman} we argued that the above parametrization should
be matched to a simple pole-dominance formula $f_+\sim 1/(m_{B^*}^2 -
q^2)$ for $q^2$ above a certain threshold $q_0^2\sim 15\,{\rm GeV}^2$,
defined as the value of $q^2$ that would allow
a smooth transition\footnote{That is equality of both the
  parametrization formulas and their first derivatives in $q_0^2$.}
from one parametrization to the other. This procedure unfortunately
does not work for our new formfactors, as the optimum $q_0^2$ turns
out to be far outside the physical regime. We therefore decide to
follow, as far as possible, the procedure advocated by Becirevic and Kaidalov
\cite{BecKai}, who suggested to write the formfactor $f_+$ as dispersion
relation in $q^2$ with a lowest-lying pole plus a contribution from
multiparticle states, which in turn is to be replaced by an effective
pole at higher mass:
\begin{eqnarray}
f_+(q^2) &=& \frac{r_1}{1-q^2/m_1^2} + \int_{(m_B+m_P)^2}^\infty
ds\,\frac{\rho(s)}{s-q^2}\label{eq:xyz}\\
&\to&
\frac{r_1}{1-q^2/m_1^2} + \frac{r_2}{1-q^2/m_{\rm fit}^2} \,.\label{eq:para}
\end{eqnarray}
The lowest-lying resonance in the $b\bar u$ channel 
is well known experimentally: it the $B^*(1^-)$ vector meson with mass
5.32~GeV; this is also the mass to be used for the $\eta$, as the
$B\to\eta$ formfactors calculated in this paper refer to a $b\to u$
transition. For the $K$ we have calculated the mass of the $B_s^*$
resonance in the heavy-quark limit and find
$$m_{B_s^*}^2-m_{B_s}^2 = m_{B^*}^2 - m_B^2 \quad  \to \quad
m_1^K=m_{B_s^*} 
= 5.41\,{\rm GeV}. 
$$

For Eq.~(\ref{eq:para}) to describe all $f_+$
and also $f_T$, which feature the same $1^-$ resonance, 
in terms of three fit parameters, $r_1$, $r_2$ and $m_{\rm
  fit}$, it is crucial that the position of the lowest pole is
sufficiently below the
two-particle cut starting at $(m_B+m_P)^2$. We find that indeed most
$f^\pi_{+,T}$ formfactors, with the exception of $f^{\pi{\rm
    (set~4)}}_T$, are described very well by (\ref{eq:para}). For $f^{\pi{\rm
    (set~4)}}_T$, however, and all $f_{+,T}^{K,\eta}$, $m_{\rm fit}$
gets too close to $m_1$, so that the fit becomes numerically
unstable. In this case, it is appropriate to expand (\ref{eq:para}) to
first order in $m_{\rm fit}-m_1$, which yields 
\begin{equation}\label{eq:paraKeta}
f_{+,T}^{K,\eta}(q^2) = \frac{r_1}{1-q^2/m_1^2} +
\frac{r_2}{(1-q^2/m_1^2)^2}
\end{equation}
with fit parameters $r_1$ and $r_2$, and $m_1 = m_{B^*,B^*_s}$ fixed.

For $f_0$, one can write a decomposition similar to
(\ref{eq:xyz}), but here the lowest-lying pole with quantum numbers
$0^+$ lies either above the two-particle threshold (for $\pi$ and
$\eta$) or is very close to it (for $K$, cf.\ Tab.~\ref{tab:masses}),
so that
the pole is effectively hidden under the cut and only the dispersive
term survives in (\ref{eq:xyz}). 
We again follow the suggestion of Becirevic and
Kaidalov and replace this term by an effective pole, i.e.\ we set
\begin{equation}\label{eq:paraf0}
f_0(q^2) = \frac{r_2}{1-q^2/m_{\rm fit}^2}.
\end{equation}

\begin{table}[tb]
\renewcommand{\arraystretch}{1.2}
\addtolength{\arraycolsep}{3pt}
$$
\begin{array}{|c|rrcr|}
\hline
& r_1 & r_2 & (m_1)^2 & m_{\rm fit}^2\\\hline
f_+^{\pi} &0.744 & -0.486 & (m_{1}^\pi)^2 & 40.73\\
f_0^{\pi} & 0 & 0.258 & - & 33.81\\
f_T^{\pi} & 1.387 & -1.134 & (m_{1}^\pi)^2 & 32.22\\
f_+^{K} & 0.162 & 0.173 & (m_1^K)^2 & -\\
f_0^{K} & 0 & 0.330 & - & 37.46\\
f_T^{K} & 0.161 & 0.198 & (m_1^K)^2 & -\\
f_+^{\eta} & 0.122 & 0.155 & (m_1^\eta)^2 & -\\
f_0^{\eta}& 0 &0.273  & - & 31.03\\
f_T^{\eta} & 0.111 & 0.175 & (m_1^\eta)^2 & -\\
\hline
\end{array}
$$
\renewcommand{\arraystretch}{1}\addtolength{\arraycolsep}{-3pt}
\caption[]{Fit parameters for Eq.~(\ref{eq:para}) for set 2 in 
Tab.~\ref{tab:fB} and central values of the input parameters of the
DAs, Eqs.~(\ref{eq:center}), (\ref{a1def}) and Tab.~\ref{tab:t3}. 
$m_1$ is the vector meson mass in the corresponding channel:
$m_1^{\pi,\eta} = m_{B^*} = 5.32\,$GeV and $m_1^K = m_{B^*_s} = 5.41\,$GeV. 
The scale of $f_T$ is $\mu = 4.8\,$GeV.}\label{tab:fitpars}
\end{table}

\begin{table}[tb]
\addtolength{\arraycolsep}{2pt}
\renewcommand{\arraystretch}{1.2}
$$
\begin{array}{|c|cccc|}
\hline
& {\rm set~1}& {\rm set~2}& {\rm set~3}& {\rm set~4}\\\hline
{\rm fit~1} & 0.97 & 1\phantom{.00} & 1.01 & 1.05\\
{\rm fit~2} & 0.97 & 0.98 & 0.99 & 1.00\\
{\rm fit~3} & 0.95 & 0.98 & 1.00 & 1.04\\\hline
\end{array}
$$
\addtolength{\arraycolsep}{-2pt}
\renewcommand{\arraystretch}{1}
\caption[]{Total semileptonic decay rates 
$\Gamma(B\to\pi e \nu)$ normalised to
  1 for set 2, fit 1, for different formfactor parametrizations and
  input parameter sets.}\label{tab:semi}
\end{table}
\begin{figure}
$$\epsfxsize=0.45\textwidth\epsffile{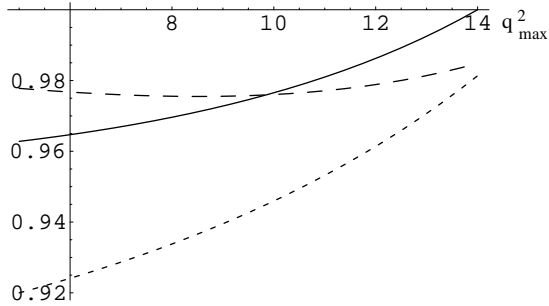}$$
\caption[]{Variation of the total semileptonic rate $\Gamma(B\to\pi
  e\nu)$ as function of $q^2_{\rm max}$, the maximum $q^2$ for which
  LCSR results are included in the fits. The rate is normalized to 1
  for $q^2_{\rm max}=14\,{\rm GeV}^2$ and fit 1. Solid line: fit 1,
  long dashes: fit 2, short dashes: fit 3. Input parameters: set
  2.}\label{fig:semi}
\end{figure}

The accuracy of the fits of the LCSR results to the above
parametrizations is generally very high and best for sets 1 to 3 of
Tab.~\ref{tab:fB} with
$m_b = (4.80\pm 0.05)\,$GeV, with a maximum 1.2\% deviation; set 4
fares slightly worse with an accuracy of 2\% or better. The
quality of the fits is discussed in more detail in
App.~\ref{app:C}. The uncertainty introduced by fitting is much
smaller than the actual uncertainty of the sum rule calculation, which
we have found to be around 10\% at $q^2=0$, and also
much smaller than the intrinsic and irreducible sum rule uncertainty,
which we have estimated to be $\sim 7\%$. Nevertheless it is
legitimate to ask 
whether the extrapolation of the fits to $q^2>14\,{\rm GeV}^2$,
or the variation of the ``cutoff'' $q^2_{\rm max}=14\,{\rm GeV}^2$,
introduce an additional uncertainty. In answering this question, we
first would like to recall that for most applications it is actually
 sufficient to know the formfactors for $q^2<14\,{\rm
  GeV}^2$ only --- these include in particular nonleptonic $B$ decays
treated in QCD factorization, and also the rare decays
$B\to(\pi,K,\eta)\ell^+\ell^-$, as the spectrum for invariant lepton
masses above the $c\bar c$ threshold, i.e.\ $q^2\geq m_{J/\psi}^2
\approx 10\,{\rm GeV}^2$, is dominated by long-distance processes
unrelated to $B\to (\pi,K,\eta)$ formfactors. The only, but very
important case where the formfactor is needed over the full range
of $q^2$ is the semileptonic decay $B\to \pi \ell \nu$, which depends
on $f_+^\pi$ and (for decays into $\tau$) on $f_0^\pi$. We discuss
the effect of the extrapolation on this decay by studying three different
parametrizations of $f_+^\pi$:
\begin{itemize}
\item[fit 1:] Eq.~(\ref{eq:para}), our standard parametrization;
\item[fit 2:] a modified version of (\ref{eq:paraold}), with one zero
  of the denominator fixed at $m_1^2=m_{B^*}^2$:
$$
f_+^\pi(q^2) = \frac{f_+^\pi(0)}{(1-q^2/m_1^2)(1-q^2/m_{\rm
    fit}^2)}\,;
$$
\item[fit 3:] a parametrization 
similar to (\ref{eq:paraKeta}), but with the pole mass as fit
  parameter:
$$f_+^\pi(q^2) = \frac{r_1}{1-q^2/m_{\rm fit}^2} + 
\frac{r_2}{(1-q^2/m_{\rm fit}^2)^2}\,.
$$
\end{itemize}
We quantify the difference between these parametrizations by
calculating the semileptonic decay rate, the integral of
Eq.~(\ref{eq:spectrum}) over  $q^2$ from 0 to $(m_B-m_\pi)^2$, normalizing
to our central values, set 2 and fit 1. The results are
collected in Tab.~\ref{tab:semi}. 
It is evident that the dependence of the rate on the fit is rather
mild, despite the double-pole of fit 3, which is however sufficiently
far away from the endpoint of the spectrum, 
$m_{\rm fit} = (5.6\pm 0.1)\,$GeV, and hence has only moderate 
impact on the rate. We conclude that the extrapolation of $f_+^\pi$ causes an
uncertainty in the total semileptonic decay rate $\Gamma(B\to \pi e \nu)$ which
is considerably
less than the expected intrinsic sum rule uncertainty of $\sim 14\%$. 

We conclude the discussion of the
uncertainty of the extrapolations by studying the effect of changing
the maximum value of $q^2$ for which the sum rules results are
included in the fits. Our default value $q^2_{\rm max}$ is $14\,{\rm GeV}^2$;
lowering $q^2_{\rm max}$ changes the fit parameters of all
three parametrizations and hence the predictions for the total
semileptonic decay rate. Fig.~\ref{fig:semi} shows the
corresponding change in the rate, normalised to our central values fit
1 and $q^2_{\rm max} = 14\,{\rm GeV}^2$. Again the dependence of the
rate on  $q^2_{\rm max}$ is mild, which corroborates
our conclusion that the precise shape of the formfactor is not that
relevant, as long as it does not exhibit too strong a singularity 
at $q^2 = (5.32\,{\rm GeV})^2$.

There are also other tests and checks for the validity of the
extrapolation of (\ref{eq:para}) to the full physical regime
$q^2<(m_B-m_P)^2$:
firstly, the coefficient $r_1$ for $f_+^\pi$ is related to the coupling 
$g_{BB^*\pi}$ as
\begin{equation}
\label{eq:r1}
r_1 = \frac{f_{B^*} g_{BB^*\pi}}{2m_{B^*}} \quad .
\end{equation}
At the upper end of the physical range in $q^2$ we can expect
vector-meson 
dominance to be effective
and therefore the fit-parameter should be close to the above value.
In fact lattice \cite{gBBpi_lat} and meson-loop calculations
(cf.\ the first reference in \cite{LCSRs:reviews}) yield 
$r_1 \approx 0.8$, but are at variance with a determination of
$g_{BB^*\pi}$ from LCSRs which yields $r_1\approx 0.44$ \cite{gBBpi}.   
The lattice and meson-loop calculations are  further supported by
the agreement of their predictions for $g_{DD^*\pi}$ with experimental 
measurements, whereas LCSRs again give a value that is too low by
almost a factor of two. The author of \cite{qhd_double}
speculates that this discrepancy may be explained by a
violation of quark-hadron duality in the LCSRs used for the
determination of $g_{DD^*\pi}$ and $g_{BB^*\pi}$, which would preclude 
a clean determination of these couplings from LCSRs. 
Another possible solution of the problem was suggested in 
Ref.~\cite{BecLC}, where it was shown that 
the value of $r_1$ from LCSRs increases once a 
radial excitation with negative residue is included in the hadronic
parametrization of the correlation function.\footnote{
Note that the corresponding spectral function is not 
positive definite.} If we interpret our fit results as determinations
of $g_{BB^*\pi}$, we get the following values of $r_1$ for the
sets 1 to 4: (0.73,0.74,0.77,0.94) 
(cf.~Tab.~\ref{tab:ffit}), which is in reasonable agreement with
lattice and meson-loop calculations.

Secondly, there is one further constraint on the formfactor $f_0$. 
As first pointed out in Ref.~\cite{softpion}, in
the soft-pion limit $p \to 0$ and $m_{\pi}^2 \to 0$ (i.e.\ $q^2 = m_B^2$) 
$f_0^\pi(m_B^2)$ is related to the decay constants of the $B$ and
$\pi$ as
\begin{equation}
\label{eq:simple}
f_0^\pi(m_B^2) = \frac{f_B}{f_\pi}.
\end{equation}
We can compare this relation with our parametrization by solving it
for $f_B$. For the four parameter sets of Tab.~\ref{tab:fB},
we get from  Eq.~\eqref{eq:simple} 
$f_B^{\rm set 1} = 201\,$MeV,
$f_B^{\rm set 2} = 193\,$MeV, $f_B^{\rm set 3} = 190\,$MeV and $f_B^{\rm
  set 4}= 207\,$MeV, which
is in good agreement with lattice and sum rule calculations.   

Let us conclude with one more remark.
In LEET, $f_+$ and $f_0$ are related as \cite{LEET}:
\begin{equation}
\label{eq:LEET} 
f_0 = \frac{2E}{m_B}\,f_+\,,
\end{equation}
which is valid in the combined limits $m_B\to\infty$ and $E\to\infty$.
This constraint was used in Ref.~\cite{BecKai} to reduce the number of
fit parameters to two as necessitated by the limited accuracy of the lattice
formfactors. We do not impose this constraint
explicitly, but find that it is valid to 4\% accuracy for our
formfactors, for not too large $q^2$.

Summarizing, we conclude that, for all formfactors, 
the three-parameter formula
(\ref{eq:para}) provides both an excellent fit to the LCSR results for
$q^2<14\,{\rm GeV}^2$ and  a smooth extrapolation to $14\,{\rm
  GeV}^2<q^2< (m_B-m_P)^2$, and is consistent with all known constraints.

\section{Summary \& Conclusions}

In this paper we have given a thorough and careful examination of the
predictions of QCD sum rules on the light-cone for the formfactors
$f_+$, $f_0$ and $f_T$ for the decays $B\to\pi,K,\eta$. We have not
discussed $B\to\eta'$, which is not accessible within the method due
to its large mass. 

The main improvements of our results with respect to our previous
publications \cite{PB98,ballroman} are:
\begin{itemize}
\item predictions for all formfactors of $B\to\pi,K,\eta$ transitions
  to $O(\alpha_s)$ accuracy for twist-2 and 3 two-particle
  contributions;
\item a well-defined and precise method for fixing sum rule specific
  parameters (cf.\ Sec.~\ref{sec:borelfix});
\item a careful assessment of uncertainties at zero momentum transfer (cf.\
  Sec.~\ref{sec:hadronic} and \ref{sec:4.3});
\item a detailed breakdown of the contributions of different
  Gegenbauer moments $a_i$ to the formfactors (cf.\ App.~\ref{app:C}), which
\begin{itemize}
\item renders straightforward the
  implementation of future updates of these parameters;
\item allows the assessment of the impact of nonasymptotic twist-2
  distribution amplitudes on QCD factorised nonleptonic B decays in a
  coherent way, to 4th order in the conformal
  expansion;
\end{itemize}
\item a parametrization of the $q^2$-dependence of formfactors valid
  in the full physical regime of momentum transfer that reproduces all
  relevant analytical properties of the formfactors (cf.\
  Sec.~\ref{sec:fitting}).
\end{itemize}

Our main results for $q^2=0$ are collected in Tab.~\ref{tab:final} and
Eq.~(\ref{eq:resq2at0}). They depend crucially on the values of the
Gegenbauer moments describing the twist-2 distribution amplitudes of
$\pi$, $K$ and $\eta$, cf.\ App.~\ref{app:A}. We have determined these
parameters as discussed in Sec.~\ref{sec:hadronic}, but a better
determination from an independent source, e.g.\
lattice calculations, would be extremely useful. This applies in
particular to the SU(3) breaking parameter $a_1^K$, whose size and
even sign is under discussion (cf.\ Ref.~\cite{SU(3)breaking}). 
Once more precise values for these parameters will be available, it is
straightforward to obtain the corresponding formfactors from the
data collected in App.~\ref{app:C}. Setting aside $a_1$,
the total theoretical uncertainty of the formfactors at $q^2=0$ is
10\% to 13\%, which includes 
a variation of all input parameters. It can be further
improved by reducing the uncertainties of, in particular, $a_2$, $a_4$,
the quark condensate and $\eta_3$, the dominant quark-quark-gluon
matrix element. A reduction of the uncertainty of $a_{2,4}$ by a
factor of two will give a $\sim 2\%$ gain in accuracy, reducing the
uncertainty of the quark condensate and $\eta_3$ by the same factor
will give another $2\%$. The uncertainty due to
the variation of only the sum rule specific parameters is 7\%, which cannot
be reduced any further and hence sets the minimum theoretical
uncertainty that can be achieved within this method. Comparing with the
uncertainties quoted in our previous publications, 
we have achieved a reduction of the global
estimate $\sim 15\%$ quoted in \cite{PB98} and also of the 20\%
uncertainty for $f_+^\pi(0)$ quoted in \cite{ballroman}. This
is partially due to a reduction of the uncertainties of the hadronic input
parameters, in particular $m_b$, and partially due to a refinement
of the assessment of sum rule specific uncertainties as discussed in
Sec.~\ref{sec:borelfix}. 

We have also calculated all formfactors for $0\leq q^2\leq 14\,{\rm
  GeV}^2$; the upper bound on $q^2$ is due to the limitations of the
  light-cone expansion which requires the final-state meson to have
  energies $E\gg \Lambda_{\rm QCD}$: for $q^2_{\rm max} = 14\,{\rm
  GeV}^2$ the meson energy is $E = 1.3\,$GeV. In order to allow a
  simple implementation of our results, we have given a
  parametrisation that includes the main features of the analytical
  properties of the formfactors and is valid in the full physical
  regime $0\leq q^2\leq (m_B-m_P)^2$. The corresponding results for
  our preferred set of input parameters are given in
  Tab.~\ref{tab:fitpars}; a detailed breakdown of the contributions of
  different parameters to the full formfactors is given in
  App.~\ref{app:C}. The main features of the results are that the
  formfactors $f_+$ and $f_T$ are nearly equal as predicted by
  LEET
  and that $f_0$ is very well described by a single-pole formula.
The uncertainty induced by the extrapolation of the parametrization to
  larger momentum transfers is an issue only for the semileptonic
  decay $B\to\pi e \nu$; we have checked that the change of the total
  rate is at most 5\% for three different extrapolations of the
  light-cone sum rule results.

Our approach is complementary to standard lattice calculations, in the
sense that it works best for large energies of the final state meson
(i.e.\ small $q^2$), whereas lattice calculations work best for small
energies -- a situation that may change in the
future with the implementation of moving NRQCD \cite{davies}.
Previously, the LCSR results for $f_{+,0}^\pi$ at small and 
moderate $q^2$ were found to 
nicely match the lattice results obtained for large $q^2$
\cite{match}. The situation will have to be reassessed in view of our
new results and it will be very interesting to see if and how it will 
develop with 
further progress in both lattice and LCSR calculations.

\section*{Acknowledgements}
R.Z.\ is greatful to M.B.~Voloshin for clarifications; he is 
supported by the Swiss National Science Foundation.
P.B.\ would like to thank the W.I.\ Fine
Theoretical Physics Institute at the University of Minnesota in Minneapolis for
hospitality while this paper was semi-finalised.

\appendix

\section*{Appendix}
\setcounter{equation}{0}
\renewcommand{\theequation}{A.\arabic{equation}}
\renewcommand{\thetable}{\Alph{table}}
\setcounter{section}{0}
\setcounter{table}{0}

\section{Fit Parameters and Comments}
\label{app:C}

\begin{table}[tb]
\setlength{\extrarowheight}{2.0pt}
\addtolength{\arraycolsep}{2pt}
$$
\begin{array}{|l|rlrrr| rlrrr|}
\hline
& \multicolumn{5}{c|}{\text{set 2, $m_b=4.8\,$GeV}} & 
\multicolumn{5}{c|}{\text{set 4, $m_b=4.6\,$GeV}} \\
\hline
& r_1 & m_1^2 & r_2 & m_{\rm fit}^2 & \Delta & r_1 & m_1 & r_2 & 
m_{\rm fit}^2 & \Delta \\
\hline
f_+^{\pi} &0.744 & (m_{1}^\pi)^2 & -0.486 & 40.73 &0.3 &0.944 & 
(m_{1}^\pi)^2 & -0.669 & 34.27 &0.3 \\
f_0^{\pi} & 0 & - & 0.258 & 33.81 &  0.1  & 0 & - & 0.270 & 33.63  &1.2 \\
f_T^{\pi} & 1.387 &  (m_{1}^\pi)^2 & -1.134 & 32.22 &0.5 &
\multicolumn{5}{l|}{\mbox{use (\ref{eq:Keta}) with~}r_1 = 0.152,}\\
& & & & & &\multicolumn{5}{l|}{r_2 = 0.122, m_1 = m_1^\pi, \Delta = 0.4} \\
\hline
f_+^{\pi,as} &0.918 & (m_{1}^\pi)^2 & -0.675 & 38.20 &0.1 &0.711 & 
(m_{1}^\pi)^2 & -0.441 & 44.31 &0.1 \\
f_0^{\pi,as} & 0 & - &0.244 & 30.46 &0.8 & 0 & - & 0.270 & 31.93 & 0.1 \\
f_T^{\pi,as} & 1.556 & (m_{1}^\pi)^2 & -1.321 & 32.56 &0.2 & 1.331 & 
(m_{1}^\pi)^2 & -1.061 & 33.43 &0.4 \\
\hline
\end{array}
$$
\addtolength{\arraycolsep}{-2pt}
\caption[]{Fit parameters for the $\pi$ Eq.~(\ref{eq:doubledouble})
for both the full formfactors and the asymptotic ones, $f^{as}$, 
Eq.~\eqref{eq:formform}, using the sets 2 and 4 in Tab.~\ref{tab:fB}. 
The formfactor $f_0$ is fitted to the parametrization (\ref{eq:singlesingle}).
The mass parameters $m_1^x$ are given in Tab.~\ref{tab:masses}. 
$\Delta$ is a measure of the quality of the fit and 
is defined in \eqref{eq:Delta}.}
\label{tab:ffit}
\end{table}

\begin{table}[tb]
\addtolength{\arraycolsep}{2pt}
$$
\begin{array}{|l|rrlr| rrlr|}
\hline
& \multicolumn{4}{c|}{\text{set 2, $m_b=4.8\,$GeV}} & 
\multicolumn{4}{c|}{\text{set 4, $m_b=4.6\,$GeV}} \\
\hline
& \multicolumn{1}{c}{r_1} & \multicolumn{1}{c}{r_2} & 
\multicolumn{1}{c}{m_{\rm fit}(m_1)} & \multicolumn{1}{c|}{\Delta} & 
 \multicolumn{1}{c}{r_1} & \multicolumn{1}{c}{r_2} & 
\multicolumn{1}{c}{m_{\rm fit}(m_1)} & \multicolumn{1}{c|}{\Delta}\\
\hline
f_+^{K} & 0.1616 & 0.1730 & (m_1^K)^2   & 1.2 & 0.1903 & 
 0.1478 & (m_1^K)^2   & 1.0 \\
f_0^{K} & 0 & 0.3302 & 37.46 & 1.0 & 0 & 0.3338 & 38.98 & 1.9 \\
f_T^{K} & 0.1614 & 0.1981 & (m_1^K)^2  & 0.5 & 0.1851 & 
 0.1905 & (m_1^K)^2 & 1.7 \\
f_+^{\eta} & 0.1220 & 0.1553 & (m_1^\eta)^2  & 1.0 & 0.1380 & 
 0.1462 &  (m_1^\eta)^2  & 0.9 \\
f_0^{\eta}& 0 &0.2734  & 31.03 & 0.5 & 0 & 0.2799 & 30.46 & 2.0 \\
f_T^{\eta} & 0.1108 & 0.1752 & (m_1^\eta)^2  & 0.5 & 
0.1160 & 0.1841 & (m_1^\eta)^2   & 1.6 \\
\hline
f_+^{K,as} & 0.0541 & 0.2166 & (m_1^K)^2  & 0.2 & 0.0991 & 0.2002
 &  (m_1^K)^2   & 0.6 \\
f_0^{K,as} & 0&0.2719  & 30.33 & 0.7 & 0&0.2984 & 31.99  & 0.5 \\
f_T^{K,as} & 0.0244  & 0.2590 & (m_1^K)^2 
& 0.8 & 0.0660 & 0.2621 &   (m_1^K)^2   & 1.3 \\
f_+^{\eta,as} & 0.0802  & 0.1814 &(m_1^\eta)^2   & 1.0 & 0.1201 
 & 0.1636 & (m_1^\eta)^2   & 0.6 \\
f_0^{\eta,as} & 0 & 0.2604 & 28.80 & 0.5 & 0 & 0.2803 & 29.59 & 0.8 \\
f_T^{\eta,as} & 0.0570  & 0.2115 &  
 (m_1^\eta)^2  & 0.3 & 0.0914  & 0.2096 &  (m_1^\eta)^2  & 1.0 \\
\hline
\end{array}
$$
\addtolength{\arraycolsep}{-2pt}
\caption[]{Fit parameters for $K$ and $\eta$ for Eq.~(\ref{eq:Keta}), 
for both the full formfactors and the asymptotic ones, $f^{as}$, 
Eq.~\eqref{eq:formform}, using the sets 2 and 4 in Tab.~\ref{tab:fB}. 
The formfactor $f_0$ is fitted to the parametrisation \ref{eq:singlesingle}.
The mass parameters $m_1$ are given in Tab.~\ref{tab:masses}. 
$\Delta$ is a measure of the quality of the fit and 
is defined in \eqref{eq:Delta}.}
\label{tab:ffit2}
\end{table}

\begin{table}[tb]
\addtolength{\arraycolsep}{2pt}
\renewcommand{\arraystretch}{1.1}
$$
\begin{array}{|l|rrrrr| rrrrr|}
\hline
& \multicolumn{5}{c|}{\text{set 2,    $m_b$=4.8 GeV}} & 
\multicolumn{5}{c|}{\text{set 4,    $m_b$=4.6 GeV}} \\
\hline
& \multicolumn{1}{c}{a} & \multicolumn{1}{c}{b\times 10^2} &
\multicolumn{1}{c}{c\times 10^2} & \multicolumn{1}{c}{d\times 10^3} &
\multicolumn{1}{c|}{\delta} & \multicolumn{1}{c}{a} & 
\multicolumn{1}{c}{b\times 10^2} &
\multicolumn{1}{c}{c\times 10^2} & \multicolumn{1}{c}{d\times 10^3} &
\multicolumn{1}{c|}{\delta}\\\hline
f_+^K(a_1)&        0.310     &   0.930     &   0.139    &   -0.083   & 0.3    &   0.276    &   
0.060     &   0.151    &   -0.157   &   0.7   \\
f_0^K(a_1)&        0.308    &   0.106    &   0.026    &   -0.048   & 0.2   &   0.273    &   
-0.433   &   0.0001   &   -0.051   &   0.2   \\
f_T^K(a_1)&        0.381    &   1.056    &   0.167    &   -0.108   & 0.3    &   0.354    &   
0.027    &   0.178    &   -0.194   &   0.7   \\
\hline
f_+^{\pi}(a_2)&    0.187    &   -0.517   &   0.014    &   -0.117   & 0.5    &   0.040     &   
-0.762   &   -0.201   &   0.050     &   1.5   \\
f_0^{\pi}(a_2)&    0.185    &   -0.841   &   -0.075   &   -0.005   & 0.4    &   0.041    &   
-1.078   &   -0.123   &   0.068    &   1.2   \\
f_T^{\pi}(a_2)&    0.203    &   -0.659   &   -0.008   &   -0.118   & 0.3    &   0.038    &   
-0.944   &   -0.244   &   0.073    &   1.5   \\
f_+^K(a_2)&        0.228    &   -0.632   &   0.017    &   -0.143   & 0.5    &   0.049    &
   -0.931   &   -0.245   &   0.061    &   1.5   \\
f_0^K(a_2)&        0.226    &   -1.031   &   -0.092   &   -0.005   & 0.4    &   0.050     & 
  -1.32\phantom{0}    &   -0.150    &   0.083    &   1.2   \\
f_T^K(a_2)&        0.264    &   -0.858   &   -0.011   &   -0.153   & 0.3    &   0.049    &
   -1.228   &   -0.318   &   0.095    &   1.5   \\
f_+^{\eta}(a_2)&   0.185    &   -0.514   &   0.014    &   -0.116   & 0.5    &   0.039    &
   -0.757   &   -0.199   &   0.049    &   1.5   \\
f_0^{\eta}(a_2)&   0.183    &   -0.829   &   -0.076   &   -0.002   & 0.4    &   0.041    & 
  -1.068   &   -0.122   &   0.069    &   1.2   \\
f_T^{\eta}(a_2)&   0.216    &   -0.722   &   -0.007   &   -0.128   & 0.3    &   0.040     & 
  -1.019   &   -0.259   &   0.076    &   1.4   \\
\hline
f_+^{\pi}(a_4)&    -0.141   &   -0.775   &   0.004    &   0.161    & 0.7    &   -0.054   &
   -0.506   &   0.621    &   -0.326   &   5.2   \\
f_0^{\pi}(a_4)&    -0.139   &   -0.687   &   0.170     &   0.002    & 1.5    &   -0.061   & 
  0.703    &   0.323    &   -0.209   &   2.9   \\
f_T^{\pi}(a_4)&    -0.167   &   -0.895   &   0.077    &   0.143    & 1.1    &   -0.047   &
   -0.327   &   0.698    &   -0.394   &   4.9   \\
f_+^K(a_4)&        -0.173   &   -0.947   &   0.005    &   0.196    & 0.7    &   -0.067   &
   -0.618   &   0.759    &   -0.398   &   5.2   \\
f_0^K(a_4)&        -0.170    &   -0.838   &   0.209    &   0.001    & 1.5    &   -0.075   &
   0.871    &   0.392    &   -0.254   &   2.9   \\
f_T^K(a_4)&        -0.217   &   -1.165   &   0.101    &   0.187    & 1.1    &   -0.061   &   
-0.426   &   0.909    &   -0.513   &   4.9   \\
f_+^{\eta}(a_4)&   -0.140    &   -0.770    &   0.004    &   0.159    & 0.7    &   -0.054   &   
-0.502   &   0.616    &   -0.323   &   5.2   \\
f_0^{\eta}(a_4)&   -0.138   &   -0.681   &   0.170     &   0.0005   & 1.5    &   -0.061   &   
0.710     &   0.318    &   -0.206   &   2.9   \\
f_T^{\eta}(a_4)&   -0.178   &   -0.955   &   0.083    &   0.153    & 1.1    &   -0.050    &   
-0.349   &   0.745    &   -0.421   &   4.9   \\
\hline
\end{array}
$$
\addtolength{\arraycolsep}{-2pt}
\renewcommand{\arraystretch}{1}
\caption[]{Fit parameters for Eq.~(\ref{eq:poly}) for the functions 
$f^{a_i}$ defined in (\ref{eq:formform}). $\delta$ is a measure of 
the quality of the fit and defined in \eqref{eq:delta}.}\label{tab:afit}
\end{table}

This appendix extends the discussion of Sec.~\ref{sec:fitting}.
\paragraph{Full formfactors.}
As discussed in Sec.~\ref{sec:fitting}, we fit the LCSR results to 
the following parametrizations:
\begin{itemize}
\item for $f_{+,T}^\pi$:\footnote{Apart from $f_T^\pi$ for set 4,
  which shows the same behavior as $f_{+,T}^{K,\eta}$ and hence is
  parametrised the same way, i.e.\ according to (\ref{eq:Keta}).}
\begin{equation}
\label{eq:doubledouble}
f(q^2) = \frac{r_1}{1-q^2/m_1^2} + \frac{r_2}{1-q^2/m_{\rm fit}^2} \quad ,
\end{equation}
where $m_1^\pi$ is the mass of $B^*(1^-)$, $m_1^\pi=5.32\,$GeV; the
fit parameters are $r_1$, $r_2$ and $m_{\rm fit}$;
\item for $f_{+,T}^{K,\eta}$ and $f_T^\pi$ (set 4):
\begin{equation}
\label{eq:Keta}
f(q^2) = \frac{r_1}{1-q^2/m_{1}^2} + \frac{r_2}{(1-q^2/m_{1}^2)^2},
\end{equation}
where $m_1$ is the mass of the $1^-$ meson in the corresponding
channel, cf.\ Tab.~\ref{tab:masses}; the fit parameters are $r_1$ and $r_2$; 
\item for $f_0$:
\begin{equation}
\label{eq:singlesingle}
f_0(q^2) = \frac{r_2}{1-q^2/m_{\rm fit}^2}\,,
\end{equation}
the fit parameters are $r_2$ and $m_{\rm fit}$.
\end{itemize}
The fit parameters are collected in the upper halves of
Tabs.~\ref{tab:ffit} and \ref{tab:ffit2}. $\Delta$ is a measure of
the quality of the fit  and defined as
\begin{equation}
\label{eq:Delta}
\Delta = 100\,\max_{t}\, \left| \frac{f(t)-f^{\rm fit}(t)}{f(t)} \right| 
\,, \quad t \in \{0,\tfrac{1}{2},\dots,\tfrac{27}{2}, 14\}\,\text{GeV}^2,
\end{equation}
i.e.\ it gives, in per cent, 
the maximum deviation of the fitted formfactors from the
original LCSR result for $q^2<14\,{\rm GeV}^2$.
From the $\Delta$ given in the table we conclude that the overall
quality of the fits is very good and best for the pion and also 
that they work better for our preferred set 2 than for set 4. 
\paragraph{Split formfactors.}
As discussed in Sec.~\ref{sec:hadronic}, the values of the Gegenbauer moments
$a_{1,2,4}$ are not very well known. In Sec.~\ref{sec:fitting} and
Tabs.~\ref{tab:ffit}, \ref{tab:ffit2} we have presented results only
for our preferred 
choice of these parameters, i.e.\
$$
\begin{array}{r@{\:=\:}l@{\,,\quad}r@{\:=\:}l@{\,,\quad}r@{\:=\:}l}
a_1^K(1\,{\rm GeV}) & 0.17 & a_2^{\pi,K,\eta}(1\,{\rm
  GeV})&0.115 & a_4^{\pi,K,\eta}(1\,{\rm GeV}) & -0.015,\\[5pt]
a_1^K(2.2\,{\rm GeV}) & 0.135 & a_2^{\pi,K,\eta}(2.2\,{\rm
  GeV}) & 0.080 & a_4^{\pi,K,\eta}(2.2\,{\rm GeV}) &
 -0.0089;
\end{array}
$$
for set 4, the $a_i$ are scaled up to $\mu_{\rm IR}= 2.6\,$GeV.
In order to allow the inclusion of future updates of these
values, we split the formfactors into contributions from different
Gegenbauer moments. We define\footnote{Note that this splitting is
  exact and valid for arbitrarily large $a_i$ --- there are no nonlinear terms
  in $a_i$.}
\begin{equation}
\label{eq:formform}
f(q^2) = f^{as}(q^2)+a_1(\mu_{\rm IR}) 
f^{a_1}(q^2)+a_2(\mu_{\rm IR}) f^{a_2}(q^2) + a_4(\mu_{\rm IR}) f^{a_4}(q^2),
\end{equation}
where $f^{as}$ contains twist-2 contributions from the asymptotic DA
and also all higher-twist contributions not proportional to
$a_{1,2,4}$. The task is now to fit all functions
$f^{as,a_1,a_2,a_4}$, in the 
interval $0 < q^2 < 14 \text{GeV}^2$,
to appropriate parametrizations. 

For $f^{as}$, which gives the dominant contribution to all
formfactors, we use the same parametrisation as for the full formfactors.
The results are collected in the lower halves of  
Tabs.~\ref{tab:ffit} and \ref{tab:ffit2}. Again, the fits
are very good and best for the pion and set 2. 
\begin{table}
\setlength{\extrarowheight}{2.0pt}
\renewcommand{\arraystretch}{1.1}\addtolength{\arraycolsep}{3pt}
$$
\begin{array}{|l|l@{\hspace*{1pt}}c|l@{\hspace*{1pt}}c|c|}
\hline
& \;m_1^2& (1^-)              & \;m_{1*}^2 & (0^+) & \; q^2_{\text{max}} \\
\hline 
\pi\,(\eta)  & \; 5.32^2 & = 28.4 & \; 5.63^2 & = 31.7 & \;26.4\: (22.8)\\  
K    & \; 5.41^2 & = 29.3 & \; 5.72^2  & = 32.7 & \;23.8\\\hline
\end{array}
\renewcommand{\arraystretch}{1}\addtolength{\arraycolsep}{-3pt}
$$
\caption[]{Masses of $1^-$  and $0^+$ resonances in the $b\bar u$ and
  $b\bar s$ channels. The $1^-$ masses are obtained from experiment
  and heavy-quark relations, the $0^+$ masses from a potential model
  \cite{bardeen}. All numbers in units 
$\text{GeV}^2$.}\label{tab:masses}
\end{table}

The $f^{a_i}$ turn out to be slowly varying functions of $q^2$, which 
can be fitted by a polynomial of 3rd degree:
\begin{equation}\label{eq:poly}
f^{a_i}(q^2) = a + b\,(q^2)+c\,(q^2)^2+d\,(q^2)^3 \quad .
\end{equation}
The measure of the quality of the fit has now to be defined in a slightly
different way, as the $f^{a_i}$ have zeros in the fit interval. We define
the fit-quality $\delta$ as
\begin{equation}
\label{eq:delta}
\delta = 100\:\frac{\sum_t|f(t)-f^{\rm fit}(t)|}{\sum_{t}|f(t)|} 
\,,\quad t \in \{0,\tfrac{1}{2}, \dots,\tfrac{27}{2},14\}\,\text{GeV}^2\,,
\end{equation}
i.e.\ as the average deviation of the fit from the true value, in per cent.
The fit parameters are given in Tab.~\ref{tab:afit}. As one can read
off from the $\delta$'s, the fits are best for $f^{a_1}$, 
still good for $f^{a_2}$ and worst for
$f^{a_4}$. The limited quality of the fits for $f^{a_4}$ is due to a
change of sign of its derivative at the upper end of the
fit interval, which cannot be reliably reproduced by a polynomial of
3rd degree. 

We would like to stress that none of the split-formfactor
parametrizations must be used for $q^2$ larger than 14$\,{\rm
  GeV}^2$. For calculating the full formfactors for arbitrary
$a_{1,2,4}$, the following procedure should be followed:
\begin{itemize}
\item determine $a_{1,2,4}$ at the scale $\mu_{\rm IR}^2 =
  m_B^2-m_b^2$; the scaling factors from $\mu =1\,$GeV up to $2.2\,$GeV
  (i.e.\ $m_b = 4.8\,$GeV) are $(0.793,0.696,0.590)$ for $(a_1,a_2,a_4)$;
\item choose set 2 (preferred) or set 4;
\item calculate $f^{as}$ from the appropriate formula 
(\ref{eq:doubledouble}), (\ref{eq:Keta}) or (\ref{eq:singlesingle}),
   using the fit parameters
  from Tab.~\ref{tab:ffit} or \ref{tab:ffit2};
\item calculate $f^{a_{1,2,4}}$ from (\ref{eq:poly}), using the fit parameters
  from Tab.~\ref{tab:afit};
\item calculate the total formfactor from (\ref{eq:formform});
\item extend the formfactor to the full kinematical regime by fitting
  it to (\ref{eq:doubledouble}), (\ref{eq:Keta}) or
  (\ref{eq:singlesingle}). 
\end{itemize}

\section{Distribution Amplitudes}\label{app:A}

In this appendix we collect explicit expressions for all the DAs that
enter the formfactors. These expressions are well known and have been
taken from Ref.\ \cite{wavefunctions}.

The key point is that, to leading order in QCD, DAs can be expressed
as a partial wave expansion in terms of contributions of increasing
conformal spin, the so-called conformal expansion. The coefficients of
different partial waves renormalize multiplicatively to LO in QCD, but
mix at NLO, the reason being that the symmetry underlying the
conformal expansion, the conformal symmetry of massless QCD, is
anomalous and broken by radiative corrections. 

The two-particle twist-2 amplitude \eqref{eq:22} is expanded as
\begin{equation}\label{eq:da}
\phi(u,\mu) = \phi_{\text{as}}(u)\sum_{n \geq 0} a_n(\mu)
C^{3/2}_n(\zeta)
\end{equation}
with $\zeta \equiv 2u-1$ and $a_0=1$ from 
normalization:
$$\int_0^1 du\,\phi(u,\mu) = 1.$$
The $C^{3/2}_n(\zeta)$ are 
Gegenbauer polynomials. The conformal spin of the term in $C_n^{3/2}$ is
$j=n+2$. For the $\pi$ and $\eta$ one has $a_{2n+1}=0$ due to
G-parity, but $a_1^{K} \sim (m_s-m_q)$ for the $K$
\cite{SU(3)breaking}, which is one source of SU(3) breaking for the
formfactors. 

As only the first few  Gegenbauer moments $a_n$ are known numerically,
we truncate the series at $n=4$; 
the values of the  conformal spins included are listed in 
Tab.~\ref{tab:overview}, whereas the numerical values of the $a_i$ are
discussed in Sec.~\ref{sec:4}. The truncation is justified as long as the
perturbative kernels $T$ with which the DAs are convoluted are
slowly varying functions of $u$, so that the rapidly oscillating
Gegenbauers suppress terms with high $n$. In our case the $T$ are
nonsingular for all $u$, including the endpoints $u=0,1$, so the
truncation of the series is justified.
The term labeled $\phi_{\text{as}}$ in (\ref{eq:da}) is the asymptotic
DA which is reached for large scales $\mu\to\infty$; it is completely
determined by perturbation theory and given by
$$\phi_{\text{as}}(u)=6u(1-u);$$
it is the same for all mesons.
The Gegenbauer moments $a_n$ become relevant at moderate scales and
depend on the hadron in question.

Let us now define the three-particle DAs.
To twist-3 accuracy, there
is only one:
\begin{eqnarray}
\lefteqn{\langle 0 | \bar u(x) \sigma_{\mu\nu}\gamma_5
  gG_{\alpha\beta}(vx) d(-x)| \pi^-(p)\rangle\ =}\hspace*{0.5cm}\nonumber\\
& = & i\,\frac{f_\pi m_\pi^2}{m_u+m_d} \left(p_\alpha p_\mu
  g_{\nu\beta} - p_\alpha p_\nu
  g_{\mu\beta} - p_\beta p_\mu g_{\nu\alpha} + p_\beta
  p_\nu g_{\alpha\mu} \right) {\cal T}(v,p\cdot x) +
  \dots,\label{eq:3pT3}
\end{eqnarray}
where the ellipses stand for Lorentz structures of twist-5 and higher
and where we used the following short-hand notation for the integral
defining the three particle DA:
\begin{equation}
{\cal T}(v,p\cdot x) = \int {\cal D}\underline{\alpha} \, e^{-ip\cdot
  x(\alpha_u
  -\alpha_d + v\alpha_g)} {\cal T}(\alpha_d,\alpha_u,\alpha_g).
\end{equation}
Here $\underline{\alpha}$ is the set of three momentum fractions
$\alpha_d$ ($d$ quark), $\alpha_u$ ($u$ quark) and $\alpha_g$
(gluon). The integration measure is defined as
$$
\int {\cal D}\underline{\alpha} = \int_0^1 d\alpha_d d\alpha_u
d\alpha_g \delta(1-\alpha_u-\alpha_d-\alpha_g).
$$
There are also four three-particle DAs of twist-4, defined as
\begin{eqnarray}
\lefteqn{\langle 0 | \bar u(x)\gamma_\mu\gamma_5
gG_{\alpha\beta}(vx)d(-x)|\pi^-(p)\rangle\ =}\hspace*{0.5cm}\nonumber\\
& = & p_\mu (p_\alpha x_\beta - p_\beta x_\alpha)\, \frac{1}{p\cdot x}\, f_\pi
m_\pi^2 {\cal A}_\parallel(v,p\cdot x) + (p_\beta g_{\alpha\mu}^\perp -
p_\alpha g_{\beta\mu}^\perp) f_\pi m_\pi^2 {\cal A}_\perp(v,p\cdot x),\\
\lefteqn{\langle 0 | \bar u(x)\gamma_\mu i
g\widetilde{G}_{\alpha\beta}(vx)d(-x)|\pi^-(p)\rangle\
=}\hspace*{0.5cm}\nonumber\\
& = & p_\mu (p_\alpha x_\beta - p_\beta x_\alpha)\, \frac{1}{p\cdot x}\, f_\pi
m_\pi^2 {\cal V}_\parallel(v,p\cdot x) + (p_\beta g_{\alpha\mu}^\perp -
p_\alpha g_{\beta\mu}^\perp) f_\pi m_\pi^2 {\cal V}_\perp(v,p\cdot x);
\end{eqnarray}
$g^\perp_{\mu\nu}$ is defined as
$$
g^\perp_{\mu\nu} = g_{\mu\nu} -\frac{1}{p\cdot x}(p_\mu x_\nu+ p_\nu x_\mu).
$$
To next-to-leading
conformal spin ($j=7/2,9/2$), the twist-3 three-particle distribution
amplitude ${\cal T}$ is given 
by\footnote{In the literature the notation $f_{3\pi}=f_{\pi}\eta_3$ is
  also widely used.} 
\begin{equation*}
{\cal T}(\alpha_u,\alpha_d,\alpha_g) = 360 \eta_3 \alpha_u\alpha_d\alpha_g^2
\{1+\omega_3\frac{1}{2}(7\alpha_g-3)\} \quad .
\end{equation*}
The two-particle twist-3 distribution amplitudes $\phi_p$ and
$\phi_\sigma$ in Eqs.~\eqref{eq:32a} and 
\eqref{eq:32b} depend on ${\cal T}$
through the equations of motions \cite{wavefunctions},\footnote{An
  explicit expression for $\phi_p$ in terms of ${\cal T}$ is given in 
Ref.~\cite{PBSCET}, Eq.~(16).} 
which implies that their coefficients are not independent
from each other.
The expansion up to NNL order ($j=3/2,7/2,9/2$) reads\footnote{At 
first glance it seems that 
$\phi_p$ is taken to a higher order in conformal expansion than 
$\phi_{\sigma}$, but as discussed in the first reference of 
\cite{wavefunctions}, $\phi_p$ and $\phi_{\sigma}$
are not pure spin projections, which means that the coefficients of a
given Gegenbauer polynomial contain contributions from different
partial waves.} 
\begin{eqnarray*}
\phi_p(u)
&=& 1+\{30\eta_3-\frac{5}{2}\rho_{\pi}^2\}C^{1/2}_2(\zeta)+\{-3\eta_3\omega_3
-\frac{27}{20}\rho_{\pi}^2-\frac{81}{10}\rho_{\pi}^2a_2\}C^{1/2}_4(\zeta), \\
\phi_{\sigma}(u)&=& 6u(1-u)\big\{1+\{5\eta_3-\frac{1}{2}\eta_3\omega_3-
\frac{7}{20}\rho_{\pi}^2-\frac{3}{5}\rho_{\pi}^2a_2\} C^{3/2}_2(\zeta) \quad .
\end{eqnarray*}
The two-particle twist-4 corrections  $g_{\pi}$ and ${\mathbb A}$
in Eq.~\eqref{eq:22} are
given to NNL conformal spin ($j=1,3,5$) by\footnote{Note that,
  contrary to appearances, the contributions of $g_\pi$ and ${\mathbb
    A}$ to
  \eqref{eq:22} do not vanish for zero meson mass: $\eta_4$
  implicitly contains a factor $1/m_\pi^2$ and survives in the limit
  $m_\pi^2\to 0$.}
\begin{eqnarray*}
g_{\pi}(u)&=& 1+\{1+\frac{18}{7}a_2+60\eta_3+\frac{20}{3}\eta_4\}
C^{1/2}_2(\zeta)
+\{-\frac{9}{28}a_2-6\eta_3\omega_3\}C^{1/2}_4(\zeta),\\
{\mathbb A}(u) & = & 6u\bar u \left\{ \frac{16}{15} + \frac{24}{35} \,
  a_2 + 20 \eta_3 + \frac{20}{9} \,\eta_4 \right.\\
& &\left. + \left( -\frac{1}{15} + \frac{1}{16}\, - \frac{7}{27}\, \eta_3
  \omega_3 - \frac{10}{27}\, \eta_4 \right) C_2^{3/2}(\xi) + \left
  ( -\frac{11}{210} \, a_2 - \frac{4}{135}\, \eta_3\omega_3 \right)
  C_4^{3/2}(\xi) \right\} \\
& & {}+ \left(-\frac{18}{5}\, a_2 + 21\eta_4\omega_4 \right) \left\{ 2
  u^3 (10-15 u + 6 u^2)\ln u + 2\bar u^3 (10-15\bar u + 6 \bar u^2)
  \ln\bar u\right.\\
& & \hspace*{4.3cm}\left. + u \bar u (2 + 13u\bar u)\right\}.
\end{eqnarray*}
Finally the three-particle twist-4 DAs
are to NL spin ($j=3,5$) given by
\begin{alignat*}{2}
&{\cal A}_{\parallel}(\underline{\alpha})&\;=\;& 120
  \alpha_u\alpha_d
\alpha_g(
a_{10} (\alpha_d-\alpha_u)\}, \\
&{\cal V}_{\parallel}(\underline{\alpha})&\;=\;& 120
  \alpha_u\alpha_d
\alpha_g(v_{00}+
v_{10}(3\alpha_g-1)\}, \\
&{\cal A}_{\perp}(\underline{\alpha})&\;=\;&
  30\alpha_g^2(\alpha_u-
\alpha_d)[h_{00}+
h_{01}\alpha_g 
+h_{10}(5\alpha_g-3)/2\}, \\
&{\cal V}_{\perp}(\underline{\alpha})&\;=\;& -30\alpha_g^2\{h_{00}
\overline{\alpha}_g+
h_{01}[\alpha_g\overline{\alpha}_g-6\alpha_u\alpha_d] 
+h_{10}[\alpha_g\overline{\alpha}_g-3/2
(\alpha_u^2+\alpha_d^2)]\} \, ,
\end{alignat*}
where $\overline{\alpha}=1-\alpha$ and the $a_{ij}$, $v_{ij}$ and $h_{ij}$
are related to hadronic matrix elements $\eta_4$, $\omega_4$ and $a_2$ as
$$
\begin{array}{r@{~=~}l@{\,,\quad}r@{~=~}l@{\,,\quad}r@{~=~}l}
a_{10} & \frac{21}{8}\eta_4\omega_4-\frac{9}{20}a_2 &
v_{10} & \frac{21}{8}\eta_4\omega_4 &
v_{00} & -\frac{1}{3}\eta_4\,, \\[5pt]
h_{01} & \frac{7}{4}\eta_4\omega_4-\frac{3}{20}a_2 & 
h_{10} & \frac{7}{2}\eta_4\omega_4+\frac{3}{20}a_2 &
v_{00} & -\frac{1}{3}\eta_4 \,.
\end{array}
$$
Taking everything together, 
we have 7 hadronic parameters $\{c_i\}=\{a_1,a_2,a_4,\eta_3,
\omega_3,\eta_4,\omega_4\}$ which parametrize all DAs to twist-4 and NLO in
conformal spin.
The $c_i$ are scale-dependent and are usually given at the
scale $1\,\text{GeV}$. To LO in QCD, they do not mix under
renormalisation, so that the scaling up to 
$\mu_{\text{IR}} = \sqrt{m_B^2-m_b^2}$ is given by
\begin{equation*}
c_i(\mu_{\text{IR}}) = L^{\gamma_{c_i}/\beta_0} c_i(1\,{\rm GeV}) ,
\end{equation*}
with $L = \alpha_s(\mu_{\text{IR}})/\alpha_s(1\,{\rm GeV})$,  
$\beta_0=11-2/3N_f$. The one-loop anomalous dimensions $ \gamma_{c_i}$
are given in Tab.~\ref{tab:anomalous}. 
\begin{table}[bt]
\begin{center}
\begin{tabular}{|l|c|c|c|c|c|c|c|c|c|c|}
\hline
& \multicolumn{5}{c|}{tree} & \multicolumn{5}{c|}{$O(\alpha_s)$} \\
\hline
twist & 2  & \multicolumn{2}{c|}{3} & \multicolumn{2}{c|}{4} & 2 & 
\multicolumn{2}{c|}{3} & \multicolumn{2}{c|}{4}  \\ 
\hline
x-particle & 2  &  2              & 3              &  2 & 3 & 2  &  2  
            & 3              &  2 & 3 \\
\hline
$j_L$      & 2  &  $\frac{3}{2}$  & $\frac{7}{2}$  &  1 & 3 & 2  &  
$\frac{3}{2}$  &        -      &  - & - \\
$j_{NL}$   & 4  &  $\frac{7}{2}$  & $\frac{9}{2}$  &  3 & 5 & 4  &  
$\frac{7}{2}$  &        -       &  - & - \\
$j_{NNL}$  & 6  &  $\frac{9}{2}$ &        -        &  5 & - & 6  &  
-              &        -       &  - & - \\ 
\hline
\end{tabular}
\end{center}
\caption{Overview of the contributions included
in the calculations. For the $K$ we also include conformal spin $j=3$ for
twist-2 which explicitly parametrizes  SU(3) flavor 
breaking.}\label{tab:overview} 
$$
\addtolength{\arraycolsep}{3pt}
\renewcommand{\arraystretch}{1.4}
\begin{array}{|c|c|c|c|c|}
\hline
\ds\gamma_{a_n} & \ds\gamma_{\eta_3} & \ds\gamma_{\omega_3} & 
\ds\gamma_{\eta_4} & \ds\gamma_{\omega_4} \\[5pt]
\hline &&&&\\[-15pt]
\ds C_F\left(1-\frac{2}{(n+1)(n+2)}-\sum_{m=2}^{n+1}\frac{1}{m}\right) &
\ds\frac{16}{3}C_F+C_A & 
\ds-\frac{25}{6}C_F+\frac{7}{3}C_A & \ds\frac{8}{3}C_F &
\ds-\frac{8}{3}C_F+\frac{10}{3}C_A \\[12pt]
\hline
\end{array}
\addtolength{\arraycolsep}{-3pt}
\renewcommand{\arraystretch}{1}
$$
\caption[]{One-loop anomalous dimensions of hadronic parameters in 
DAs.}\label{tab:anomalous}
\begin{center}
\begin{tabular}{|c|ccc|}
\hline
& $\pi$ & $K$ & $\eta$\\ \hline
$\eta_3$ & 0.015 & 0.015 & 0.013\\
$\omega_3$ &  $-3$ & $-3$ & $-3$\\
$\eta_4$ & 10 & 0.6 & 0.5\\
$\omega_4$ & 0.2 & 0.2 & 0.2\\ \hline
\end{tabular}
\end{center}
\caption{Input parameters for twist-3 and 4  DAs, 
calculated from QCD sum rules.
The accuracy is about 50\%. Renormalization scale is
1~GeV.}\label{tab:t3}
\end{table}
Note  that  the anomalous 
dimension  increases with increasing conformal spin, 
 $\gamma \sim \log j$,
which implies that the truncation of the conformal expansion becomes
the better the high the scale. The numerical values for all these parameters at
the scale $\mu=1\,$GeV are collected in Tab.~\ref{tab:t3}, taken from
Ref.~\cite{wavefunctions}.

\section{Spectral Densities for $f_+$}\label{app:B}

The total spectral density of $\Pi_+$ is obtained as sum of all the
contributions listed below, i.e.\
$$\rho_{\Pi_+} = \rho_{\rm T2} + \rho_{\rm T3} + \rho_\sigma + \rho_p
  + \rho_{\rm T4}^{\rm 2p} + \rho_{\rm T4}^{\rm 3p}. 
$$
$\rho_{\rm T2}$ is the contribution from the twist-2 DA, $\rho_{\rm
    T3}$ from the twist-3 three-particle DA, $\rho_{\sigma(p)}$ from
  the twist-3 two-particle DA $\phi_{\sigma(p)}$ and $\rho_{\rm
    T4}^{\rm 2(3)p}$ from the two(three)-particle DAs of twist-4.
There is also one constant term,  ${\rm
  T4}_c$, which is due to twist-4 corrections that cannot be expressed
via a dispersion relation, so that the total Borel-transformed $\Pi_+$
is given as
$$\hat{B} \Pi_+ = \int_{m_b^2}^\infty ds\,\rho_{\Pi_+}(s)\,e^{-s/M^2}
+ {\rm T4}_c.
$$

{\footnotesize
\begin{alignat*}{2}
&\rho_{\rm T2}\, &\,=\;&
\frac{3\,f_{\pi}\,m_b}{{( q^2 - s ) }^7}  ( m_b^2 - q^2 ) \,
     ( m_b^2 - s ) \,
     ( 15\,a_4\,( 42\,m_b^8 + 
          q^8 + 10\,q^6\,s + 
          20\,q^4\,s^2 + 10\,q^2\,s^3 + s^4 \\&\,&-&\, 
          84\,m_b^6\,( q^2 + s ) + 
          28\,m_b^4\,( 2\,q^2 + s ) \,
           ( q^2 + 2\,s )  - 
          14\,m_b^2\,( q^2 + s ) \,
           ( q^4 + 4\,q^2\,s + s^2 )  )  +   
       {( q^2 - s ) }^2\, \\&\,&\times&\,
        ( 6\,a_2\,( 5\,m_b^4 + q^4 + 3\,q^2\,s + s^2 - 
             5\,m_b^2\,( q^2 + s )  )  + 
          ( q^2 - s ) \,
           ( a_0\,( q^2 - s )  + 
             3\,a_1\,( -2\,m_b^2 + q^2 + s )  )  ) ) \\&\,&+&\,  
 \mathfrak{a_s} \big\{ \frac{3\,{\bf a_0}\,f_{\pi}\,m_b}{{s( q^2 - s ) }^3}
       ( ( m_b^2 - s ) \,
             ( -2\,m_b^2\,q^2 + 2\,q^4 + 
               m_b^2\,( 4 + {\pi }^2 ) \,s - 
               ( 1 + {\pi }^2 ) \,q^2\,s - 3\,s^2 ) +  
            ( m_b^2 - q^2 ) \\&\,&\times&\,
             ( s\,(s-m_b^2) \,
                {\log (1 - \frac{q^2}{m_b^2})}^2  + 
               s\,\log (\frac{s}{m_b^2})\,
                ( -2\,s + 
                  ( m_b^2 - s ) \,
                   \log (\frac{s}{m_b^2}) )  + 
               2\,( m_b^2 - s ) \,
                ( 2\,m_b^2 - 5\,s \\&\,&+&\,
                  2\,s\,\log (\frac{s}{m_b^2}) ) \,
                \log (\frac{s}{m_b^2}-1) + 
               2\,s\,(s-m_b^2) \,
                {\log (\frac{s}{m_b^2}-1)}^2 - 
               2\,( m_b^2 - s ) \,
                \log (1 - \frac{q^2}{m_b^2})\,
                ( m_b^2 \\&\,&+&\, s + 
                  s\,\log (\frac{s}{m_b^2}) - 
                  2\,s\,\log (\frac{s}{m_b^2}-1) )  ) 
+ 2\,( m_b^2 - q^2 ) \,( m_b^2 - s ) \,
          ( \text{Li}_2(\frac{q^2}{q^2-m_b^2}) + 
            \text{Li}_2(1 - \frac{m_b^2}{s})  \\&\,&+&\,
            4\,\text{Li}_2(1 - \frac{s}{m_b^2}) )  ) +  \frac{{\bf a_1}\,f_{\pi}\,m_b}{s{( q^2 - s ) }^4}
       ( ( m_b^2 - s ) \,
             ( 6\,q^2\,( 6\,m_b^4 - 6\,m_b^2\,q^2 + 
                  q^4 )  + 
               ( -2\,m_b^4\,
                   ( 95 + 9\,{\pi }^2 ) \\&\,&+&\,   
                  27\,m_b^2\,( 6 + {\pi }^2 ) \,q^2 + 
                  ( 3 - 9\,{\pi }^2 ) \,q^4 ) \,s + 
               ( m_b^2\,( 158 + 9\,{\pi }^2 )  - 
                  3\,( 50 + 3\,{\pi }^2 ) \,q^2 )
                s^2 - 13\,s^3 )  + 
            3\,( 8\,( m_b^2 - s )  \\&\,&\times&\,
                ( q^2- m_b^2) \,
                ( 2\,m_b^2 - q^2 - s ) \,s\,
                \log (\frac{{\mu}^2}{m_b^2}) + 
               3\,( m_b^2 - q^2 ) \,
                ( m_b^2 - s ) \,
                ( 2\,m_b^2 - q^2 - s ) \,s\,
                {\log (1 - \frac{q^2}{m_b^2})}^2  \\&\,&+&\,
               s\,\log (\frac{s}{m_b^2})\,
                ( 12\,m_b^4\,( q^2 + s )  + 
                  14\,q^2\,s\,( q^2 + s )  - 
                  2\,m_b^2\,
                   ( 4\,q^4 + 19\,q^2\,s + 7\,s^2 )  - 
                  3\,( m_b^2 - q^2 ) \,
                   ( m_b^2 - s ) \\&\,&\times&\,
                   ( 2\,m_b^2 - q^2 - s ) \,
                   \log (\frac{s}{m_b^2}) )  - 
               2\,( m_b^2 - q^2 ) \,
                (s-m_b^2) \,
                ( -2\,m_b^2 + q^2 + s ) \,
                ( 6\,m_b^2 - 23\,s + 
                  6\,s\,\log (\frac{s}{m_b^2}) ) \\&\,&\times&\,
                \log (\frac{s}{m_b^2}-1) + 
               6\,( m_b^2 - q^2 ) \,s\,
                (s-m_b^2) \,
                ( -2\,m_b^2 + q^2 + s ) \,
                {\log (\frac{s}{m_b^2}-1)}^2 + 
               6\,( m_b^2 - q^2 ) \\&\,&\times&\,
                ( m_b^2 - s ) \,
                ( 2\,m_b^2 - q^2 - s ) \,
                \log (1 - \frac{q^2}{m_b^2})\,
                ( m_b^2 - s + 
                  s\,\log (\frac{s}{m_b^2}) - 
                  2\,s\,\log (\frac{s}{m_b^2}-1) )  )
 - 18\,( m_b^2 - q^2 ) \\&\,&\times&\,( m_b^2 - s ) \,
          ( 2\,m_b^2 - q^2 - s ) \,
          ( \text{Li}_2(\frac{q^2}{q^2-m_b^2}) + 
            \text{Li}_2(1 - \frac{m_b^2}{s}) + 
            4\,\text{Li}_2(1 - \frac{s}{m_b^2}) )  )  \\&\,&+&\, \frac{{\bf a_2}\,f_{\pi}\,m_b}{4\,
       {s( q^2 - s ) }^5}
       (( m_b^2 - s ) \,
             ( -24\,( m_b^2 - q^2 ) \,q^2\,
                ( 30\,m_b^4 - 15\,m_b^2\,q^2 + q^4 )  \
+ ( 5\,m_b^6\,
                   ( 1183 + 72\,{\pi }^2 )  \\&\,&-&\, 
                  20\,m_b^4\,( 407 + 36\,{\pi }^2 ) \,
                   q^2 + 12\,( 5 - 6\,{\pi }^2 ) \,
                   q^6 + 
                  216\,m_b^2\,( 11 + 2\,{\pi }^2 ) \,q^4
 ) \,s - ( 5\,m_b^4\,
                   ( 1525 + 72\,{\pi }^2 )  \\&\,&-&\, 
                  16\,m_b^2\,( 575 + 36\,{\pi }^2 ) \,
                   q^2 + 36\,( 73 + 6\,{\pi }^2 ) \,q^4 \
) \,s^2 + ( m_b^2\,
                   ( 2083 + 72\,{\pi }^2 )  - 
                  8\,( 260 + 9\,{\pi }^2 ) \,q^2 ) \,
                s^3 - 61\,s^4 )  
\end{alignat*}
\newpage
\begin{alignat*}{2}
&\,&+&\, 
            12\,( 25\,( m_b^2 - q^2 ) \,
                ( m_b^2 - s ) \,s\,
                ( 5\,m_b^4 + q^4 + 3\,q^2\,s + s^2 - 
                  5\,m_b^2\,( q^2 + s )  ) \,
                \log (\frac{{\mu}^2}{m_b^2}) - 
               6\,( m_b^2 - q^2 ) \\&\,&\times&\,
                ( m_b^2 - s ) \,s\,
                ( 5\,m_b^4 + q^4 + 3\,q^2\,s + s^2 - 
                  5\,m_b^2\,( q^2 + s )  ) \,
                {\log (1 - \frac{q^2}{m_b^2})}^2 + 
               s\,\log (\frac{s}{m_b^2})\,
                ( -60\, m_b^6 \,
                   ( q^2 + s )  \\&\,&+&\, 
                  37\,q^2\,s\,( q^4 + 3\,q^2\,s + s^2 )  + 
                  30\,m_b^4\,
                   ( 3\,q^4 + 8\,q^2\,s + 3\,s^2 )  - 
                  m_b^2\,
                   ( 25\,q^6 + 237\,q^4\,s + 
                     261\,q^2\,s^2 + 37\,s^3 )  \\&\,&+&\, 
                  6\,( m_b^2 - q^2 ) \,
                   ( m_b^2 - s ) \,
                   ( 5\,m_b^4 + q^4 + 3\,q^2\,s + s^2 - 
                     5\,m_b^2\,( q^2 + s )  ) \,
                   \log (\frac{s}{m_b^2}) )  + 
               2\,( m_b^2 - q^2 ) \,
                ( m_b^2 - s ) \\&\,&\times&\,
                ( 5\,m_b^4 + q^4 + 3\,q^2\,s + s^2 - 
                  5\,m_b^2\,( q^2 + s )  ) \,
                ( 12\,m_b^2 - 55\,s + 
                  12\,s\,\log (\frac{s}{m_b^2}) ) \,
                \log (\frac{s}{m_b^2}-1) + 
               12\,s \\&\,&\times&\, ( m_b^2 - q^2 ) \,
                (s-m_b^2) \,
                ( 5\,m_b^4 + q^4 + 3\,q^2\,s + s^2 - 
                  5\,m_b^2\,( q^2 + s )  ) \,
                {\log (\frac{s}{m_b^2}-1)}^2 - 
               12\,( m_b^2 - q^2 ) \\&\,&\times& \,
                ( m_b^2 - s ) \,
                ( 5\,m_b^4 + q^4 + 3\,q^2\,s + s^2 - 
                  5\,m_b^2\,( q^2 + s )  ) \,
                \log (1 - \frac{q^2}{m_b^2})\,
                ( m_b^2 - s + 
                  s\,\log (\frac{s}{m_b^2}) - 
                  2\,s \\&\,&\times&\,  \log (\frac{s}{m_b^2}-1) )  ) 
+ 144\,( m_b^2 - q^2 ) \,( m_b^2 - s ) \,
          ( 5\,m_b^4 + q^4 + 3\,q^2\,s + s^2 - 
            5\,m_b^2\,( q^2 + s )  ) \\&\,&\times&\,
          ( \text{Li}_2(\frac{q^2}{q^2-m_b^2}) + 
            \text{Li}_2(1 - \frac{m_b^2}{s}) + 
            4\,\text{Li}_2(1 - \frac{s}{m_b^2}) )  ) \\&\,&+&\,
  \frac{{\bf a_4}\,f_{\pi}\,m_b\,} {10\,{s ( q^2 - s ) }^7}
       (-( ( m_b^2 - s ) \,
               ( 30\,( m_b^2 - q^2 ) \,q^2\,
                  ( 1260\,m_b^8 - 
                    1890\,m_b^6\,q^2 - 
                    105\,m_b^2\,q^6 + 
                    2\,q^8 \\&\,&+&\,  840\,m_b^4\,q^4)  
- ( 21\,m_b^{10}\,
                     ( 23207 + 900\,{\pi }^2 )  - 
                    63\,m_b^8\,
                     ( 18827 + 900\,{\pi }^2 ) \,q^2 - 
                    700\,m_b^4\,
                     ( 439 + 45\,{\pi }^2 ) \,
                     q^6 \\&\,&+&\, 
                    150\,m_b^2\,
                     ( 176 + 45\,{\pi }^2 ) \,
                     q^8 + 
                    30\,( 19 - 15\,{\pi }^2 ) \,
                     q^{10} + 
                    175\,m_b^6\,
                     ( 5603 + 360\,{\pi }^2 ) \,q^4 ) \
\,s +  84\,( m_b^8 \\&\,&\times&\,
                     ( 13051 + 450\,{\pi }^2 ) -
                    112\, m_b^6\,
                     ( 22157 + 900\,{\pi }^2 ) \,q^2 - 
                    800\,m_b^2\,
                     ( 616 + 45\,{\pi }^2 ) \,
                     q^6 + 
                    150\,( 283 + 30\,{\pi }^2 ) \,
                     q^8 \\&\,&+&\, 
                    525\,m_b^4\,
                     ( 3523 + 180\,{\pi }^2 ) \,q^4 )
\,s^2   ( -7\,m_b^6\,
                     ( 119363 + 3600\,{\pi }^2 )  + 
                    756\,m_b^4\,
                     ( 2131 + 75\,{\pi }^2 ) \,q^2 + 
                    200\,( 848 \\&\,&+&\, 45\,{\pi }^2 ) \,
                     q^6 - 
                    10125\,m_b^2\,( 89 + 4\,{\pi }^2 ) \,
                     q^4 ) \,s^3 + 
                 ( 7\,m_b^4\,
                     ( 34967 + 900\,{\pi }^2 )  - 
                    6\,m_b^2\,
                     ( 57989 + 1800\,{\pi }^2 ) \,q^2 \\&\,&+&\, 
                    125\,( 1075 + 36\,{\pi }^2 ) \,q^4 ) \,s^4 + ( -2\,m_b^2\,
                     ( 10553 + 225\,{\pi }^2 )  + 
                    ( 21461 + 450\,{\pi }^2 ) \,q^2 )
\,s^5 + 181\,s^6 )  )  + 
            30\, \\&\,&\times&\,
( 91\,( m_b^2 - q^2 ) \,
                ( m_b^2 - s ) \,s\,
                ( 42\,m_b^8 + 
                  q^8 + 
                  10\,{ q^6  }\,s + 20\,q^4\,s^2 + 
                  10\,q^2\,s^3 + s^4 - 
                  84\,m_b^6\,
                   ( q^2 + s )  \\&\,&+&\, 
                  28\,m_b^4\,( 2\,q^2 + s ) \,
                   ( q^2 + 2\,s )  - 
                  14\,m_b^2\,( q^2 + s ) \,
                   ( q^4 + 4\,q^2\,s + s^2 )  ) \,
                \log (\frac{{\mu}^2}{m_b^2}) - 
               15\,( m_b^2 - q^2 ) \,
                ( m_b^2 - s ) \,s\, \\&\,&\times&\,
                ( 42\,  m_b^8  +  
                   q^8 + 
                  10\, q^6\,s + 20\,q^4\,s^2 + 
                  10\,q^2\,s^3 + s^4 - 
                  84\,m_b^6 ) \,
                   ( q^2 + s )  + 
                  28\,m_b^4\,( 2\,q^2 + s ) \,
                   ( q^2 + 2\,s )  - 
                  14\,m_b^2 \\&\,&\times&\, ( q^2 + s ) \,
                   ( q^4 + 4\,q^2\,s + s^2 )  ) \,
                {\log (1 - \frac{q^2}{m_b^2})}^2 + 
               s\,\log (\frac{s}{m_b^2})\,
                ( -1260\,m_b^{10}\,
                   ( q^2 + s )  + 
                  630\,m_b^8\,
                   ( 5\,q^4 
\end{alignat*}
\newpage
\begin{alignat*}{2}
&\,&+&\,  12\,q^2\,s + 5\,s^2 )  - 
                  210\,m_b^6\,
                   ( q^2 + s ) \,
                   ( 13\,q^4 + 48\,q^2\,s + 13\,s^2 )  + 
                  121\,q^2\,s\,
                   (  q^8  + 
                     10\,q^6\,s + 20\,q^4\,s^2 \\&\,&+&\, 
                     10\,q^2\,s^3 + s^4 )  + 
                  105\,m_b^4\,
                   ( 9\,q^8 + 
                     82\, q^6\,s + 160\,q^4\,s^2 + 
                     82\,q^2\,s^3 + 9\,s^4 )  - 
                  m_b^2\,
                   ( 91\,q^{10} \\&\,&+&\, 
                     2305\, q^8\,s + 
                     9400\,q^6\,s^2 + 
                     9700\,q^4\,s^3 + 2575\,q^2\,s^4 + 121\,s^5 )
+ 15\,( m_b^2 - q^2 ) \,( m_b^2 - s ) \\&\,&\times&\,
                   ( 42\, m_b^8 + 
                      q^8  + 
                     10\, q^6 \,s + 20\,q^4\,s^2 + 
                     10\,q^2\,s^3 + s^4 - 
                     84\,m_b^6\,
                      ( q^2 + s )  + 
                     28\,m_b^4\,( 2\,q^2 + s ) \,
                      ( q^2 + 2\,s )  \\&\,&-&\, 
                     14\,m_b^2\,( q^2 + s ) \,
                      ( q^4 + 4\,q^2\,s + s^2 )  ) \,
                   \log (\frac{s}{m_b^2}) )  + 
               4\,( m_b^2 - q^2 ) \,
                ( m_b^2 - s ) \,
                ( 42\,m_b^8 + 
                  q^8 + 
                  10\,q^6\,s \\&\,&+&\,  20\,q^4\,s^2 + 
                  10\,q^2\,s^3 + s^4 - 
                  84\,m_b^6\,
                   ( q^2 + s )  + 
                  28\,m_b^4\,( 2\,q^2 + s ) \,
                   ( q^2 + 2\,s )  - 
                  14\,m_b^2\,( q^2 + s ) \\&\,&\times&\,
                   ( q^4 + 4\,q^2\,s + s^2 )  ) \,
                ( 15\,m_b^2 - 83\,s + 
                  15\,s\,\log (\frac{s}{m_b^2}) ) \,
                \log (\frac{s}{m_b^2}-1) + 
               30\,( m_b^2 - q^2 ) \,s\,
                (s-m_b^2) \\&\,&\times&\,
                ( 42\, m_b^8  + 
                   q^8  + 
                  10\, q^6\,s + 20\,q^4\,s^2 + 
                  10\,q^2\,s^3 + s^4 - 
                  84\,m_b^6\,
                   ( q^2 + s )  + 
                  28\,m_b^4\,( 2\,q^2 + s ) \,
                   ( q^2 + 2\,s )  \\&\,&-&\, 
                  14\,m_b^2\,( q^2 + s ) \,
                   ( q^4 + 4\,q^2\,s + s^2 )  ) \,
                {\log (\frac{s}{m_b^2}-1)}^2 - 
               30\,( m_b^2 - q^2 ) \,
                ( m_b^2 - s ) \,
                ( 42\,m_b^8 + 
                  q^8 \\&\,&+&\, 
                  10\, q^6\,s + 20\,q^4\,s^2 + 
                  10\,q^2\,s^3 + s^4 - 
                  84\,m_b^6\,
                   ( q^2 + s )  + 
                  28\,m_b^4\,( 2\,q^2 + s ) \,
                   ( q^2 + 2\,s )  - 
                  14\,m_b^2\\&\,&\times&\,( q^2 + s ) \,
                   ( q^4 + 4\,q^2\,s + s^2 )  ) \,
                \log (1 - \frac{q^2}{m_b^2})\,
                ( m_b^2 - s + 
                  s\,\log (\frac{s}{m_b^2}) - 
                  2\,s\,\log (\frac{s}{m_b^2}-1) )  ) 
\\&\,&+&\,  900 s \,( m_b^2 - q^2 ) \,( m_b^2 - s ) \,
          ( 42\, m_b^8  + 
            q^8 + 10\,q^6\,s + 
            20\,q^4\,s^2 + 10\,q^2\,s^3 + s^4 - 
            84\,m_b^6\,( q^2 + s )  \\&\,&+&\, 
            28\,m_b^4\,( 2\,q^2 + s ) \,
             ( q^2 + 2\,s )  - 
            14\,m_b^2\,( q^2 + s ) \,
             ( q^4 + 4\,q^2\,s + s^2 )  ) \,
          ( \text{Li}_2(\frac{q^2}{q^2-m_b^2}) + 
            \text{Li}_2(1 - \frac{m_b^2}{s}) \\&\,&+&\,  4\,\text{Li}_2(1 - \frac{s}{m_b^2})
 )  \big\}  
\\&\rho_{\rm T3}&\,=\;&  \frac{-15\,\eta_3\,{\mu_{\pi}}^2}{{( q^2 - 
        s ) }^6} \,{( m_b^2 - q^2 ) }^2\,
     ( m_b^2 - s ) \,
     ( 2\,( q^2 - s ) \,
        ( -5\,m_b^2 + 2\,q^2 + 3\,s )  + 
       ( 7\,m_b^4 - 6\,m_b^2\,q^2 + q^4 - 8\,m_b^2\,s \\&\,&+& 
          4\,q^2\,s + 2\,s^2 ) \,\omega_3 )  
\end{alignat*}
\newpage
\begin{alignat*}{2}
&\rho_p\, &\,=\;&  \frac{a_0{\mu_{\pi}}^2\,( 3\,m_b^2\,s - 3\,q^2\,s ) }
   {6\,{( q^2 - s ) }^2\,s} + \frac{5 \eta_3 \mu_{\pi}^2} {{( q^2 - s ) }^4\,s} ( 18\,m_b^6\,s - 
       36\,m_b^4\,q^2\,s - 3\,q^6\,s + 
       21\,m_b^2\,q^4\,s - 18\,m_b^4\,s^2 \\&\,&+&\, 30\,m_b^2\,q^2\,s^2 - 
       12\,q^4\,s^2 +  3\,m_b^2\,s^3 - 3\,q^2\,s^3 ) +
\mathfrak{a_s} \big\{ \frac{{\bf a_0}\,{\mu_{\pi}}^2\,} {6\,{( q^2 - s ) }^2\,s} 
( (q^2+s)\,(s-3\,m_b^2)   \\&\,&-&\,
       3\,( m_b^2 - 4\,q^2 ) \,s\,
        {\log (\frac{q^2}{m_b^2})}^2 -
       18\,s\,( 3\,m_b^2 - 3\,q^2 + s ) \,
        {\log (1 - \frac{q^2}{m_b^2})}^2 
       \,( m_b^2 - 4\,q^2 ) \,s\,
        \log (\frac{q^2}{m_b^2}) \\&\,&\times&\,
        ( \log (1 - \frac{q^2}{m_b^2})
 - 
          \log (\frac{s-q^2}{m_b^2}) )  + 
       6\,\log (1 - \frac{q^2}{m_b^2})\,
        ( m_b^2\,q^2 + 3\,m_b^2\,s - 
          {q}^2\,s + 
          s\,( ( 2\,m_b^2 - 3\,q^2 ) \\&\,&\times&\,
              \log (\frac{{\mu}^2}{m_b^2}) + 
             q^2\,\log (\frac{s}{m_b^2}) 
             ( m_b^2 + q^2 + s ) \,
              \log (\frac{s-q^2}{m_b^2}) - 
             ( m_b^2 - 2\,q^2 )
              \log (\frac{s}{m_b^2}-1) )  )  + 
       3\,( m_b^2\,s \\&\,&\times&\, {\log (\frac{s}{m_b^2})}^2 - 
           {\log (\frac{s-q^2}{m_b^2})}^2 \, s\,( 5\,q^2 + s ) \,  
             +\log (\frac{s}{m_b^2})\,
           ( 2\,m_b^2\,q^2 + m_b^2\,s - 
             3\,q^2\,s - 4\,q^2\,s\,\log (\frac{s}{m_b^2}-1) )  
\\&\,&+&\, 2\,\log (\frac{s-q^2}{m_b^2})\,
           ( 2\,q^2\,s - 
             m_b^2\,( q^2 + s ) -  
             s\,( 2\,m_b^2 - 3\,q^2 + s ) \,
              \log (\frac{{\mu}^2}{m_b^2}) + 
             s\,( 3\,m_b^2 - 2\,q^2 + 2\,s ) \\&\,&\times&\,
              \log (\frac{s}{m_b^2}-1) )  + 
          2\,\log (\frac{s}{m_b^2}-1)\, 
           ( 2\,q^2\,s - m_b^2\,
              ( q^2 + 4\,s )  + 
             s\,( s\,\log (\frac{{\mu}^2}{m_b^2}) - 
                ( m_b^2 + 2\,s ) \,
                 \log (\frac{s}{m_b^2}-1) )  )  )  \\&\,&+&\,
 6\,s\,( m_b^2\,\text{Li}_2(\frac{q^2}{q^2-m_b^2}) + 
          ( 2\,m_b^2 - 3\,q^2 + s ) \,
           \text{Li}_2(1 - \frac{q^2}{m_b^2}) + 
          m_b^2\,\text{Li}_2(1 - \frac{m_b^2}{s}) - 
          ( m_b^2 + q^2 + s ) \\&\,&\times&\,\text{Li}_2(1 - \frac{q^2}{s}) -  
          ( 5\,q^2 + s ) 
           \text{Li}_2(\frac{m_b^2 - s}{m_b^2 - q^2}) - 
          ( m_b^2 + q^2 + s ) \,
           \text{Li}_2(\frac{m_b^2 - s}{q^2 - s}) - 
          ( m_b^2 + q^2 + s ) \\&\,&\times&\,
           \text{Li}_2(\frac{(q^2-m_b^2) \,s}
             {m_b^2\,( q^2 - s ) }) + 
          ( 2\,m_b^2 - 3\,q^2 + s ) \, 
           \text{Li}_2(\frac{m_b^2 - q^2}{s-q^2}) + 
          2\,( m_b^2 - q^2 ) \,
           \text{Li}_2(1 - \frac{s}{m_b^2}) + 
          ( m_b^2 - 4\,q^2 ) \\&\,&\times&\, \text{Li}_2(1 - \frac{s}{q^2}) )  ) 
{6\,{( q^2 - s ) }^2\,s} + 
  \frac{5\,{\bf \eta_3}\,{\mu_{\pi}}^2}{{( q^2 - s ) }^4\,s}
     ( -3\,m_b^2\,( q^2-12\,m_b^2 ) \,q^4 - 
       6\,m_b^6\,(  {\pi }^2-47 ) \,s - 
       18\,m_b^4\,( 24 \\&\,&+&\, {\pi }^2 ) \,q^2\,s + 
       ( 3 + 2\,{\pi }^2 ) \,q^6\,s + 
       3\,m_b^2\,( 8\,{\pi }^2-17 ) \,q^4\,s + 
       6\,m_b^4\,( {\pi }^2-60 ) \,s^2 + 
       3\,m_b^2\,( 205 + 6\,{\pi }^2 ) \,q^2\,s^2 \\&\,&-&\, 
       ( 3 +26\,{\pi }^2 ) \,q^4\,s^2 + 69\,m_b^2\,s^3 - 
       (147 +2\,{\pi }^2 ) \,q^2\,s^3 +
       ( -9 + 2\,{\pi }^2 ) \,s^4 + 
       3\,( m_b^2 - 4\,q^2 ) \,{( q^2 - s ) }^2\,
        s\,{\log (\frac{q^2}{m_b^2})}^2 \\&\,&+&\, 
       6\,( m_b^2 - 4\,q^2 ) \,{( q^2 - s ) }^2\,
        s\,\log (\frac{q^2}{m_b^2})\,
        ( \log (1 - \frac{q^2}{m_b^2}) - 
          \log (\frac{s-q^2}{m_b^2}) )  + 
       6\,s\,\log (\frac{{\mu}^2}{m_b^2})\,
        ( 15\, m_b^6 - 
          15\,m_b^4\,( 2\,q^2 + s )  \\&\,&-&\, 
          3\,q^2\,s\,( q^2 + 4\,s )  + 
          m_b^2\,
           ( 9\,q^4 + 30\,q^2\,s + 6\,s^2 ) + 
          ( 6\, m_b^6 - 6\,m_b^4\,s - 
             3\,q^2\,( q^4 - 4\,q^2\,s + s^2 )  + 
             m_b^2\,
              ( -4\,q^4 - 4\,q^2\,s   
\end{alignat*}
\newpage
\begin{alignat*}{2}
&\,&+&\, 2\,s^2 )  ) \,
           \log (1 - \frac{q^2}{m_b^2}) - 
          {( {q}^2 - s ) }^2\,
           ( 2\,m_b^2 - 3\,q^2 + s ) \,
           \log (\frac{s-q^2}{m_b^2}) + 
          ( -6\,m_b^6 + 
             6\,m_b^2\,q^4 + 6\,m_b^4\,s + 
             s\,( -5\,q^4 \\&\,&-&\, 2\,q^2\,s + s^2 )  \
) \,\log (\frac{s}{m_b^2}-1) )  + 
       3\,( -( s\,( 6\,m_b^6 - 
               {( q^2 - s ) }^2\,( 3\,q^2 - s )  - 
               6\,m_b^4\,( q^2 + s )  + 
               3\,m_b^2\,( q^4 + s^2 )  ) \, 
\\&\,&-&\,  {\log (1 - \frac{q^2}{m_b^2})}^2 )  + 
          m_b^2\,s\, ( 6\,m_b^4 + q^4 + 4\,q^2\,s + s^2 
          6\,m_b^2\,( q^2 + s )  ) \,
           {\log (\frac{s}{m_b^2})}^2 - 
          {( q^2 - s ) }^2\,\log (\frac{s-q^2}{m_b^2}) \\&\,&\times&\,
           ( 2\,( -2\,q^2\,s + 
                m_b^2\,( q^2 + s )  )  + 
             s\,( 5\,q^2 + s ) \,
              \log (\frac{s-q^2}{m_b^2}) ) 
          2\,( 6\,m_b^6\,
              ( 2\,q^2 + 9\,s )  - 
             6\,m_b^4\,( 2\,q^4 + 18\,q^2\,s + 9\,s^2 )  \\&\,&-&\, 
             2\,q^2\,s\,( q^4 + 7\,q^2\,s + 13\,s^2 )  + 
             m_b^2\,( q^6 + 44\,q^4\,s + 
                89\,q^2\,s^2 + 16\,s^3 ) 
             ( -3\,m_b^2 + 2\,q^2 - 2\,s ) \,
              {( q^2 - s ) }^2\,s \\&\,&\times&\,
              \log (\frac{s-q^2}{m_b^2}) ) \,
           \log (\frac{s}{m_b^2}-1) - 
          2\,s\,( -6\,m_b^6 - 
             6\,m_b^4\,( q^2 - s )  + 
             2\,s\,( -5\,q^4 - 2\,q^2\,s + s^2 ) 
             m_b^2\,( 13\,q^4 + 4\,q^2\,s \\&\,&+&\,  s^2 )  ) \
\,{\log (\frac{s}{m_b^2}-1)}^2 + 
          \log (\frac{s}{m_b^2})\,
           ( -12\,m_b^4\,s^2 - 
             3\,q^2\,s\,{( q^2 + 3\,s ) }^2 + 
             m_b^2\,
              ( 2\,q^6 + 21\,q^4\,s + 
                12\,q^2\,s^2 +  13\,s^3 )  \\&\,&-&\, 
             4\,q^2\,s\,( 6\,m_b^4 + q^4 + 4\,q^2\,s + s^2 - 
                6\,m_b^2\,( q^2 + s )  ) \,
              \log (\frac{s}{m_b^2}-1) )  + 
          2\,\log (1 - \frac{q^2}{m_b^2})\,
           ( 6\,m_b^6\,
              ( q^2 + 2\,s )  \\&\,&-&\, 
             {q}^2\,s\,( q^4 - 8\,q^2\,s + s^2 )  - 
             6\,m_b^4\,( q^4 + 2\,q^2\,s + 2\,s^2 )  + 
             m_b^2\,
              ( q^6 + q^4\,s + q^2\,s^2 + 
                3\,s^3 )  + 
             s\,( q^2\,
                 ( 6\,m_b^4 + q^4 \\&\,&+&\,  4\,q^2\,s + s^2 - 
                   6\,m_b^2\,( q^2 + s )  ) \,
                 \log (\frac{s}{m_b^2}) + 
                {( q^2 - s ) }^2\,
                 ( m_b^2 + q^2 + s ) \,
                 \log (\frac{s-q^2}{m_b^2}) + 
                ( -12\,m_b^4\,q^2 + 
                   m_b^2\,( 11\,q^4 \\&\,&+&\,  14\,q^2\,s - s^2 ) + 
                   2\,q^2\,( q^4 - 8\,q^2\,s + s^2 )  ) \,
                 \log (\frac{s}{m_b^2}-1) )  )  )  \
+ 6\,s\,( m_b^2\,( 6\,m_b^4 + q^4 + 4\,q^2\,s + s^2 - 
             6\,m_b^2\, \\&\,&\times&\,  ( q^2 + s )  ) \,
           \text{Li}_2(\frac{q^2}{q^2-m_b^2}) + 
          {( q^2 - s ) }^2\,
           ( 2\,m_b^2 - 3\,q^2 + s ) \,
           \text{Li}_2(1 - \frac{q^2}{m_b^2}) + 
          m_b^2\,( 6\,m_b^4 + q^4 + 4\,q^2\,s + s^2 \\&\,&-&\, 
             6\,m_b^2\,( q^2 + s )  ) \,
           \text{Li}_2(1 - \frac{m_b^2}{s}) 
          {( q^2 - s ) }^2\,( m_b^2 + q^2 + s ) \,
           \text{Li}_2(1 - \frac{q^2}{s}) - 
          {( q^2 - s ) }^2\,( 5\,q^2 + s ) \,
           \text{Li}_2(\frac{m_b^2 - s}{m_b^2 - q^2}) \\&\,&-&\, 
          {( q^2 - s ) }^2\,( m_b^2 + q^2 + s ) \,
           \text{Li}_2(\frac{m_b^2 - s}{q^2 - s}) -
          {( q^2 - s ) }^2\,( m_b^2 + q^2 + s ) \,
           \text{Li}_2(\frac{(q^2-m_b^2) \,s}
             {m_b^2\,( q^2 - s ) }) + 
          {( q^2 - s ) }^2\,
           ( 2\,m_b^2 \\&\,&-&\,  3\,q^2 + s ) \,
           \text{Li}_2(\frac{m_b^2 - q^2}{s-q^2}) + 
          2\,( m_b^2 - q^2 ) \,
           ( 6\,m_b^4 + q^4 + 4\,q^2\,s + s^2 - 
             6\,m_b^2\,( q^2 + s )  ) \,
           \text{Li}_2(1 - \frac{s}{m_b^2}) + 
          ( m_b^2 \\&\,&-&\,  4\,q^2 ) \,{( q^2 - s ) }^2\,
           \text{Li}_2(1 - \frac{s}{q^2}) ) \big\}
\end{alignat*}
\newpage
\begin{alignat*}{2}
&\rho_{\sigma}\, &\,=\;&   \frac{\mu_{\pi}^2} {2\,{( q^2 - s ) }^5} \,( a_0\,{( q^2 - s ) }^2\,
        ( -( m_b^2\,( 3\,q^2 + s )  )  + 
          q^2\,( q^2 + 3\,s )  )  - 
       30\,\eta_3\,( 10\,m_b^8 + 
          10\,m_b^6\,( q^2 - s )  \\&\,&-&\, 
          30\,m_b^4\,q^2\,( q^2 + s )  + 
          m_b^2\,( 13\,q^6 + 45\,q^4\,s + 21\,q^2\,s^2 + s^3 )  
- q^2\,( q^6 + 11\,q^4\,s + 15\,q^2\,s^2 + 3\,s^3 )  )
  )  \\&\,&+&\,  
\mathfrak{a_s} \big\{ \frac{{\bf a_0}\,{\mu_{\pi}}^2\,}{6\,
       {( q^2 - s ) }^3\,s}
       ( 3\,m_b^2\,q^4 + 
         q^2\,( -4\,m_b^2\,( -9 + {\pi }^2 )  + 
            ( -3 + 2\,{\pi }^2 ) \,q^2 ) \,s + 
         ( m_b^2\,( 21 - 4\,{\pi }^2 )  \\&\,&+&\,  
            4\,( -15 + {\pi }^2 ) \,q^2 ) \,s^2 + 
         ( 3 + 2\,{\pi }^2 ) \,s^3 + 
         6\,s\,( -4\,q^2\,s + 
            m_b^2\,( 3\,q^2 + s )  ) \,
          {\log (1 - \frac{q^2}{m_b^2})}^2 + 
         6\,s\,\log (\frac{{\mu}^2}{m_b^2})\\&\,&\times&\, 
          ( 3\,( -2\,q^2\,s + 
               m_b^2\,( q^2 + s )  )  + 
            ( {q}^2 - s ) \,
             ( -( ( m_b^2 + q^2 ) \,
                  \log (1 - \frac{q^2}{m_b^2}) )  - 
               ( m_b^2 + q^2 + s ) \,
                \log (\frac{m_b^2 - s}{q^2 - s}) \\&\,&-&\,  
               ( m_b^2 - 4\,q^2 ) \,
                \log (\frac{m_b^2 - q^2}{s-q^2}) + 
               ( m_b^2 + q^2 ) \,
                \log (\frac{s}{m_b^2}-1) )  )  + 
         3\,( -2\,( q^2 - s ) \,s\,
             ( m_b^2 + q^2 + s ) \,
             {\log (\frac{s}{m_b^2})}^2 \\&\,&+&\,  
            2\,( q^2 - s ) \,
             \log (\frac{m_b^2 - s}{q^2 - s})\,
             ( -( m_b^2\,
                  ( q^2 + 2\,s )  )  + 
               {s}\,( m_b^2 + q^2 + s ) \,
                \log (\frac{s-q^2}{m_b^2}) )  + 
            ( q^2 - s ) \,s\,
             ( ( m_b^2 \\&\,&+&\,  q^2 + s ) \,
                {\log (\frac{m_b^2 - q^2}{s-q^2})}^2 - 
               ( m_b^2 + q^2 + s ) \,
                {\log (\frac{( m_b^2 - q^2 ) \,s}
                    {m_b^2\,(s-q^2) })}^2 - 
               2\,( m_b^2 + q^2 + s ) \,
                \log (\frac{( m_b^2 - q^2 ) \,s}
                  {m_b^2\,(s-q^2) }) \\&\,&\times&\, 
                \log (\frac{s-q^2}{m_b^2}) + 
               2\,( 2\,m_b^2 - 3\,q^2 + s ) \,
                {\log (\frac{s-q^2}{m_b^2})}^2 + 
               2\,\log (\frac{m_b^2 - q^2}{s-q^2})\,
                ( -3\,m_b^2 + 6\,q^2 + 
                  ( 2\,m_b^2 \\&\,&-&\,  3\,q^2 + {s} ) \,
                   \log (\frac{s-q^2}{m_b^2}) )  )  - 
            2\,( 4\,m_b^4\,( q^2 + s )  + 
               2\,q^2\,s\,( q^2 + 3\,s )  - 
               2\,m_b^2\,q^2\,( q^2 + 7\,s )  + 
               ( q^2 - s ) \,s\\&\,&\times&\, 
                ( 5\,m_b^2 - 5\,q^2 + 3\,s ) \,
                \log (\frac{s-q^2}{m_b^2}) ) \,
             \log (\frac{s}{m_b^2}-1) + 
            2\,s\,( -2\,s\,( q^2 + s )  + 
               m_b^2\,( 3\,q^2 + s )  ) \,
             {\log (\frac{s}{m_b^2}-1)}^2 \\&\,&+&\,  
            \log (\frac{s}{m_b^2})\,
             ( {{q}}^2\,s\,( 3\,q^2 + s )  - 
               m_b^2\,( q^2 - s ) \,
                ( 2\,q^2 + 5\,s )  - 
               2\,( q^2 - s ) \,s\,
                ( m_b^2 + q^2 + s ) \,
                ( \log (\frac{m_b^2 - q^2}{s-q^2}) \\&\,&-&\,  
                  \log (\frac{( m_b^2 - q^2 ) \,s}
                    {m_b^2\,(s-q^2) }) - 
                  \log (\frac{s-q^2}{m_b^2}) )  + 
               2\,s\,( -5\,m_b^2\,q^2 + q^4 - 3\,m_b^2\,s + 
                  6\,q^2\,s + s^2 ) \,\log (\frac{s}{m_b^2}-1)
 )  )  \\&\,&+&\,  6\,\log (1 - \frac{q^2}{m_b^2})\,
          ( -( {{q}}^2\,( q^2 - 7\,s ) \,
               s )  + 2\,m_b^4\,( q^2 + 2\,s )  - 
            m_b^2\,
             ( q^4 + 9\,q^2\,s + 2\,s^2 )  + 
            {s}\,( ( 4\,q^2\,s - 
                  m_b^2\,( 3\,q^2 \\&\,&+&\,  s )  ) \,
                \log (\frac{q^2}{m_b^2}) + 
               ( 2\,m_b^2\,( q^2 + s )  - 
                  q^2\,( q^2 + 3\,s )  ) \,
                \log (\frac{s}{m_b^2}) + 
               ( q^2 - s ) \,
                ( -( ( m_b^2 + q^2 + s ) \,
                     ( \log (\frac{m_b^2 - q^2}{s-q^2}) \\&\,&-&\,  
                       \log (\frac{( m_b^2 - q^2 ) \,s}
                         {m_b^2\,(s-q^2) }) )  )  - ( m_b^2 - 4\,q^2 ) \,
                   \log (\frac{s-q^2}{m_b^2}) )  - 
               2\,( q^2\,( 3\,q^2 - 7\,s )  + 
                  m_b^2\,( q^2 + 3\,s )  ) \,
                \log (\frac{s}{m_b^2}-1) )  )  \\&\,&+&\,  
         6\,s\,( ( 4\,q^2\,s - 
               m_b^2\,( 3\,q^2 + s )  ) \,
                \text{Li}_2(1 - \frac{q^2}{m_b^2}) - 
            ( m_b^2 - 4\,q^2 ) \,( q^2 - s ) \,
             ( 2\,      \text{Li}_2(\frac{m_b^2 - s}{m_b^2 - q^2}) + 
                \text{Li}_2(\frac{m_b^2 - s}{q^2 - s})  
\end{alignat*}
\newpage
\begin{alignat*}{2}
&\,&-&\,       \text{Li}_2(\frac{q^2\,( m_b^2 - s ) }
                 {m_b^2\,( q^2 - s ) }) )  - 
            ( 3\,m_b^2\,( 3\,q^2 + s )  - 
               ( q^2 + s ) \,( 5\,q^2 + s )  \
) \,    \text{Li}_2(1 - \frac{s}{m_b^2}) )  )  + 
    \frac{5\,{\bf \eta_3}\,{\mu_{\pi}}^2\,} {6\,
       {( q^2 - s ) }^5\,s}
       ( 18\,m_b^2\,q^4 \\&\,&\times&\,
          ( 40\,m_b^4 - 40\,m_b^2\,q^2 + q^4 )  + 
         ( 5\,m_b^8\,
             ( -359 + 24\,{\pi }^2 )  - 
            30\,m_b^6\,( 127 + 16\,{\pi }^2 ) \,q^2 + 
            30\,m_b^4\,( 129 + 20\,{\pi }^2 ) \,q^4 \\&\,&+&\, 
            24\,m_b^2\,( 40 - 11\,{\pi }^2 ) \,q^6 + 
            6\,( -3 + 2\,{\pi }^2 ) \,q^8 ) \,s + 
         2\,( 5\,m_b^6\,( 295 - 36\,{\pi }^2 )  + 
            30\,m_b^4\,( 129 + 16\,{\pi }^2 ) \,q^2 - 
            6\,m_b^2\\&\,&\times&\, ( 609 + 68\,{\pi }^2 ) \,q^4 + 
            6\,( -27 + 20\,{\pi }^2 ) \,q^6 ) \,s^2 + 
         6\,( 5\,m_b^4\,( -25 + 8\,{\pi }^2 )  - 
            19\,m_b^2\,( 39 + 4\,{\pi }^2 ) \,q^2 + 
            9\,( 55 \\&\,&+&\, 4\,{\pi }^2 ) \,q^4 ) \,s^3 - 
         4\,( m_b^2\,( 71 + 6\,{\pi }^2 )  - 
            66\,q^2 ) \,s^4 + 
         ( -37 + 12\,{\pi }^2 ) \,s^5 + 
         36\,s\,( 10\, m_b^8  - 
            10\,m_b^6\,s - 
            10\,m_b^4\,q^2\\&\,&\times&\,  ( q^2 + 2\,s )  - 
            4\,q^2\,s\,( q^4 + 3\,q^2\,s + s^2 )  + 
            m_b^2\,( 3\,q^6 + 25\,q^4\,s + 21\,q^2\,s^2 + s^3 )
  ) \,{\log (1 - \frac{q^2}{m_b^2})}^2 + 
         12\,s\\&\,&\times&\,\log (\frac{{\mu}^2}{m_b^2})\,
          ( -( m_b^2\,
               ( 5\,m_b^2 - 2\,q^2 ) \,
               ( 5\,m_b^4 + 20\,m_b^2\,q^2 - 28\,q^4 )  )
  + ( 40\,m_b^6 + 135\,m_b^4\,q^2 - 
               249\,m_b^2\,q^4 + 47\,q^6 ) \,{s} \\&\,&-&\, 
            3\,( 5\,m_b^4 + 18\,m_b^2\,q^2 - 32\,q^4 ) \,
             s^2 + 9\,( m_b^2 - 2\,q^2 ) \,s^3  - 
            3\,( 10\, m_b^8 - 
               20\,m_b^6\,( q^2 + s )  + 
               10\,m_b^4\,( q^2 + s ) \\&\,&\times&\,
                ( 2\,q^2 + s )  + 
               q^2\,( q^6 + 7\,q^4\,s + 3\,q^2\,s^2 - s^3 )
  - m_b^2\,( 9\,q^6 + 23\,q^4\,s + 7\,q^2\,s^2 + s^3 )
  ) \,\log (1 - \frac{q^2}{m_b^2}) \\&\,&+&\, 
            3\,{( q^2 - s ) }^3\,
             ( -( ( m_b^2 + q^2 + s ) \,
                  \log (\frac{m_b^2 - s}{q^2 - s}) )  - 
               ( m_b^2 - 4\,q^2 ) \,
                \log (\frac{m_b^2 - q^2}{s-q^2}) )  + 
            3\,( 10\, m_b^8 - 
               20\,m_b^6\,( q^2 + s )  \\&\,&+&\, 
               10\,m_b^4\,( q^2 + s ) \,
                ( 2\,q^2 + s )  + 
               q^2\,( q^6 + 7\,q^4\,s + 3\,q^2\,s^2 - s^3 )  - 
               m_b^2\,( 9\,q^6 + 23\,q^4\,s + 7\,q^2\,s^2 + 
                  s^3 )  ) \\&\,&\times&\,\log (\frac{s}{m_b^2}-1) )
  + 36\,\log (1 - \frac{q^2}{m_b^2})\,
          ( 20\,m_b^8\,q^2 - 
            10\,m_b^6\,q^2\,( 3\,q^2 + 5\,s )  + 
            2\,m_b^4\,( q^2 + 2\,s ) \,
             ( 6\,q^4 + 13\,q^2\,s \\&\,&+&\, s^2 )  - 
            {{q}}^2\,s\,
             ( q^6 + q^4\,s - 15\,q^2\,s^2 - 7\,s^3 )  - 
            m_b^2\,
             ( q^8 + 7\,q^6\,s + 35\,q^4\,s^2 + 35\,q^2\,s^3 + 
               2\,s^4 )  + 
            {s}\,( ( -10\,
                    m_b^8 \\&\,&+&\, 10\,m_b^6\,s + 
                  10\,m_b^4\,q^2\,( q^2 + 2\,s )  + 
                  4\,q^2\,s\,( q^4 + 3\,q^2\,s + s^2 )  - 
                  m_b^2\,( 3\,q^6 + 25\,q^4\,s + 21\,q^2\,s^2 + 
                     s^3 )  ) \,\log (\frac{q^2}{m_b^2}) \\&\,&+&\, 
               ( 10\,m_b^6\,( 2\,q^2 + s )  - 
                  10\,m_b^4\,
                   ( 3\,q^4 + 5\,q^2\,s + s^2 )  + 
                  2\,m_b^2\,
                   ( 6\,q^6 + 24\,q^4\,s + 14\,q^2\,s^2 + s^3 )
  - q^2\,( q^6 + 11\,q^4\,s \\&\,&+ &\, 15\,q^2\,s^2 + 3\,s^3 )  )
 \,\log (\frac{s}{m_b^2}) + 
               {( q^2 - s ) }^3\,
                ( -( ( m_b^2 + q^2 + s ) \,
                     ( \log (\frac{m_b^2 - q^2}{s-q^2}) - 
                       \log (\frac{( m_b^2 - q^2 ) \,s}
                         {m_b^2\,(s-q^2) }) )  )
  - ( m_b^2 \\&\,&-&\,  4\,q^2 ) \,                                       
                   \log (\frac{s-q^2}{m_b^2}) )  - 
               2\,( 10\,m_b^6\,( 2\,q^2 + s )  - 
                  10\,m_b^4\,
                   ( 3\,q^4 + 5\,q^2\,s + s^2 )  + 
                  q^2\,( 3\,q^6 - 23\,q^4\,s - 3\,q^2\,s^2 - 
                     7\,s^3 )  \\&\,&+&\, 
                  m_b^2\,( 11\,q^6 + 51\,q^4\,s + 25\,q^2\,s^2 + 
                     3\,s^3 )  ) \,
                \log (\frac{s}{m_b^2}-1) )  )  + 
         6\,( -6\,{( q^2 - s ) }^3\,s\,
             ( m_b^2 + q^2 + s ) \,
             {\log (\frac{s}{m_b^2})}^2 \\&\,&+&\, 
            6\,{( q^2 - s ) }^3\,
             \log (\frac{m_b^2 - s}{q^2 - s})\,
             ( -( m_b^2\,
                  ( q^2 + 2\,s )  )  + 
               {s}\,( m_b^2 + q^2 + s ) \,
                \log (\frac{s-q^2}{m_b^2}) )  + 
            3\,{( q^2 - s ) }^3\,s\,
             ( ( m_b^2 + q^2 + s ) \,
\end{alignat*}
\newpage
\begin{alignat*}{2}
&\,&\times&   {\log (\frac{m_b^2 - q^2}{s-q^2})}^2 - 
               ( m_b^2 + q^2 + s ) \,
                {\log (\frac{( m_b^2 - q^2 ) \,s}
                    {m_b^2\,(s-q^2) })}^2 - 
               2\,( m_b^2 + q^2 + s ) \,
                \log (\frac{( m_b^2 - q^2 ) \,s}
                  {m_b^2\,(s-q^2) })\,
                \log (\frac{s-q^2}{m_b^2}) \\&\,&+&\, 
               2\,( 2\,m_b^2 - 3\,q^2 + s ) \,
                {\log (\frac{s-q^2}{m_b^2})}^2 + 
               2\,\log (\frac{m_b^2 - q^2}{s-q^2})\,
                ( -3\,m_b^2 + 6\,q^2 + 
                  ( 2\,m_b^2 - 3\,q^2 + {s} ) \,
                   \log (\frac{s-q^2}{m_b^2}) )  )  \\&\,&+&\, 
            2\,(  m_b^8 \,
                ( -120\,q^2 + 170\,s )  + 
               20\,m_b^6\,
                ( 9\,q^4 + 24\,q^2\,s - 10\,s^2 )  - 
               6\,m_b^4\,( 12\,q^6 + 148\,q^4\,s + 
                  123\,q^2\,s^2 - 3\,s^3 )  \\&\,&-&\, 
               2\,q^2\,s\,( 3\,q^6 + 98\,q^4\,s + 195\,q^2\,s^2 + 
                  9\,s^3 )  + 
               2\,m_b^2\,q^2\,
                ( 3\,q^6 + 140\,q^4\,s + 531\,q^2\,s^2 + 
                  216\,s^3 )  + 
               3\,s\,{(s-q^2)}^3\\&\,&\times&\,
                ( 5\,m_b^2 - 5\,q^2 + 3\,s ) \,
                \log (\frac{s-q^2}{m_b^2}) ) \,
             \log (\frac{s}{m_b^2}-1) - 
            6\,s\,( 10\,m_b^8 - 
               10\,m_b^6\,( 4\,q^2 + 3\,s )  + 
               2\,s\,( q^2 + s ) \\&\,&\times&\,
                ( 11\,q^4 - 2\,q^2\,s + s^2 )  + 
               10\,m_b^4\,
                ( 5\,q^4 + 8\,q^2\,s + 2\,s^2 )  - 
               m_b^2\,( 23\,q^6 + 65\,q^4\,s + 41\,q^2\,s^2 + 
                  s^3 )  ) \,
             {\log (\frac{s}{m_b^2}-1)}^2 \\&\,&+&\, 
            \log (\frac{s}{m_b^2})\,
             ( -6\,m_b^2\,q^8 + 
               {{q}}^4\,
                ( 120\,m_b^4 - 157\,m_b^2\,q^2 + 9\,q^4 )
 \,s + ( 60\,m_b^6 + 60\,m_b^4\,q^2 - 
                  243\,m_b^2\,q^4 + 205\,q^6 ) \,
                {{s}}^2 \\&\,&-&\, 
               3\,( 10\,m_b^4 + 43\,m_b^2\,q^2 - 121\,q^4 )
 \,s^3 + 3\,( 5\,m_b^2 + q^2 ) \,s^4 - 
               6\,{( q^2 - s ) }^3\,s\,
                ( m_b^2 + q^2 + s ) \,
                ( \log (\frac{m_b^2 - q^2}{s-q^2}) \\&\,&-&\, 
                  \log (\frac{( m_b^2 - q^2 ) \,s}
                    {m_b^2\,(s-q^2) }) - 
                  \log (\frac{s-q^2}{m_b^2}) )  + 
               6\,s\,( q^8 + 24\,q^6\,s + 30\,q^4\,s^2 + 
                  4\,q^2\,s^3 + s^4 - 
                  20\,m_b^6\,( 2\,q^2 + s )  \\&\,&+&\, 
                  20\,m_b^4\,
                   ( 3\,q^4 + 5\,q^2\,s + s^2 )  - 
                  m_b^2\,( 25\,q^6 + 93\,q^4\,s + 59\,q^2\,s^2 + 
                     3\,s^3 )  ) \,
                \log (\frac{s}{m_b^2}-1) )  )  + 
         36\,s\,( ( -10\, m_b^8 \\&\,&+&\, 
               10\,m_b^6\,s + 
               10\,m_b^4\,q^2\,( q^2 + 2\,s )  + 
               4\,q^2\,s\,( q^4 + 3\,q^2\,s + s^2 )  - 
               m_b^2\,( 3\,q^6 + 25\,q^4\,s + 21\,q^2\,s^2 + 
                  s^3 )  ) \,
                \text{Li}_2(1 - \frac{q^2}{m_b^2}) \\&\,&-&\, 
            ( m_b^2 - 4\,q^2 ) \,
             {( q^2 - s ) }^3\,
             ( 2\,      \text{Li}_2(\frac{m_b^2 - s}{m_b^2 - q^2}) + 
                \text{Li}_2(\frac{m_b^2 - s}{q^2 - s}) - 
                \text{Li}_2(\frac{q^2\,( m_b^2 - s ) }
                 {m_b^2\,( q^2 - s ) }) )  - 
            ( 10\,m_b^8 - 5\,q^8 - 
               16\,q^6\,s \\&\,&-&\,  54\,q^4\,s^2 - 4\,q^2\,s^3 - s^4 + 
               10\,m_b^6\,( 4\,q^2 + s )  - 
               10\,m_b^4\,
                ( 7\,q^4 + 12\,q^2\,s + 2\,s^2 )  + 
               m_b^2\,( 29\,q^6 + 115\,q^4\,s + 83\,q^2\,s^2 \\&\,&+&\, 
                  3\,s^3 )  ) \,
                \text{Li}_2(1 - \frac{s}{m_b^2}) )  \big\}
\\&\rho_{\rm T4}^{2p}&\,=\;& \frac{f_{\pi}\,m_b\,\mu_{\pi}^2\,} {120\,
     {( q^2 - s ) }^7}
     \,( 27\,a_2\,
        ( 2880\,m_b^{10} - 
          25\,m_b^8\,
           ( 215\,q^2 + 229\,s )  + 
          20\,m_b^6\,( 149\,q^4 + 447\,q^2\,s + 184\,s^2 )
  \\&\,&-&\, 10\,m_b^4\,( 38\,q^6 + 352\,q^4\,s + 427\,q^2\,s^2 + 
             83\,s^3 )  +  
          5\,q^2\,( q^8 + 17\,q^6\,s + 42\,q^4\,s^2 + 22\,q^2\,s^3 + 
             2\,s^4 )  \\&\,&+&\, 
          m_b^2\,( -76\,q^8 + 84\,q^6\,s + 584\,q^4\,s^2 + 
             584\,q^2\,s^3 + 24\,s^4 )  )  + 
       5\,( 16128\,\eta_3\,m_b^{10}\,\omega_3 - 
          2520\,\eta_3\, m_b^8\,
           ( 15\,q^2 \\&\,&+&\, 13\,s ) \,\omega_3 + 
          8\,q^2\,( ( 3 + 90\,\eta_3 + 10\,\eta_4 ) \,
              {( q^2 - s ) }^2\,
              ( q^4 + 5\,q^2\,s + 2\,s^2 )  - 
             9\,\eta_3\,( q^8 + 17\,q^6\,s + 42\,q^4\,s^2
\end{alignat*}
\newpage
\begin{alignat*}{2} 
&\,&+&   22\,q^2\,s^3 + 2\,s^4 ) \,\omega_3 )  + 
          3\,m_b^2\,( 20\,q^4\,s^2\,
              ( 3 - 16\,\eta_4 + 
                144\,( \eta_3 + 2\,\eta_3\,\omega_3 )  )  + q^6\,s\,( 69 + 320\,\eta_4 + 
                240\,\eta_3\,( 6 + 19\,\omega_3 )  \\&\,&-&\, 
                420\,\eta_4\,\omega_4 )  + 
             3\,q^2\,s^3\,( -9 + 
                240\,\eta_3\,( -2 + 3\,\omega_3 )  - 
                140\,\eta_4\,\omega_4 )  + 
             5\,q^8\,( -15 - 16\,\eta_4 + 
                144\,\eta_3\,( -3 + \omega_3 )  + 
                84\,\eta_4\,\omega_4 )  \\&\,&+&\, 
             s^4\,( -27 + 80\,\eta_4 + 
                240\,\eta_3\,( -3 + \omega_3 )  + 
                420\,\eta_4\,\omega_4 )  )  + 
          5\,m_b^6\,( s^2\,
              ( -33 + 352\,\eta_4 + 
                144\,\eta_3\,( -8 + 33\,\omega_3 )  + 
                756\,\eta_4\,\omega_4 )  \\&\,&+&\, 
             q^4\,( -33 + 352\,\eta_4 + 
                144\,\eta_3\,( -8 + 47\,\omega_3 )  + 
                756\,\eta_4\,\omega_4 )  + 
             2\,q^2\,s\,( 33 + 
                576\,\eta_3\,( 2 + 11\,\omega_3 )  - 
                4\,\eta_4\,( 88 + 189\,\omega_4 )  )  )  \\&\,&+&\, 3\,m_b^4\,( -15\,q^2\,s^2\,
              ( 1 + 48\,( \eta_3 + 16\,\eta_3\,\omega_3 )  - 12\,\eta_4\,( 4 + 7\,\omega_4 )  )  - 
             3\,q^4\,s\,( 53 + 
                1680\,( \eta_3 + 3\,\eta_3\,\omega_3 )  - 
                20\,\eta_4\,( 4 + 21\,\omega_4 )  )  \\&\,&+&\, 
             q^6\,( 111 + 
                240\,\eta_3\,( 15 - 19\,\omega_3 )  - 
                20\,\eta_4\,( 20 + 63\,\omega_4 )  )  + 
             s^3\,( 240\,\eta_3\,( 9 - 10\,\omega_3 )  - 
                7\,( -9 + 
                   20\,\eta_4\,( 4 + 9\,\omega_4 )  )  )  )  )  \\&\,&+&\, 1080\,m_b^2\,
        ( m_b^2 - q^2 ) \,( m_b^2 - s ) \,
        ( 2\,m_b^2 - q^2 - s ) \,( q^2 - s ) \,
        ( 6\,a_2 - 35\,\eta_4\,\omega_4 ) \,
        ( -\log (\frac{m_b^2 - s}{q^2 - s}) + 
          \log (\frac{m_b^2 - q^2}{s-q^2}) )  ) 
\\&\rho_{\rm T4}^{3p}&\,=\;&\frac{- f_{\pi}\,m_b\,\mu_{\pi}^2} {24\,{( q^2 - s ) }^6}
       ( m_b^2 - q^2 ) \,
       ( 80\,\eta_4\,( q^2 - s ) \,
          ( -8\,m_b^4 + q^4 - 3\,q^2\,s - 6\,s^2 + 
            m_b^2\,( q^2 + 15\,s )  )  + 
         27\,a_2\,( 45\,m_b^6 + q^6 - 4\,q^4\,s \\&\,&-&\, 30\,q^2\,s^2 - 
            12\,s^3 - 5\,m_b^4\,( 7\,q^2 + 20\,s )  + 
            m_b^2\,( q^4 + 68\,q^2\,s + 66\,s^2 )  )  )   
\\&{\rm T4}_c&\,=\;& \frac{- f_{\pi}\,m_b^2\,m_b\,\mu_{\pi}^2} {160{( m_b^2 - q^2 ) }^2} e^{\frac{-m_b^2}{M^2}}\,     
     ( 12\,a_2 + 5\,( -25 + 
          48\,\eta_3\,( -10 + \omega_3 )  )  ) 
\end{alignat*}

{
}
\end{document}